\magnification=\magstep1
\baselineskip=18pt
\overfullrule=0pt

\font\first=cmr5 scaled\magstep0

\def\Section#1 #2{\S#1 {\rm\bf #2 }}
\long\def\Theorem#1{{\bf Theorem} {\it #1 }}
\long\def\Proposition#1{{\bf Proposition} {\it #1 }}
\long\def\Lemma#1{{\bf Lemma} {\it #1 }}
\long\def\Corollary#1{{\bf Corollary} {\it #1 }}

\def\Definition{{\rm\bf Definition}}
\def\Remark{{\rm\bf Remark}}
\def\Remarks{{\rm\bf Remarks}}
\def\Example {{\rm\bf Example}}
\def\Proof{{\rm\bf Proof}}
\def\Introduction{{\rm\bf Introduction}}
\def\pri{{\prime}}

\def\blsq{\hfill
          \hbox{\vrule height4pt width3pt depth2pt}}

\def\gb{\frak b}
\def\gg{\frak g}
\def\gh{\frak h}
\def\gp{\frak p}
\def\gq{\frak q}

\def\gr{\frak r}
\def\gs{\frak s}
\def\gt{{\underline {t}}}
\def\gu{\frak u}
\def\gU{\frak U}
\def\unV{\frak V}
\def\gv{{\underline {v}}}

\def\gW{\frak W}
\def\unW{\frak W }


\def\unU{\frak U}
\def\unGG{\frak g}
\def\unV{\frak V}
\def\unW{\frak W }
\def\un0{{\underline {0}}}

\font\caligraphic=cmsy10 at 10pt
\def\cal#1{{\caligraphic #1}}

\def\gg{\frak g}
\def\GG{\frak g}
\def\HH{{\hbox{\cal H}}}

\def\SS{{\hbox{\cal S}}}

\def\RR{{\hbox{\cal R}}}

\def\blsq{\hfill
          \hbox{\vrule height4pt width3pt depth2pt}}
\def\circA{{ {A}\kern-.40em{^{^{^\circ}}}}}
\def\circB{{ {B}\kern-.40em{^{^{^\circ}}}}}
\def\circa{{ {a}\kern-.40em{^{^\circ}}}}
\def\circe{{ {e}\kern-.40em{^{^\circ}}}}
\def\circE{{ {E}\kern-.40em{^{^\circ}}}}
\def\circF{{ {F}\kern-.40em{^{^\circ}}}}
\def\circf{{ {f}\kern-.42em{^{^\circ}}}}
\def\circG{{ {G}\kern-.45em{^{^{^\circ}}}}}
\def\circgg{{ {\gg }\kern-.40em{^{^\circ}}}}
\def\circK{{ {K}\kern-.45em{^{^{^\circ}}}}}
\def\circP{{ {P}\kern-.45em{^{^{^\circ}}}}}
\def\cfrakG{{ {\frak G}\kern-.60em{^{^{^\circ}}}}}

\def\reals{\Bbb R}
\def\ints{\Bbb Z}
\def\cp{{\mathrel > \joinrel \kern -.24em\triangleleft\, }}
\def\C{\Bbb C}
\def\ssubC{{\rm \kern.20em \vrule width.02em height.96ex depth-.05ex
            \kern-.20em C}}
\def\natnums{{{\rm I}\kern -.13em {\rm N}}}
\def\Q{{\rm \kern.24em \vrule width.02em height1.4ex depth-.05ex
          \kern-.26em Q}}

\font\tenmsb=msbm10
\font\sevenmsb=msbm7
\font\fivemsb=msbm5
\newfam\msbfam
\textfont\msbfam=\tenmsb
\scriptfont\msbfam=\sevenmsb
\scriptscriptfont\msbfam=\fivemsb
\def\Bbb#1{\fam\msbfam\relax#1}
\font\teneufm=eufm10
\font\seveneufm=eufm7
\font\fiveeufm=eufm5
\newfam\eufmfam
\textfont\eufmfam=\teneufm
\scriptfont\eufmfam=\seveneufm
\scriptscriptfont\eufmfam=\fiveeufm
\def\frak#1{{\fam\eufmfam\relax#1}}

\font\eightrm=cmr8
\font\eightbf=cmbx8
\font\eightit=cmti8
\font\eightsl=cmsl8
\font\eightmus=cmmi8
\def\smalltype{\let\rm=\eightrm \let\bf=\eightbf \let\it=\eightit 
\let\sl=\eightsl \let\mus=\eightmus \baselineskip=9.5pt minus .75pt
\rm}

\bigskip

\centerline {\bf Systems of PDEs obtained from factorization in loop groups}
\bigskip\bigskip
\centerline {{\bf J. Dorfmeister} \footnote{$^1$}{\smalltype Work supported in part by
the Deutsche Forschungsgemeinschaft and NSF Grant DMS-9205293.}}
\centerline{\it Department of Mathematics}
\centerline{\it University of Kansas} 
\centerline{\it Lawrence, KS 66045-2142, U.S.A.}
\centerline{\bf  H. Gradl}
\centerline{\it Mathematisches Institut}
\centerline{\it TU M$\ddot{u}$nchen} 
\centerline{\it D-80333 M$\ddot{u}$nchen, Germany}
\centerline{{\bf J. Szmigielski} \footnote{$^2$}{ \smalltype Work supported in part by the Natural Sciences and Engineering Research Council of Canada.}}
\centerline{\it Department of Mathematics and Statistics}
\centerline{\it University of Saskatchewan} 
\centerline{\it Saskatoon, SK S7N OWO, Canada}
\bigskip
\baselineskip=12truept

\centerline{{\bf ABSTRACT}} 

{\narrower\smallskip We propose a generalization of a Drinfeld-Sokolov scheme
of attaching integrable systems of PDEs to affine Kac-Moody algebras.  With
every affine Kac-Moody algebra $\gg$ and a parabolic subalgebra $\gp$, we 
associate two hierarchies of PDEs.  One, called positive, is a generalization
of the KdV hierarchy, the other, called negative, generalizes the Toda hierarchy.  We prove a coordinatization theorem, which establishes that the number of functions needed to express all PDEs of the the total hierarchy equals the rank of $\gg$.  The choice of functions, however, is shown to depend in a 
noncanonical way on $\gp$.  We employ a version of the Birkhoff decomposition and a ``2-loop'' formulation which allows us to incorporate geometrically 
meaningful solutions to those hierarchies.  We illustrate our formalism for 
positive hierarchies with a generalization of the Boussinesq system and for the negative hierarchies with the stationary Bogoyavlenskii equation.  }
\bigskip
\noindent {\bf Table of Contents}:
\medskip

\item{\S 0. }  Introduction \medskip

\item{\S 1. }  Banach loop groups and algebras \smallskip

\item\item{1.1 }  Weights, Wiener algebras \smallskip

\item\item{1.2 }  Affine Lie algebras, Banach loop groups and algebras
\smallskip

\item\item{1.3 }  Basics of Kac-Moody algebras \bigskip

\item{\S 2. }  Borel and parabolic subgroups and subalgebras  \smallskip

\item\item{2.1 }  Borel subalgebras \smallskip

\item\item{2.2 }  Parabolic subalgebras and their natural complements \smallskip

\item\item{2.3 }  $Q P$ open and dense, $P\cap Q$ trivial \smallskip

\item\item{2.4 }  Gradings \bigskip

\item{\S 3. }  Birkhoff and Bruhat decompositions \smallskip

\item\item{3.1 }  $U_+$ and $U_-$ - infinite unipotent groups \smallskip

\item\item{3.2 }   Birkhoff and Bruhat decompositions for $G^{fin}$ and standard Borel subgroups\smallskip

\item\item{3.3 }   Birkhoff decompositions for $G_w$ \smallskip

\item\item{3.4 }  Birkhoff decompositions for $G_w$ \smallskip

\item\item{3.5 }  Birkhoff decompositions for $G_w$ - density of $Q P$ \smallskip

\item\item{3.6 }  ``$e^{tE} \in G_{-} G_{+}$'' - meromorphicity of factorization  Theorem 
\smallskip

\item\item{3.7 } Abelian flows on $G_w$
\bigskip

\item{\S 4. }  More on Banach Lie groups and subgroups \smallskip

\item\item{4.1 } Definitions of  $G^R, G^r, \GG ^R, \GG^r, \HH, \HH_+, \HH_-$ 
\smallskip

\item\item{4.2 } Basic facts about $\HH$.
\bigskip
\item{\S 5. } Factorization  \smallskip

\item\item{5.1 } Heisenberg subalgebra, cyclic element \smallskip

\item\item{5.2 } Abelian action on $\HH$ \smallskip

\item\item{5.3 } Definitions of potentials $\Omega _j$ and $ZCC$. \bigskip

\item{\S 6. }  Systems of PDEs obtained from factorization \smallskip

\item\item{6.1 } Formulas for $\Omega _j$ \smallskip

\item\item{6.2 } ``$\partial _j e^he^{-h}$ '' formula \smallskip

\item\item{6.3 }  Positive potentials are differential polynomials in
$\Omega _1$. \smallskip

\item\item{6.4 }  Various results on dimensions of $\gW$, $ \tilde{\gW }$, $\gU$,
$\gb$. \smallskip

\item\item{6.5 }  $\Omega _1$ is a $\partial _1$ - differential polynomial
of $\dim \GG _0$ ``basic
functions''.
\item\item{6.6} Examples-cousins of the Boussinesq system
\bigskip

\item{\S 7. }  Negative Potentials 
\smallskip

\item\item{7.1 } All potentials are differential polynomials in ``$a$''
\smallskip

\item\item{7.2 } $\partial _1\Omega _{-j}$ is a $\partial _1 - \partial _j
$ - differential
polynomials in $\Omega _1$
\smallskip

\item\item{7.3 } $ \ker ad\, E \cap \gq^{(-1)} = \{ 0\} \Rightarrow \Omega
_1$ determines $\Omega
_{-1}$ \smallskip
\item\item{7.4} An example of a negative potential-stationary Bogoyavlenskii equation \bigskip

\item\item{Appendix A} Proofs of Propositions (3.3.1) and (3.5.2)
\smallskip
\item\item{Appendix B}  Injectivity of $ad\, E |_{\gg^{(0)}}$
\bigskip
\item{References}

\vfil\eject

\noindent

\Introduction

In recent years loop groups have been successfully used in the investigation of geometric objects, like
surfaces  of constant mean curvature in ${\Bbb R}^3$, harmonic maps into compact
symmetric spaces, isometric immersions from space forms into space forms. 
At the heart of all these uses is the construction of solutions to certain nonlinear partial
differential equations as compatibility conditions of a system of matrix
equations.  In this context it is an outgrowth of soliton theory. However,
the geometric applications also yield some new features. While the
classical uses of loop groups for finding solutions to certain nonlinear
partial differential equations only use ``positive flows'', the geometric
applications require also ``negative flows''. An application of this yields 
an extension of the potential KDV hierarchy by an equation investigated
 by Bogoyavlenskii [1] and also by Hirota-Satsuma [2].  It is therefore natural to
extend to negative flows what is classical and well established for
positive flows.  In this paper we present such an extension.
 The general setting is as follows: Let $G$ be a Banach Lie  loop group, 
i.e. a certain group of maps
from the unit circle $S^1$ into $G\ell(n,{\Bbb C})$.  We also consider 
two subgroups
$G^+$ and $G^-$ of $G$ and assume that $G^-G^+$ is open and dense in $G$. 
Finally we
consider an abelian subgroup of $G$ given by  $\exp\{ \dots tE_{-1} + xE_{1} \dots\}$. For the
groups considered in this paper we prove that for $h$ in $G$ and all
sufficiently small $x \ne 0$ we have
$x.h= \exp\{x E_1\}h$ is in $G^-G^+$ . If we denote by $h_-$ the part of $x.h$ which is in $G^-$ then we can split 
$ t.h_- =\exp\{t E_{-1}\}h_-=\ g^-g^+$.
Differentiating both sides one obtains a system of matrix equations for $g^-$
and $g^+$. The compatibility conditions for this system of equations then yield the (system) of scalar
nonlinear partial differential equations we are primarily interested in.
This scheme is well known to practitioners of soliton theory.  Yet, by admitting a large class of choices of the subgroups $G ^+ $ and $G^-$, and a large abelian group we arrive at a framework unifying
many of the known formulations into one theory. 
 
This paper contains a reformulation of the theory due mainly to Drinfeld
and Sokolov [3] (see also Wilson's paper [4]), relating affine
Kac-Moody algebras to certain classes of non-linear PDEs.  The
starting point for our reformulation is the idea of factorization
outlined above.  This point of view is not originally discussed in
[3], even though it figures prominently in places where
representation theory of affine Kac-Moody algebras is actually used to
generate solutions to soliton equations.  Using this reformulation we
propose an extension of the formalism to include new hierarchies of
equations.  The basic result in [3] is that, given the Dynkin diagram
of an affine algebra, one can associate to it a hierarchy of
non-linear PDEs.  This hierarchy is called the generalized modified
KdV hierarchy.  It is then pointed out that another hierarchy is
attached to the Dynkin diagram with all points except for one removed,
and this is what is called the Gelfand-Dikii hierarchy.  One can
reinterpret the way two hierarchies are attached to the root system by
saying that both hierarchies share the same Dynkin diagram, however,
there is another piece of information setting apart those two
hierarchies, namely, in the former case one chooses a Borel subalgebra
as an additional data (no points removed from the Dynkin diagram),
whereas in the latter case it is a maximal parabolic subalgebra that
plays that role (all except for one points removed).  The other
observation of Drinfeld and Sokolov is that they point out that also
the sine-Gordon equation can be attached to the full diagram of
$A_1^{(1)}$, so in some sense this equation is in the same category as
the modified KdV equation.  This generalizes to the Toda equations for
other affine algebras.  So one can loosely say that when no points are
removed from the Dynkin diagram, there are two natural hierarchies of
PDEs to consider, namely the generalized modified KdV and the Toda
hierarchy.  On the other hand there does not seem to have been known
if for the Dynkin diagrams with a removed node one has pairs of PDE
hierarchies.

We have extended the original picture of [3] in the following way.  We
consider a pair $(\cal S,\gp)$ consisting of a Dynkin diagram $\cal S$
of an affine Kac-Moody Lie algebra and an arbitrary parabolic $\gp$
related to the same graph.  With $(\cal S,\gp)$ we associate two
hierarchies of differential equations in two different matrix
variables, called $\Omega _1$ and $\Omega_{-1}$.  They generate what
we call the positive and the negative hierarchies respectively. Each
$\Omega$ appearing in the sequel generates a one parameter flow, and
all these flows commute.  The assignment of flow variables is that
$t_{-1}\in {\Bbb C}$ is assigned to $\Omega _{-1}$, $x\in {\Bbb C}$ is
assigned to $\Omega _1$, in general, $t_{j}\in {\Bbb C} $ is the flow
variable for $\Omega _j, \, j \in {\Bbb Z}$.  $\Omega _1$ and
$\Omega_{-1}$ parametrize in the sense explained in the course of the
paper all other flows.  These two form a natural generalization of the
pair giving the modified KdV and the sine-Gordon equation.  The new
feature here is that in the case when at least one point is removed
from the Dynkin diagram one can in fact relate the two in a
differential fashion, that is, $\Omega_{-1}$ can be expressed in terms
of derivatives of elements of $\Omega _1$.  This rather surprising
fact we prove for all nontwisted affine algebras and all choices of
points of their Dynkin diagrams except for those cases specified in
Proposition (7.3.2) and Theorem (7.3.3).  This result has interesting
ramifications even for the well known case of the (potential) KdV
equation, which comes from $A_1^{(1)}$ and a maximal parabolic
subalgebra (one point removed, as the diagram of $A_1^{(1)}$ has only
two nodes).  It turns out that as a result of the differential
dependence of $\Omega _{-1}$ on $\Omega _1$ the KdV variable
$v=v(t_{-1},x,t_3)$, which parametrizes $\Omega _1$, satisfies another
differential equation with two independent variables, namely the
stationary Bogoyavlenskii equation [1]. This equation is not an
evolution equation, in complete analogy to the case of the modified
KdV and the sine-Gordon equation.

We would like to point out that in our approach 
we consistently use group factorizations to both generate equations as
well as to provide, in principle, solutions to these equations.  This
should be contrasted with the approach of Drinfeld and Sokolov who use
formal pseudodifferential operators to formulate the Gelfand-Dikii
hierarchy and its modified version, the modified Gelfand-Dikii hierarchy.
  Our approach is therefore somewhat more in a spirit of
classical Lie theory. It is not transparent at all, however, how to
bring out a hamiltonian aspect of theory, something quite prominent
in [3].  On the other hand, we arrive naturally at numerous ``cousins''
of the Gelfand-Dikii hierarchy.

Our approach originates from the use  of Kac-Moody algebras and groups and
Grassmann like manifolds for the description of solutions to certain
nonlinear partial differential equations as it was pioneered by Sato [5] and
Segal-Wilson [6].
It is quite natural to work in this context with "full" affine Kac-Moody
algebras , i.e. central extensions of loop algebras, additionally augmented
by a degree derivation . In a completely algebraic context, the
corresponding groups have been defined and investigated by
Garland [7], Peterson-Kac [8] and Tits [9].  However, 
for a description of solutions to differential
equations a "completion" of these groups and Lie algebras is
more than just a matter of aesthetics.  
Indeed, it is well known by now that for those algebraic groups
one obtains a sector of rational solutions only. The finite gap solutions 
on the other hand require at least a type of completion we are considering.
  This happens despite the fact that, with some extra work (see [10, chap14]),
one can include in the algebraic approach based on those ``thin groups'' 
special finite gap solutions, namely solitons.  
Since the transition from problems involving PDEs, even if they are geometric in character, is not sufficiently  refined yet to tell us the most convenient 
topology in which to work, we have chosen to work with 
a Banach topology.   We have therefore been interested in Banach structures on
Kac-Moody Lie algebras and the corresponding groups. Here one considers
first loop algebras, i.e. on considers the derived algebra of an affine
Kac-Moody algebra modulo its center. Following Goodman-Wallach [11] we 
obtain a
large class of Banach structures that allow us to complete loop algebras
and to obtain this way Banach Lie algebras. It is not difficult to find the 
associated Banach Lie groups.
 For our purposes we need a few additional specific features of the groups
used. To prove those we rely mostly on the
work of Goodman-Wallach [11] and Pressley-Segal [12]. Therefore we restrict our
attention throughout this paper to nontwisted  affine Kac-Moody algebras.
We would hope that eventually the results of this paper will be extended to
twisted affine Kac-Moody algebras, perhaps to even more general Kac-Moody
algebras. But it would be equally interesting to carry out the
investigations of this paper with different Banach structures: the Banach
structures used in this paper are all related with solution spaces
containing only meromorphic solutions. It was shown, however, in [13]
that by considering completely different function spaces, like the Fourier
transforms of functions from $L^1({\Bbb R})$, one can obtain solution spaces consisting  of 
$L^1$
functions only.

In addition to using positive as well as negative flows and to using
Banach structures we extend the setting of [3] in a third aspect: it
turns out that one needs to use not only Kac-Moody algebras and
groups, but one also needs to consider double loop algebras and
groups. This was first noticed in [14].  There, an effort was made to
describe the standard "completely integrable" nonlinear partial
differential equations in terms of the loop group/Grassmannian
picture. It turned out that for the sine-Gordon equation one is
naturally led to consider double loop groups and natural "positive"
and "negative" subgroups . This was enhanced by [15]: while
investigating constant mean curvature tori in ${\Bbb R}^3$,
i.e. special (real) solutions to the sinh-Gordon equation, it was
observed that not only should one consider double loop groups $G
\times G$, but one should even consider $G^r \times G^R$, where
$0<r<1<R<\infty$ and $G^r$ and $G^R$ are defined on a circle of radius
$r$ and $R$ respectively. The reason for this is that it is impossible
to construct the solutions describing constant mean curvature tori in
the "1-loop" setting by the procedure above [16, Theorem 2.7].  Yet,
it is possible to construct such solutions in the "2-loop"
setting. For more details on this see [15] and [17].  We should
perhaps add that the term used there is an r-loop approach rather than
a ``2-loop'' approach used in the present paper.

Interestingly enough the hierarchies of PDEs we are considering can be
used to obtain solutions to the self dual Yang-Mills equations [18].
From that perspective the theory we are putting forward comprises a
part of the theory of the self dual Yang-Mills equations.

Here is a short description of the paper. In \S 1 and 2 we present
some elementary facts about affine Kac-Moody algebras, their
completions as well as basic properties of parabolic algebras and
groups.  The latter topic is further developed in \S 3.  In particular
in that chapter we prove Theorem (3.5.1), Corollary (3.5.3) and
Theorem (3.6.1), crucial for the whole paper. In \S 4 we define our
``2-loop'' group setting. \S 5 contains a description of the Zero
Curvature formulation of the systems of PDEs corresponding to our
``2-loop'' formulation.  This is further developed in \S 6, where in
particular we show, that all systems of PDEs appearing in the positive
hierarchy depend on dim$\gg_0$ basic functions.  This result is proven
in Theorem (6.5.1).  \S 6 ends with examples illustrating this part of
the theory. The negative hierarchy is studied in \S 7.  The main
result here is Proposition (7.3.2). We give an example of the first
flow in the negative hierarchy.  This turns out to be an equation
studied in [1] by Bogoyavlenskii.  The appendices contain omitted
proofs of two propositions from \S 3 and some results, used in the
paper, regarding the map $ad\, E$.

There are some very interesting and important questions that have not
been addressed in this paper. One is the question of hamiltonicity of
the solution spaces and to what extent the solution spaces, as Banach
manifolds, are completely integrable in some rigorous sense. We feel
that it is possibly advantageous to restrict the above questions to
the dressing orbits on the solution spaces.  This way one is perhaps
able to make contact with the AKS (Adler-Kostant-Symes) method used by
many authors for the construction of solutions to completely
integrable nonlinear partial differential equations.  Another question
is to what extent one can find Miura like transformations from the
solution space relative to one parabolic $\gp_1$ to the solution space
of another parabolic $\gp_2$.  We would like to point out that such
Miura maps on the level of our solution spaces may not exist.  As an
example, in [19], it has been shown that there does not exist a Miura
map from the solution space to the MKdV equation to the solution space
to the potential KdV equation.  Related with this is the question of
how to define and to describe B$\ddot{a}$cklund transformations for
the equations considered in this paper. From their origin,
B$\ddot{a}$cklund transformations should be diffeomorphisms of the
solutions spaces or at least maps from solution spaces to solution
spaces. The actual use of B$\ddot{a}$cklund transformations, however,
seems to be different.  It will be very interesting to pursue the
above questions in detail.

Finally, the referee has kindly pointed out to us that there already exists a 
generalization of the Drinfeld-Sokolov systems [20].  
The Hamiltonian theory is discussed in [21] and [22].
The generalization we propose for positive flows is formally included in 
the other generalization.  However, our way of parametrizing the occurring 
potential $\Omega _1$ is quite different from that of other authors.  
We illustrate the difference on the example of the potential KdV in 
\S 6.  We also believe that the concept of the negative hierarchies 
for cases other than the minimal parabolic case (the Toda equations) is  a 
key new element of our perspective, and this sets apart our approach from 
that of [20].

\vfill
\eject
{\bf\S 1.  Banach loop groups and algebras}
\medskip

In this section we will define the main objects of the entire paper:  loop
groups and algebras.
Most of the definitions and results are taken from [11], but they are listed
here for the convenience of the reader, and to fix notation.  
\bigskip

\noindent
{\bf 1.1 }\quad A function $w:{\ints} \to (0,\infty )$ is called a
\underbar{weight}, if $w(k + l)
\le w(k) w(l)$ for all $k,l\in \ints $.  A weight $w$ is called
\underbar{symmetric}, if $w(-k) =
w(k)$ for all $k\in \ints $.  Two classes of examples are:

$\bullet $ symmetric exponential-polynomial weights:
$$
w_{a, t} (k) = (1 + \vert k\vert )^a \cdot e^{t\vert k\vert}\quad\hbox
{for}\quad a,t\ge 0\ .
$$
\bigskip

$\bullet $ ``Gevrey class''
$$
w_{t,s} (k) = \exp (t\cdot \vert k\vert ^s)\quad\hbox{for}\quad t > 0,\ 0 <
s< 1\ .
$$
\medskip

For a symmetric weight $w$ define
\medskip

$$
A_w:\ = \{ f:S^1 \to {\C},
\lambda \mapsto \sum\limits _{n\in\ints } a_n\lambda ^n
,\Vert f\Vert
_w < \infty \} \leqno (1.1.1)
$$
where
$$
\Vert f\Vert _w = \sum _{n\in \ints } \vert a_n\vert \cdot w(n)\ .
$$

One easily verifies that $\Vert\cdot \Vert _w$ is a norm, thus $A_w$ is a
commutative
Banach $*$-algebra, with pointwise multiplication, the $*$-operation being
complex conjugation.

We call $A_w$ the \underbar{weighted Wiener algebra} associated with the weight $w$.  In this paper we will use exclusively symmetric weights of non-analytic type, i.e. satisfying $\lim _{n \to \infty} w(n)^{1 \over n}=1$.
We note that all Gevrey class weights are of non-analytic type, while in the first example only weights with $t=0$ are non-analytic.
\bigskip

\noindent
{\bf 1.2 }\quad Now let ${\buildrel\circ\over\GG }$ be a simple
finite-dimensional complex Lie
algebra of type\hfil\break
$X_l$ (i.e. $X\in\{A,B,C,D,E,F,G\}$).  Let $\psi $ be an automorphism of
the Dynkin
diagram of order $k$.  Then, $\psi $ can be extended to an
automorphism of ${\buildrel\circ\over\GG }$.  Its order is $k$, too.  We will also call it $\psi $.
\medskip

Define
$$
\GG ^{fin } : =\left \{ 
\eqalign{  x&:
\lambda \mapsto \sum\limits _{-m\le j\le n} \lambda ^jA_j:
m,n\in\natnums _0,\,
A_j\in {\buildrel\circ\over\GG }\cr
&\hbox{such that}\qquad x(\lambda \cdot e^{2\pi i/k}) = \psi \big(
x(\lambda )\big)
\quad\hbox{for all}\quad \lambda\in S^1 \cr }\right \}.
\leqno (1.2.1)
$$
This Lie algebra $\GG^{fin}$ is called the {\it affine Lie algebra
of type}
${X^{(k)}_l}$.
It differs from the affine Kac-Moody algebra of type $X^{k}_l$ by a
one-dimensional
center.  Nevertheless we will also use the latter name for $\GG^{fin}$.  Now let
${\buildrel\circ\over G }$ be the connected and simply-connected Lie group such
that Lie
${\buildrel\circ\over G } = {\buildrel\circ\over\GG }$.  We may assume that
${\buildrel\circ\over G }$ is a subgroup of a suitable $ SL_n (\C ),\
{\buildrel\circ\over\GG }$ a
subalgebra of $ s\ell _n (\C )$.  Now recall from general Lie group theory
that any Lie algebra
automorphism can be ``exponentiated'' to a simply-connected Lie group ${\buildrel\circ\over G }$ in a
unique way.  Therefore, there is a unique automorphism $\phi $ of
${\buildrel\circ\over G }$ such
that $ d\phi (e) = \psi$.  Obviously, $\phi $ is of order $k$, too.
\medskip

Thus we may define:
\medskip

$$
G_w :\ = \left \{ \eqalign{ & g\in SL_n (A_w):g (\lambda )\in{\buildrel\circ\over G
}\quad\hbox{and}\cr
&g (\lambda e^{2\pi i/k}) = \phi \big( g(\lambda )\big) \quad\hbox{for
all}\ \ \lambda \in S^1} \right \}
 \leqno (1.2.2)$$
\medskip

$$
\GG _w : =\left\{\eqalign{ &x \in s\ell _n (A_w) : x(\lambda ) \in
{\buildrel\circ\over\GG
}\quad\hbox{and}\cr
&x (\lambda \cdot e^{2\pi i/k}) = \psi \big( x(\lambda )\big)\quad\hbox{for
all}\ \
\lambda \in S^1} \right \} . \leqno (1.2.3)
$$
\medskip

For these objects, Goodman and Wallach prove (essentially in [11; 5.1] and [12;
5.5, 6.8, 6.9]):
\medskip
\item{1. } $G_w$ is a complex Lie subgroup of $SL_n(A_w), \GG _w$
is complex Lie
subalgebra of\hfil\break
$sl_n (A_w)$.
\medskip

\item{2. } Lie $G_w = \GG _w$.
\medskip

\item{3. } $\GG _w$ is a Banach Lie algebra, the completion of
$\GG ^{fin}$ w.r.t. to
the norm defined by the symmetric weight $w$.
\medskip

\item{4. } $G_w$ is connected and simply-connected (Lemma 5.5, [11]).
\medskip

Therefore we will call $G_w$ resp. $\GG _w$ the {\it Banach loop group} resp. {\it loop algebra of type} $X^{(k)}_l$\quad (w.r.t. the weight $w$).
\bigskip

\noindent
\Remark :

\item{1. }  In [11], $G_w$ is denoted by $\widetilde{G_w}\big(\GG _w\ \hbox{by}\
[\tilde{\GG}]_w\big)$. \medskip

\item{2. }  Goodman and Wallach discuss in detail the case where $\psi $ is the
identity,  i.e. the
type $X^{(n)}_l$ (also called ``non-twisted case'', cf. [10]).  However in the 
last two sections of chapter 6 they indicate how to generalize 
the results cited above to arbitrary affine algebras.
\medskip

\item{3. } In the following we will write $G$ resp. $\GG $ instead of $\GG
_w$ and $G_w$ when there
is no ambiguity.
\bigskip

\noindent
{\bf 1.3 }\quad In this section we collect some properties of the Kac-Moody Lie algebra $\GG
^{fin}$ [10].
\medskip

\item {1. }\quad There exists a system of ``Chevalley generators'' $\{ e_i
, f_i, h_i: i = 0, \cdots l \}$, i.e.

\item\item{} $\bullet\quad e_i, f_i, h_i$ generate $ \GG^{fin} $ as a Lie
algebra

\item\item{} $\bullet\quad [h_i,h_j] = 0$

\item\item{} $\bullet\quad [e_i, f_j] = \delta _{ij}h_i$

\item\item{} $\bullet\quad [h_i, e_j] = a_{ij} e_j$

\item\item{} $\bullet\quad [h_i, f_j] = -a_{ij} f_j$

\item\item{} $\bullet\quad \hbox {Moreover, for all $i,j=0\cdots ,l,i\ne j$ we
have} (ad e_i)^{1-a_{ij}} e_j = 0,\ (adf_i)^{1-a_{ij}} f^i = 0$.
\noindent where $A = (a_{ij})$ is a generalized Cartan
matrix, i.e.

\item\item{} $\bullet\quad a_{ii} = 2$  for all  $i = 0, \cdots , l $

\item\item{} $\bullet\quad a_{ij} \le 0 $  for all  $i\not= j$

\item\item{} $\bullet\quad a_{ij} = 0\quad \Leftrightarrow\quad  a_{ji} = 0. $
\item {} For $ \gg ^{fin} $, it is known (see e.g. [10, $\S$ 6.1 ])
that $A$ is symmetrizable, that is, there
exists an invertible
diagonal matrix $D$ such that $DA$ is a symmetric matrix.
\medskip

\item {2. }\quad Let $\GG ^{fin}_-$ resp. $\GG^{fin}_0 $ resp. $\GG
^{fin}_+$ be the
subalgebras generated by the $f$'s resp. $h$'s resp. $e$'s.

Then there is a natural triangular decomposition
$$
\GG^{fin } = \GG ^{fin}_- \oplus \GG^{fin}_0 \oplus\GG ^{fin }_+\ .
$$
\medskip
\item {3.}\quad Let $\triangle = \triangle _- \cup \triangle _+$ be a
root system of $\GG
^{fin}, \triangle _{\pm }$ denoting the set of positive resp. negative roots
associated with $\gg _0 ^{fin}$ and $\gg _{\pm}^{fin}$.  Then
$ \GG ^{fin
}_{\pm } = \bigoplus\limits _{\alpha\in\triangle_{\pm}} \GG ^{fin}_{\alpha
}$. \medskip

\item {4.}\quad Denote by $ \Pi = \{ \alpha _0,\cdots , \alpha _l\}$ a
set of simple roots
corresponding to $\triangle $.
\medskip
\item {5.}\quad Finally by $G^{fin}$ we denote the  group generated by $\{\exp\,x_{\alpha}\}$, where $x_{\alpha}\in \gg _{\alpha}^{fin}$, $\alpha \in \triangle^{re}$. The subset of real roots $\triangle ^{re}$ is defined and  studied in 
[10, $\S 5$]. 
\vfil
\eject

\noindent
{\bf\S 2.  Borel and parabolic subgroups and subalgebras }
\medskip
This section contains a collection of the results about
Borel and - more generally - parabolic
subgroups and algebras used later in the paper. For more details on the
basic theory of these subalgebras see
[23, Ch. 8, Section 3.4].
\medskip

\noindent
{\bf 2.1}\quad We start with

\noindent
\Theorem {{\bf (2.1.1).}\quad ([8, Theorem 3]).\quad Every Borel
subalgebra of $\gg
^{fin}$ is $ Ad(G^{fin })$ - conjugate to $\gg^{fin} _0\oplus \gg^{fin}_+$
or $ \gg ^{fin}_0\oplus
\gg ^{fin}_-$.
}\medskip

\noindent
\Remark:

\item{(a) } A subalgebra $\gb $ of a Lie algebra $\gg $ is called a Borel
subalgebra if it is
maximal completely solvable.
\medskip

\item{(b) } A subalgebra $\gb $ of a Lie algebra $\gg $ is called
completely solvable in $\gg$ if
there is a flag
$$
\cdots\supset \gb _{-1} \supset\gb _0 \supset \gb _{+1} \supset \gb _{2}
\supset \cdots
$$
of $ ad(\gb )$-invariant subspaces of $\gg $ such that
\medskip

\item\item{} $\bullet\quad \gg = \bigcup\limits _{i\in\ints} \gb _i$
\medskip

\item\item{} $\bullet\quad \bigcap\limits _{i\in\ints} \gb _i = \{ 0 \}$
\medskip

\item\item{} $\bullet\quad \gb = \gb _0$
\medskip

\item\item{} $\bullet\quad \dim {\gb _i\over \gb _{i+1}} \le 1$

\medskip

In view of the Theorem above we may restrict ourselves to 
the case of the
{\it standard Borel subalgebra} $\gb ^{fin} := \gg^{fin}_0 \oplus \gg
^{fin}_ + $, respectively,  the
{\it standard Borel subgroup} $B^{fin}$, the corresponding 
subgroup of
$G^{fin}$.  We define the {\it standard Borel subalgebra} $\gb _w$ of
the Banach loop
algebra $\gg _w$ as the completion of $\gb ^{fin}$ in $\gg _w$.  To obtain a similar statement on
the group level we
note that $\gb _w$
has a closed complement in $\gg _w$ (namely $(\gg _w)_-$).  Therefore, by 
([23], Ch. 3,
\S6, Theorem 2) there exists a connected Banach Lie group $B_w$ 
such that Lie $ B_w = \gb
_w$ and so that $B_w \subset G_w$ is an integral subgroup of $G_w$.
\medskip

\noindent
{\bf 2.2}\quad In general, a {\it parabolic subalgebra} of a Lie algebra is
a subalgebra
containing a Borel subalgebra.  By virtue of Section 2.1 we may as well restrict ourselves to
the following class:
\medskip

\noindent
\Definition {{\bf (2.2.1).}\quad A subalgebra $\gp ^{fin} \subset \gg ^{fin}$ is
called a
{\it standard-parabolic subalgebra (spsa) } if $\gg ^{fin}_0 \oplus \gg
^{fin}_+ \subseteq
\gp^{fin}$.  For a subset $X \subset \Pi $ (the set of simple roots)
let $\gp ^{fin}_X$ be the smallest {\it spsa} containing all root spaces  $\gg
^{fin}_{-\alpha}$ for $\alpha\in
X$.
}\medskip

\noindent
\Remark :\quad 
Similarly, we define $p_w=\overline{p_X^{fin}}$.
\medskip

\noindent
The following is well known [23, Ch. 8, Section 3.4, Ch. 4, Sect.2.6]:
\medskip

\noindent
\Proposition {{\bf (2.2.2).}

\item{(a) } Let $\gp^{fin}$ be a \underbar{spsa} of $\gg ^{fin}$.  Then there
is a subset $X\,\subset \Pi $ such that $\gp^{fin} = \gp^{fin}_X$.
\medskip

\item{(b) } Let $X\subset \Pi $ and $\tilde{\triangle }_+ = \{
\sum k_i\alpha _i\in
\triangle _+ :\alpha _i\in X\}$.  Then $\gp ^{fin}
_X = \bigoplus\limits
_{\alpha\in\tilde{\triangle }_+}\, \gg^{fin}_{-\alpha }\oplus
\gg^{fin}_0\oplus\gg ^{fin}_+$.
\medskip

\item{(c) } Let $ \gq ^{fin}_X : = \bigoplus\limits
_{\alpha\in\triangle
_+\backslash\tilde{\triangle }_+}\gg ^{fin}_{-\alpha }$.  Then $\gq^{fin}
_X$ is a
subalgebra of $\gg ^{fin}$, and $\gg ^{fin} = \gq^{fin}_X \oplus \gp
^{fin}_X$.
\medskip

We call $\gq^{fin}_X$ the \underbar{natural complement}
of $\gp _X ^{fin}$.  We define $\gq _w=\overline{\gq^{fin}}$.  Then $\gg_w=
\gq _w \oplus \gp _w$.

}\medskip

\noindent
We therefore have a 1 - 1 - correspondence between {\it spsa's} and subsets
of the set of simple
roots.  Let us now fix the subset $X$ and define $\gq _w=\overline {\gq
_X ^{fin}}$.  Then $\gq _w$ is a natural complement of $\gp _w$. Now we 
turn to formulating the corresponding facts on the group level.
\medskip
\noindent
\Corollary {{\bf (2.2.3).}

\item{(a) } For every \underbar{spsa} $\gp ^{fin}\,\subset\,\gg^{fin}$ and its
natural complement $ \gq^{fin}$
there are unique connected subgroups $ P ^{fin}$ and $ Q^{fin}\, \subset\,
G^{fin}$ where  $P ^{fin}$ is generated by 
$\exp(\gp ^{fin}_{\alpha}),\gp ^{fin}_{\alpha} \subset \gp_w$ and $Q^{fin }$ 
by $\exp(\gq^{fin}_{\alpha}),\gp ^{fin}_{\alpha} \subset \gq_w, \alpha \in
\triangle ^{re}$.
\item{(b) } For every \underbar{spsa} $\gp_w\,\subset\,\gg_w$ and its natural
complement $\gq_w$ there are
unique connected Banach Lie groups $ P_w$ and $ Q_w\,\subset\, G_w$ such
that Lie $P_w = \gp_w$, Lie
$Q_w = \gq_w$ and $P_w$ and $Q_w$ are integral subgroups of $G_w$.}
\medskip

\noindent
\Proof. \quad (a) holds automatically by construction, (b) again follows from the fact
that $ \gp_w$ and $ \gg_w$ are
closed complements of each other.\blsq
\medskip

\noindent
{\bf 2.3.}\quad For the sake of simplicity we drop the superscript $(\ )^{fin}$
in this section.

Let $\gg$ be an affine Kac-Moody algebra and $\{ e_i, f_i, h_i : i = 0,
\cdots , n\}$ be a set of
canonical generators.  The assignment
$$
cdeg\, e_i := 1,\ cdeg\, f_i := -1\leqno(2.3.1)
$$
for all $i$
defines a grading on
$\gg $; we refer to it as the {\it canonical grading}.  For $ x \in \gg$
we denote the
homogeneous component of $x$ of canonical degree $k$ by $x_k$ and write:
$cdeg\, x_k = k $.  Let $
\gg _k$ be the subspace of all homogeneous elements of canonical degree
$k$.  Thus:
$$\gg = \bigoplus\limits _{k\in\ints }\gg _k.\leqno(2.3.2)$$
Let \quad $\gg _- :=
\bigoplus\limits _{k < 0} \gg _k ,\ \gg _+ := \bigoplus\limits _{k>0} \gg
_k$,\quad then:
$$ \gg = \gg_-\oplus \gg _0 \oplus\gg_+. \leqno(2.3.3)$$
Note that $\gg _0$ is generated by $\{ h_i : i = 0, \cdots , n\}$ and thus
abelian.
\medskip

For a given standard parabolic subalgebra $\gp$ of $\gg$ we define the
$p${\it -grading} of $\gg$ by assigning
$$
pdeg\, f_i : = \cases{ 0, &if $f_i\in \gp$\cr
-1, &otherwise $\ $
}$$
$$
pdeg\, e_i : = -pdeg\, f_i\quad\hbox{for all}\ \ i\ .\leqno(2.3.4)
$$
For $ x \in \gg $ let $ x^{(k)}$ be the homogeneous component of $x$ of
$p$-degree $k$ and $\gg
^{(k)}$ the subspace of all such elements. In a similar manner as above:
$$
\gg = \bigoplus\limits _{k\in\ints } \gg ^{(k)}\ .\leqno(2.3.5)
$$
We define:$$
 \gg^{(-)} := \bigoplus\limits _{k<0} \gg ^{(k)},\qquad \gg^{(+)} :=
\bigoplus\limits_{k>0} \gg^{(k)}\ ,
$$
$$
\gg = \gg ^{(-)} \oplus\gg^{(0)} \oplus \gg^{(+)}. \leqno(2.3.6)$$
\noindent
Note that the canonical grading and the $p$-grading coincide iff $\gp$ is a minimal
parabolic subalgebra, i.e.
the Borel subalgebra $\gg _0 \oplus \gg _+$.

In general, $\gg^{(0)}$ is no longer abelian.  To understand better the 
structure of $\gg^{(0)}$ we observe that 
$\gg ^{(0)}$ is generated by $\{e_i,f_i:\alpha _i \in X\} \cup \{h_i:i=0,\cdots,n\}$.  The following lemma is well known in the finite dimensional setting.
\medskip

\noindent
\Lemma {{\bf (2.3.1).}\quad Let $\gp\not= \gg$ be a parabolic subalgebra of
$\gg$.  Then $\gg ^{(0)}$
is a \hfil\break
finite-dimensional reductive subalgebra of $\gg$.
}\medskip

\noindent
\Proof.\quad Let $ X \subset \Pi$ be such that $\gp =\gp_X$.
Since $\gp\not= \gg ,\
X \not= \Pi$.  To see that $ \gg^{(0)}$ is finite-dimensional,
denote by $ \tilde{\gg}$ the subalgebra of $\gg$ 
generated by the set $\{e_i,f_i:
\alpha _i \in X \}$.  Thus, the algebra $\tilde{\gg}$ corresponds to a certain subdiagram of the
Dynkin diagram of $\gg$.
By Lemma 4.4 in [10, Ch. 4] the Dynkin diagram of $\tilde{\gg}$ is a
disjoint union of
diagrams corresponding to simple finite-dimensional Lie algebras.
Thus $\tilde{\gg}$ is
finite-dimensional.  Moreover, $\gg^{(0)}/\tilde{\gg}$ is finite dimensional by the remark above.   Thus $\gg ^{(0)} $ is finite dimensional.  Since $ [\gg ^{(0)}, \gg^{(0)}] = \tilde{\gg },\ \gg
^{(0)}$ is reductive (cf. [23, Ch. 1, \S 6.4]).
\blsq \medskip

Although not obvious, the vector spaces $ \gg ^{(k)}$ are finite dimensional
for all $ k\in\ints$.
\medskip

\noindent
\Proposition {{\bf (2.3.2).}\quad Let $\gg$ be graded with respect to a spsa:
$\gg =
\bigoplus\limits _{k\in\ints} \gg^{(k)}$.\hfil\break
Then $ \dim \gg ^{(k)} <\infty $ for all $ k\in\ints $.
}\medskip

\noindent
\Proof. \quad See Corollary to Proposition 5.11 in [3].   \blsq
\medskip

\noindent
{\bf 2.4.}\quad The entire theory is based on the following facts.
\medskip

\noindent
\Theorem {{\bf (2.4.1).}
\item{(a) } $ Q_w P_w$ is open in $G_w$

\item{(b) }$ P_w\cap Q_w$ is trivial (i.e. the identity).
}\medskip

\noindent
\Proof.

\item{(a) } It is easy to see that the map

$$
\psi :\cases{ \gq_w + \gp_w &$\to G_w$\cr
q + p &$\mapsto (\exp q)(\exp p)$
}$$

\item{\ \ \ } is a  local diffeomorphism at the identity.  From this it
follows that $ Q_w  P_w$
is open in $G_w$.
\medskip

\item {(b) } Let $ g\in Q_w \cap P_w$.  
For every $x\in \gg$ of $pdeg =k$ we have: $max(pdeg(Ad(g)x-x))\leq k-1$ and $ min(pdeg(Ad(g)x-x))\geq k$ thus $Ad(g)=I$. 
We claim however that $Ad 
\vert Q_w$ is injective.  To see that, we consider $g\in Q_w$ such that $Ad(g)=I$.  Then we write $g=(\exp(a)){\hat g}$ where $pdeg(a)=m$ and ${\hat g}$ is 
generated by $\exp\, y$ with $max(pdeg(y))\le m-1$ for some negative $m$.  
 Now applying $Ad(g)$ to an arbitrary $p$-homogeneous element $x$ implies that 
$[x,a]=0$.  Thus $[\gg,a]=0$ and $a=0$ follows. This proves the claim and 
part (b).   \blsq
\medskip

\noindent
{\bf\S 3.\quad Birkhoff and Bruhat Decompositions}
\medskip
\noindent
{\bf 3.1}  We retain the notation of \S 2.  In particular,
let $ G_w = G$ and $\gg_w = \gg $ be as before and let $P=P_w$ be
a standard parabolic
subgroup $G$ with complement $Q$ and $ \gp = $ Lie $P,\ \gq = $Lie $Q$.
From Corollary (2.2.3) we know
that $ P $ and $Q$ are integral subgroups of the
connected Banach Lie group $G$.
By $ \gg^{fin}, \gp^{fin}, \gq^{fin}$ etc. we denote the finite
linear combinations of the
canonical generators, i.e. the Lie algebra in the sense of [10] or
[8].  Recall from \S 1 and \S 2 that in our Banach topology we have
$$\gg = \overline{\gg ^{fin}},\ \gp = \overline{\gp^{fin}},\
\gq =
\overline{\gq^{fin}}. \leqno (3.1.1)$$
We also recall from 1.3 that by $G^{fin},\ P^{fin},Q^{fin}$ etc. we denote the group generated by $\{
\exp\,x_{\alpha}\}$, where $\alpha \in \triangle ^{re}$ and $x_{\alpha } \in \gg _{\alpha}$, $\gg _{\alpha} \subset 
\gg^{fin}, \gp^{fin}$ and $\gq^{fin}$ respectively.  
Then 
$$G =
\overline{G^{fin}},\, P = \overline{P^{fin}}, Q =
\overline{Q^{fin}}.\leqno (3.1.2)$$
From [23; Ch. 4, no.2.6] we know $ P^{fin} =
P^{fin}_X$ for
some $X\subset\Pi = \{ \alpha _i\}$, where $ \Pi$ is a basis for the root
system $\triangle $ of
$\gg^{fin}$.  For details and notation we refer to [23; Ch. 4] and [8].

We set $ U_{\alpha } = \exp\, \gg _{\alpha }, \alpha \in \triangle ^{re}$.  Then $ U_{\alpha }\subset G^{fin}$ and $U_{\alpha }$ is
closed in $G$.  
Indeed, if
$$
\exp\, x_n \longrightarrow A\,\in G
$$
then $\exp\,x_n$ is a Cauchy sequence and thus
$$
\exp\, x_n (\exp\, x_m)^{-1} = \exp (x_n - x_m) \to I.
$$
We would like to conclude that $x_n$ is a Cauchy sequence in $\gg_{\alpha}$, which is closed. To this end we consider $r_{mn}=Ad\,exp(x_n-x_m)\,(h)=exp\,ad(x_n-x_m)\,(h)=h+[x_n-x_m,h]+\epsilon$ where $h \in \gh$. We know that $r_{nm}\to h$.  Moreover, $\epsilon$ when expanded in terms of canonical degree has no components of degree zero or $deg(x_n)=deg(x_m)$. This implies $[x_n-x_m,h]\to 0$.
 Consequently, $x_n-x_m \to 0$.  
Thus for some $x \in g_{\alpha}$ $x_n \to x$ 
implying that 
$\exp\, x_n \to \exp\, x = A.$

As in [8] we consider the subgroup $ U^{fin}_+$ of $G^{fin}$, the group
generated by $
U_{\alpha },\ \alpha \in\triangle ^{re}_+$, and similarly we define the 
subgroup $ U^{fin}_-$.
\medskip

\noindent
\Lemma{
 {\bf(3.1.1).}\quad $ U_+ = \overline{ U^{fin}_+}$ and $U_- =
\overline{U^{fin}_-}$ are connected
closed Banach Lie groups, which are integral subgroups of $G$ with Lie algebras $\gg_+$ and
$\gg_-$ respectively.  Here
$\gg^{fin}_+$ is generated by $\gg_{\alpha },\ \alpha\in\triangle ^{re}_+$
and $ \gg^{fin}_{\, -}$ is
generated by $\gg_{\alpha },\, \alpha\in\triangle _-^{re}$.}

\medskip

\noindent
\Proof .\quad Since $e_i,\, i = 0,\, \cdots ,\, l$ belongs to a real root,
we know that the vector spaces 
$\gg_{\,\alpha }, \alpha\in\triangle ^{re}_+$, generate $
\gg^{fin}_+$.  Hence the closure
generates $ \gg_+$.  Similarly we obtain $ \gg_{\, -}$.  From 1.3. we know
that $\gg_+$ and $\gg_{\,
-}$ are closed complemented subalgebras of $ \gg$.  Therefore, by [23;
Ch. 3, \S 6, Theorem
2], there are connected integral subgroups $ \hat U_{\pm}$ such that Lie $ \hat
U_{\pm} = \gg _{\pm}$.  Next
we show that $ \hat U_{\pm}$ is closed in $G$.  Let $ u_m \in \hat U_+,\ u_m \to
a \in G$.  Then we know that $
u^{-1}_m a$ is in an arbitrary small neighborhood of $I$, provided $ m $
is sufficiently 
large.  It is easy to see that $Ad(u_m) - I$ is an operator on $\gg
$ which maps $\frak r _k=\bigoplus \limits _{j\ge k} \gg _j$ into $\frak r _{k+1}$.
Therefore $Ad (a) - I$ has the same property.  But for $ u^{-1}_m
a$ in a sufficiently
small neighborhood of $I$ we know $ u^{-1}_m a = \exp\, y_m$.  Now the
degree shifting property
mentioned above applied to $Ad (\exp y_m ) = \exp\, ady_m$ implies $ y_m \in
\gg _+$.  Therefore, $ a =
u_m\, \exp\, y_m \in \hat U_+$.  This shows $ \hat U_+$ is closed.  Similarly
one sees that $\hat
U_-$ is closed.  This shows also $ \hat U_{\pm} = \overline{U^{fin}_{\pm}}
= U_{\pm}$. \blsq
\medskip
A proof similar to the one above works for 
\medskip
\noindent
\Theorem {{\bf (3.1.2).} The groups $B_w$, $P_w$, and $Q_w$ are closed in $G_w$.}
\medskip
As a consequence we see that $B_w$, $P_w$ and $Q_w$ are closed integral subgroups of $G_w$.  We show that those groups actually are Banach Lie subgroups in the sense of [23; Ch. 3, \S 1.3].  First we prove more generally
\medskip
\noindent
\Theorem {{\bf (3.1.3).} Let $\frak a$ be a closed Lie subalgebra of $\gg _w$ and $\frak b$ a closed subspace of $\gg _w$ such that $\frak a + \frak b =\gg _w$, $\frak a \cap \frak b =0$.  Denote by $A$ the integral subgroup associated 
with $\frak a$.  Assume $A \cap \exp \, V^0=\{I\}$ for some open neighborhood $V^0$ of $0$ in $\frak b$.  Then $A$ is a Banach Lie subgroup of $G_w$.}
\medskip
\noindent
\Proof. \quad Let $U$ denote an open neighborhood of $0$ in $\frak a$, and $V
\subset V^0$
an open neighborhood of $0$ in $\frak b$.  We can assume that $\exp$ is bijective on $U$ and $V$ and that $U \times V \mapsto \exp U\,\exp V\,=R$ is a diffeomorphism.  Consider now $a\in A\cap R$.  Then $a=\exp\,\hat{a}\,\exp\, \hat {b}\, ,\hat {a} \in U, \hat {b} \in V$;  therefore $\exp (-\hat {a})a=\exp\, \hat {b}\in A\cap \exp (\frak b)$, whence $\hat {b}=0,\, a=\exp\, \hat {a}$.  This shows $A \cap R=\exp U$.  Now apply 
[23; Ch.3, \S 1, Proposition 6] and obtain the claim.  \blsq
\medskip
\noindent
\Corollary {{\bf (3.1.4).}  Let $\frak a $ and $\frak b$ be closed Lie subalgebras of $\gg _w$ such that $\frak a + \frak b =\gg$ and $\frak a \cap \frak b =0$.
Assume also that for the associated integral subgroups $A$ and $B$ we have $A\cap \exp\,V^0\, =\{I\}$ and $B\cap \exp\,U^0\,={I}$ where $U^0$ and $V^0$ are some open neighborhoods in $\frak a$ and $\frak b$ respectively.  Then $A$ and $B$ are Banach Lie subgroups of $G_w$.
If $A\cap B=\{I\}$, then $AB=\{uv: u\in A, v\in B\}\cong A\times B$.}
\medskip
\noindent
\Proof. \quad The first statement follows from the Theorem above.  For the second  statement we consider the map $A\times B \mapsto G\, , (u,v)\mapsto uv$.  
Clearly, this map is analytic, and since $\frak a + \frak b = \gg _w$ we see that $AB$ is open in $G$.  Using $U$ and $V$ as in the proof of the Theorem above we see that the multiplication map is locally, around $(I,I)$ and $I$, a diffeomorphism.  Using translations we see that this true for any $(u,v)$ and $uv$. Therefore it suffices to prove that the multiplication map is injective.  It suffices to show that $uv=I$ implies $u=I=v$; but this follows from $A\cap B=\{I\}$.
\blsq

The above result and Theorem (2.4.1) imply that $B_w$, $P_w$ and $Q_w$ are 
Banach Lie subgroups of $G_w$.  
\medskip
\noindent {\bf 3.2} 
\noindent Next we consider the
group $H^{fin}$ generated by $\exp\, h,\ h\in\gh = \gg ^{fin}_0$.  Since $
\gh$ is complemented in $\gg$, $H^{fin}$ is a finite dimensional, connected 
integral subgroup of $G$ with Lie
algebra $ \gg ^{fin}_0$.  However $H^{fin}$ is closed in $G$ and thus $H^{fin}$ is
a Lie subgroup of $G$.  This can be seen by a proof analogous to the
one given for $U_{\pm}$, or by Theorem (3.1.3).  We thus set
$H = H^{fin}$.  We denote  by $ N^{fin}$  the normalizer of $H$ in $G^{fin}$.
Subsequently we define 
$W = N^{fin}/H$ and call it  the
Weyl group of $G^{fin}$.  We will show below that this is truly the Weyl group
defined in terms of reflections acting on the affine Kac-Moody root system.  
In our setting, [8, Corollary 5] states:
\medskip

\noindent
\Theorem {{\bf (3.2.1).}

\item{(a)} $G^{fin} = U^{fin}_- N^{fin} U^{fin}_+$.

\item{(b)} $G^{fin} = U^{fin}_+ N^{fin} U^{fin}_+$. 

\item{(c)} If $ g = unu^{\prime}$, $u, u^{\prime} \in  U^{fin}_+$, 
$n \in N^{fin}$ then we can assume $ u\in n\, U_-\, n^{-1}$ and
this decomposition is
unique.} \medskip

\noindent

\noindent
\Remark .

\noindent (1) 

As mentioned in the introduction, we are primarily interested in loop algebras and loop groups.  However, for the proof of many properties of the Banach Lie algebras and Lie groups used in this paper we will be using results on Kac-Moody Lie algebras and the associated groups.  We follow [8, 24] and do not include the degree derivation in our Lie algebras.  Instead, we  
use the root 
space decomposition and the principal grading induced from the full Kac-Moody algebra.  For our purposes it would be therefore most convenient to use a Banach structure on the derived algebra of a Kac-Moody algebra, i.e. a central extension of a loop algebra.  It is indeed possible to find such Banach structures  and  associated Banach Lie groups [11].  We find it therefore more appropriate for this paper to use only algebraic results for the central extension, to 
transport them via projection to our loop algebras and loop groups, and to 
extend them to our Banach loop algebras and groups.  The only somewhat delicate point here is the definition of the projection map $\pi$ on the group level.  If one denotes by $G^{\prime}$ the group considered in [8] as opposed to 
$G^{fin}$ considered in this paper, then  

$$
\matrix{
& \quad G ^{\prime}  &   & &\cr
&& \buildrel \rho ^{\prime}\over {\searrow \quad} & &\cr
& \pi\downarrow &          & G^{\prime}/Z^{\prime}\cong  G^{fin}/Z^{fin}
\subset G\ell (\gg ^{fin}\oplus {\Bbb C}c) &  \cr
&&  \nearrow {\scriptstyle \rho}  & &\cr
& \quad G^{fin} &  &     &          }$$
where $Z^{fin}$ and $Z^{\prime}$ denote the centers of $G$ and $G^{\prime}$
respectively, $\rho ^{\prime}$ is the adjoint representation of $G^{\prime}$, 
and $\rho$ denotes the ``extended'' adjoint representation [12; Proposition (4.3.3)].  Since we consider $\buildrel \circ \over G$ simply connected, we need to 
lift results from $G^{fin}/Z^{fin}$ to $G^{fin}$.  This will be straightforward in all the cases considered in this paper.  

\noindent (2)  
Part(a) describes the Birkhoff decomposition of $G^{fin}$, whereas  part (b) describes the Bruhat decomposition of $G^{fin} $. All the decompositions appearing in Theorem (3.2.1) are obtained by taking the projection of those 
of [8].  

\medskip

\noindent
\Lemma {{\bf(3.2.2).} $W \cong $  Weyl group of the corresponding full Kac-Moody algebra.
}\medskip

\noindent
\Proof . We will use the superscript PK to distinguish between objects in our 
set up and those in [8]. Let us denote by $\Pi$ the projection $G^{PK} \to G^{
fin}$.  Then we have $\Pi(N^{PK})=N^{fin}$ and $\Pi(H^{PK})=H$ and the induced
map $W^{PK}\to W$ is surjective.  To see that it is also injective we observe
that $\Pi ^{-1}(H)=H^{PK}$.   \blsq
\medskip
We recall the definition of $B ^{fin}$ and $ B_w=\overline {B^{fin}}$ from 
\S 2.1.  In what follows we will use $B=B_+=B_w$ and define $B_-$ analogously.
  We note $B_+=HU_+$ and $B_-=HU_-$.  We will also use parabolic algebras and groups $ \gp ^{fin}$, $\gp =\overline{\gp ^{fin}}$ and $P^{fin}$, $P=\overline{P^{fin}}$ respectively.  In view of 2.2 we know $\gp ^{fin}=\gp ^{fin}_X$.  Similarly we will use $P^{fin}_X$, $\gp _X$ and $P_X$.
\medskip  
\noindent
\Corollary {{\bf (3.2.3).}\quad $ G^{fin} = \bigcup\limits _{w\in W/W_X} U^{fin}_-\,
w P^{fin}_X$, where
$W_X$ is the subgroup of $W$ \hfill\break generated by
$\{ r_{\alpha }; \alpha \in X\}$.
Moreover, the union
above is disjoint.}
\medskip

\noindent
\Proof .\quad  Let $W^{\prime}\subset W$ denote a set of representatives of
$W/W_X$.  Then $ W = W^{\prime}W_X$.
From the Theorem above we know $ G^{fin} = \bigcup\limits _{{w^{\prime}\in
W^{\prime}\atop w\ \in W_X}} U^{fin}_-\,
w^{\prime} w U^{fin}_+$.  Moreover, $ P^{fin}_X = B^{fin}_+ W_X B^{fin}_+$ by
[23, Ch. 4; \S 2.5].
Hence $ G^{fin} = \bigcup\limits _{w^{\prime}\in W^{\prime}} U^{fin}_- w^{\prime} P^{fin}_X$.
To see that this union is
disjoint, we note that for $ B^{fin}_{\pm} = H\, U^{fin} _{\pm}$ we have $
B^{fin}_- w B^{fin}_+
r_i\subset B^{fin}_- w B^{fin}_+ \cup B^{fin}_- wr_i B^{fin}_+$, where $r_i=
\exp f_i \exp(-e_i)\exp f_i$,
as
mentioned in [8].
With this, one obtains mutatis mutandis, [23,Ch.4, \S 2, Lemma 1]
and then [23, Ch.4, \S2.5, Remark 2], proving the claim.\blsq
\medskip

\noindent
\Remark .\quad We note that for $w = 1$ we obtain $ U^{fin}_- \cdot 1\cdot
P^{fin} = Q^{fin}
P^{fin}$.
\medskip

\noindent
{\bf 3.3 }
\noindent The following technical result will be useful. We 
define $ \circG _X$ as the connected subgroup of $P$ generated by 
$ x_{\pm \alpha } \in \gg^{fin}_{\pm \alpha},
\alpha\in X$, with Lie algebra $\circgg_X$.  We denote by $Q_{X}^+$ the connected  Banach subgroup of
$P=P_X$ with Lie algebra
$\gq^+_X$, where $ \gq^+_X$ is the natural complement of $\circgg_X$ in $\gp$ 
i.e.
$$
\gp = \circgg _X \oplus \gq^+ _X. 
$$
Clearly, we have
$$
\gq _X^+= \frak a _Q+\gq _X^{++},
$$
where $\frak a _Q=\frak h \cap \gq _X^+$ and $\gq _X^{++}=\gg ^{(+)}$.  
Then $\circgg_X +\frak a_Q=\gg^{(0)}$ and $\gq _X^{++}$.  Moreover, $[\circgg_X, \frak a _Q]=0$ holds.  
\medskip
\noindent
\Proposition {{\bf (3.3.1)}.\quad Let $ P = P_X$ be a parabolic subgroup of $G$ and
$w\in W$.  Then
\item{(a)}  $P = P_X = \circG _X\, Q^+_X = Q^+_X \circG _X\cong \circG _X\times Q^+_X. $
\item{(b)}  The stabilizer $ U^w_-$ in $U_- $ of $wP\in G/P$ is $ U^w_- =
U_-\cap wPw^{-1}$.
Moreover, $ \dim U^w_- < \infty $.
\item{(c)} There exists a closed subgroup $ V^w_-$ of $U_-$ such
that group
multiplication induces a diffeomorphism $ U_-\cong U^w_- \times V^w_-$.}
\medskip

\noindent
We relegate the proof of this proposition to Appendix A.  
\medskip

\noindent
{\bf 3.4}.  In this section we state  the Birkhoff decomposition of $G$.  
We refer the reader to  [12, Ch.8.6] for more details.   
\medskip
\noindent
\Theorem {{\bf (3.4.1).} \quad 
\item {(a)}  Let $ g \in G$.  Then the orbit $ B_- g B_+$
contains a unique $ w\in W$.

\item {(b)} $G = \bigcup\limits _{w\in W} B_- w\, B_+$ is a disjoint union.}
\medskip

\noindent
{\bf 3.5 }\quad  We consider the sets $ U_- wP,\ w\in W/W_X  $.  In
general, these sets are not closed in $G$.  To investigate their
closure we use the Bruhat order ``$\prec$'' on $W$, i.e.  the partial order generated
by $ r_{i_1}\cdots r_{i_{s-1}}\ r_{i_{s+1}}\cdots r_{i_k}\,\prec w\,,  1 < s\le k
$, where $ w = r_{i_1} \cdots r_{i_{k}}$ is a reduced expression.

We will also need the notion of an $X$-reduced element [23
Ch.4, \S1,
exercise 3]:  $w\in W$
is called $X$-reduced if it has minimal length in the coset $ wW_X$.  In
particular, $w$ is
$X$-reduced iff $w\, r_i$ is larger then $w$ for all $ r_i\in X$.
Moreover, any $ w\in W$ can be
written in a unique way in the form $w = w^{\prime}q$, where $ w^{\prime}$ is $X$-reduced
and $ q\in W_X$.  Hence
a set $ W^{\prime}$ of representatives  of $W/W_X$  can be chosen as the set of
$X$-reduced elements in
$W$.

With this notation we can prove for $ P= P_X$
\medskip
\noindent
\Theorem {{\bf (3.5.1).}
$$
G = \bigcup _{{w\in W\atop w\,is\,X{\first\hbox{-reduced}}}}\ B_- wP = \bigcup
_{w\in
W/W_X}\ B_- wP\ .
$$}
\medskip

\noindent
\Proof .\quad Using Corollary 3.2.3 we have
$$
G = \overline{G^{fin}} = \overline{\bigcup _{w\in W} B^{fin}_- wP^{fin}}
 \supset
\overline{\bigcup _{w\in
W} B_- wP} \supset\overline{\bigcup_{w\in W}
B_- wB_+}=\bar G=G,
$$
where we have also used Theorem 3.4.1.  This established $G=\bigcup \limits _{w\in W}
 B_-wP$.  Since $W_X \in P$ we get the claim. \blsq
\medskip
Next we want to describe the closure of $C_X(w)=B_-wP_X$.  For this we recall
the definition of $U _{\alpha}$ from 3.1.  Moreover, we set $h_i=[f_i,e_i]$
and $H_i=\exp\, {\C} h_i$.  We would also like to note that the Proposition below is most likely true for any norm given by a weight $w$.  However,  at this
point we are only able to prove it for the weights $w$ satisfying our usual 
assumptions and, in addition,
$$
\sum _{n\in {\ints}}(1+|n|)|f_n|^2 \leq C \Vert f \Vert _w, \quad f\in A _w,
$$
for some $C >0$.  

\medskip
\noindent
\Proposition {{\bf (3.5.2).}\quad For every $w\in W/W_X$ we have
$$
\overline {C_X (w)} = \bigcup _{M} C_X (w^{\prime}) 
$$
where $ \cal M = \{ w^{\prime}\in W/W_X,\ w^{\prime} \succeq w\}$.
}\medskip

\noindent
We relegate the proof of this proposition to Appendix A.  
\medskip
\noindent
\Corollary {{\bf (3.5.3).}
\item {(a)}  $QP$ is dense in $G$.
\item {(b)}  If $ w \in W, w \not= I$, then $ B_- wP$ has nonzero finite
codimension in $G$.
}\medskip

\noindent
\Proof .\quad (a)  It is easy to see that $ B_-P = QP$ holds.  
Indeed, $B_-=Q\circB_-$ as in Proposition (3.3.1).  But $\circB _-P=P$. 
Then, the Proposition above shows for $w=1$ 
$$
\overline {QP} = \overline{B_- P}=
\bigcup_{w^{\prime} \succeq  1} B_- w^{\prime}B_+ ,
$$
and this is all of $G$ by Theorem (3.4.1).
\item{(b) } Follows from Proposition (3.3.1). \blsq
\medskip

\noindent
{\bf 3.6 }  The following result will be used to see that certain functions
involved in our set-up are meromorphic in the flow variable.
We recall (see [10]) that the cyclic element $E$ of
$\gg$ is  the sum of all canonical generators of (canonical) degree 1.  We will
review more facts about $E$ in Sect.5.1.  We continue to use the notation of 3.5 and the norm restriction used in Theorem (3.5.2).
\medskip

\noindent
\Theorem {{\bf (3.6.1).} \quad  For every $g\in G$ there exists some $ \varepsilon >
0 $ such that for all $ 0
< \vert t\vert < \varepsilon $ we have
$$
e^{tE} g \in QP\ .
$$
}
\medskip
\noindent
\Proof .\quad Clearly, the statement is trivial if $ g \in QP$.
Hence we will assume $
g\not\in QP$.  Then from Theorem (3.4.1)  we know $ g = b_- wb_+ $ where $
b_{\pm}\in B_{\pm}$ and $w \ne I$.  For our purpose we
can assume $ b_+ = I$.  Moreover we can assume $\ b_-\in U_-$ and even $b_-\in V_-^w$ by
Proposition (3.3.1).    Next we
consider $e^{tE} b_-$.  Since $b_- \in B_-B_+$, for
sufficiently small $t$ we have
$ e^{tE}b_- = u_-(t) u_+ (t)$.  
We consider the closed subalgebra $\frak a=w^{-1}\gb _+w$ of $\gg$.  Then
it is easy to see that 
$$
\frak a= \frak a _- + \frak a _+,
$$
where $\frak a _-=\frak a \cap \gg _-$ and $\frak a _+=\frak a \cap \gb _+$.
  Then $\frak a_{\pm}$ are closed subalgebras of $\frak a$ satisfying $\frak
a_-\cap \frak a _+=0$.  Moreover,  for the corresponding integral subgroups
$A _{\pm}$ we have $A_-\cap A_+\subset U_- \cap B_+=\{I\}$.  Therefore 
by Corollary (3.1.4) $A_-\times A_+ \cong w^{-1}B_+w$.
Thus  $$
e^{tE}b_-w = u_-(t) u_+ (t) w = u_- (t) v_+ (t) w\, q_+ (t)\ ,
$$
where $ v_+ (t) \in w\, U_-w^{-1} \cap U_+\quad\hbox{and}\quad q_+ (t)\in
B_+$ .  Using [8, Corollary 5] and the fact that the factors of the 
factorization are analytic in $t$ as well as $ u_- (0) = b_-$,
we obtain $v_+(0) = I$ and $
q_+(0) = I$.Hence we will assume $
g\not\in QP$.  Then from Theorem (3.4.1)  we know $ g = b_- wb_+ $ where $
b_{\pm}\in B_{\pm}$ and $w \ne I$.  For our purpose we
can assume $ b_+ = I$.  Moreover we can assume $\ b_-\in U_-$ and even $b_-\in V_-^w$ by
Proposition (3.3.1).    Next we
consider $e^{tE} b_-$.  Since $b_- \in B_-B_+$, for
sufficiently small $t$ we have
$ e^{tE}b_- = u_-(t) u_+ (t)$.  
We consider the closed subalgebra $\frak a=w^{-1}\gb _+w$ of $\gg$.  Then
it is easy to see that 
$$
\frak a= \frak a _- + \frak a _+,
$$
where $\frak a _-=\frak a \cap \gg _-$ and $\frak a _+=\frak a \cap \gb _+$.
  Then $\frak a_{\pm}$ are closed subalgebras of $\frak a$ satisfying $\frak
a_-\cap \frak a _+=0$.  Moreover,  for the corresponding integral subgroups
$A _{\pm}$ we have $A_-\cap A_+\subset U_- \cap B_+=\{I\}$.  Therefore 
by Corollary (3.1.4) $A_-\times A_+ \cong w^{-1}B_+w$.
Thus  $$
e^{tE}b_-w = u_-(t) u_+ (t) w = u_- (t) v_+ (t) w\, q_+ (t)\ ,
$$
where $ v_+ (t) \in w\, U_-w^{-1} \cap U_+\quad\hbox{and}\quad q_+ (t)\in
B_+$ .  Using [8, Corollary 5] and the fact that the factors of the 
factorization are analytic in $t$ as well as $ u_- (0) = b_-$,
we obtain $v_+(0) = I$ and $
q_+(0) = I$.

We claim that there exists some $ \varepsilon > 0$ 
such that for all $t$, $ 0 <\vert t \vert < \varepsilon $, we have

\item {(1) }\quad $ v_+ (t) \not= I$,

\item{(2) }\quad $ v_+ (t)w\in B_- B_+ $.

To prove (1) we recall  that $ v_+(t)$ is analytic in $t$.  Therefore, if (1)
were wrong, then

\item {(3) }\quad $e^{tE} b_- w = u_- (t) w\, q_+ (t)$\quad for all
sufficiently small $t$.  This implies
$$
w^{-1} b^{-1}_- e^{tE} b_- w = w^{-1} b^{-1}_- u_- (t) wq_+(t),
$$
whence
\item{(4) }
$$
w^{-1} r_-(t) w\, q_+(t) = \exp\{ t\, Ad (w^{-1} b^{-1}_-)E\}
$$
where $ r_-(t) = b^{-1}_- u_-(t)$.
In particular, we have $ r_- (0) = I = q_+ (0)$.

Differentiating at $t = 0$ we obtain
$
w^{-1} \dot r_- (0) w + \dot q_+(0) = Ad (w^{-1}b^{-1}_-)E\ .
$
Conjugating with $w$ yields

\item{(5) }
$$
\dot r_-(0) + w \dot q_+ (0) w^{-1} = b^{-1}_- Eb_-\ .
$$
We note that $ b^{-1}_- Eb_- = E + S$, where 
$ c\, deg S
\le~0$.

Next we decompose $ \dot q_+(0) = h + u_+$, where $ h\in \gh$ and $ u_+\in
\gg _+$.  Then (5)
implies
$
wu_+ w^{-1} = E + T\ ,
$
where $ c\, deg\, T \le 0$.

Now we write $ u_+ = \sum\limits _{\alpha \in\triangle _+} v_{\alpha }$, then
$
wu_+ w^{-1} = \sum\limits_{\alpha\in\triangle _+} w v_{\alpha } w^{-1} = E + T\ .
$
We know $ wv_{\alpha } w^{-1}\in \gg _{w(\alpha )}$.  We set  $ A =
\{\alpha\in \triangle _+;\ \
w(\alpha ) = \alpha _i$ for some $i\}$ and $ B = \triangle _+\backslash A$.
Then $ w\, u_+ w^{-1} =
\sum\limits _{\alpha \in A} c_i e_{\alpha _i} + \sum\limits_{\alpha\in B}
wv_{\alpha }w^{-1}$, and all $c_i\ne 0$.  In fact $c_i=1$.    
Therefore $u_+ = \sum\limits _{\alpha\in A} c_i\, e_{w^{-1}(\alpha _i)} +
\sum\limits _{\alpha\in B }v_{\alpha} $.
This implies $w^{-1}(\alpha _i)\in \triangle _+$ for all $i$.  But now [10, Lemma
3.11] shows $w = 1$, a
contradiction.  This proves (1).

To prove (2) note that from Theorem (3.4.1) we know that for every $t$ we have
$v_+(t)wq_+(t)\in B_-w^{\prime}B_+$ for some $w^{\prime}$.  But then obviously
$(B_+wB_+)\cap (B_-w^{\prime}B_+)\ne \emptyset$, 
whence $w \succeq w ^{\prime}$,
(see e.g. [12; Theorem (8.4.6)]).  We claim that $w\ne w^{\prime}$, if $v_+(t)
\ne I$.  Suppose $w=w^{\prime}$, then $v_+(t)w q_+(t) \in (B_+ w B_+)\cap 
(B_- w B_-)=w B_+$, where the last equality follows from [12;Theorem 8.4.5].  In particular $w^{-1}v_+(t) w \in B_+$.  But we had chosen 
$v_+$ above so that  $w^{-1}v_+(t) w \in U_-$.  This contradicts $v_+(t)\ne I$.
  As a consequence we have for all $0<|t|
\le \varepsilon $ that the corresponding $w^{\prime}$ satisfies $w \succeq w ^{\prime}$ and $w=\ne w^{\prime}$.  From the definition of $\succeq$ it is clear that there are only finitely many $w^{\prime} \in W,\,w \succeq w^{\prime}$.  Therefore, there exists a sequence $t_j \to 0$, 
$$
b_-^{-1}e^{tE}b_-w=s_-(t)w^{\prime}s_+(t), \hbox{ for all } t=t_j,\, j=1,2,\dots \hbox{ with $w^{\prime}$. }
$$  
If $w^{\prime}\ne I$, then we multiply from the left by $b_-^{-1}e^{rE}b_-$
and $s_-(t)^{-1}$.  
We obtain  
$$
s_-(t_j)^{-1}b_-^{-1}e^{rE}b_-s_-(t_j)w^{\prime}s_+(t_j)=l(r).
$$
Now we can apply the above argument to $\hat{b_-}=b_-s_-(t_j)$, 
and obtain a sequence $r_{jk} \buildrel k \over \to 0,\, w^{\prime} \succeq w^{\prime \prime}$, such that $l(r_{jk}) \in B_-w^{\prime \prime}B_+$ and $w^{\prime}\ne w^{\prime \prime}$.  
But then 
$$
s_-(t_j)^{-1}b_-^{-1}e^{r_{jk}E}e^{t_jE}b_-w \in B_-w^{\prime \prime}B_+.
$$
Hence for every $g\in G$ there
exists a sequence $z_j\to  0$ such that $e^{z_jE}g \in B_-B_+$.  We use again
the standard representation of $G$ in $Gl_{res}$.  Then  for $e^{tE}g$ to be in
$ B_-B_+$ it is necessary and sufficient that
$\chi(t)\equiv \sigma(exp(tE).g.H_+ \ne 0$, see [6, \S 5] for notation.
From what we have shown above, the holomorphic function $\chi$ vanishes for
$t=0$, but does not vanish identically.  This proves (2).  Finally, similar
to Proposition (3.3.1) one can show $B_-=Q(\circG _X)_-$.  Therefore $e^{tE}g
\in B_-B_+=QP$ for all $0<|t|<\varepsilon$. \blsq
\medskip

\noindent
{\bf 3.7 }  We present two theorems for later use.
Denote by $\Gamma $ the
connected Banach
Lie group with Lie algebra $ Ker\, ad\, E$ (this is a closed complemented
subalgebra, therefore
$\Gamma $ exists by [23, Ch.3, \S 6, Thm.2]).  $\Gamma $ is generated by $\exp\, T,\ T\in$
Lie $\Gamma = Ker\, ad\,
E$.  Thus 
$(Ad (\exp\, T)) E = \exp(ad\, T)E = E$, i.e. 
$(Ad\, g)\, E = E$ for all $g \in \Gamma$.
\medskip
\noindent \Lemma {{\bf (3.7.1).}\quad $\Gamma$ is closed. }
\medskip
\noindent \Proof. \quad Assume $ g_n\to g,\ g_n\in\Gamma$.  Then $Ad (g) E = \lim Ad(g_n) E = E$.  Moreover, since 
$g^{-1}_n\, g\to I$ we know $ g^{-1}_n g = \exp\,t_n $  for $ n
\ge n_0 $, 
$n_0$ sufficiently large.  As a consequence, $Ad(\exp (t_n))E = \exp(ad (t_n))E =
E$.  

\noindent Now consider
$ \exp(tE) \exp(t_n) \exp (-tE)= \exp(tE)\, (\exp(t_n)
\exp (- tE) \exp (- t_n))
\exp (t_n) = \exp (tE)\exp( - t (Ad( \exp t_n) E) \exp (t_n)
= \exp (tE)\exp (- tE)\exp (t_n)= \exp (t_n ).$  Therefore \break $
\exp (t_n) =
\exp Ad(\exp (tE))t_n = \exp ((\exp\, ad\, t\, E\,) t_n)$. 
Thus for $t$ sufficiently small, since $\exp$ is bijective in a small neighborhood of
$I$ we obtain
$t_n = \exp(\, ad\, t\, E)\, t_n$.  This implies $[E, t_n] = 0$, 
whence $t_n \in Ker\, ad\, E$.  As a consequence, 
$g = g_n\, \exp\, t_n \in \Gamma $. \blsq
\medskip
Finally, one defines   $\Gamma_{\pm}$ for ($Ker\, ad\, E)_{\pm}$.  Note
that by the above argument $\Gamma ,\, \Gamma
_+$ and $ \Gamma _-$ are closed in $G$.  Also note, these groups are
abelian and $ \Gamma = \Gamma
_-\Gamma _+$.

Before we state and prove the next two theorems we formulate another decomposition
of $\gg$.  Namely:
$\gg=\gs+\gu $,
where $\gs$ is the principal Heisenberg algebra and $\gu$ is its orthogonal,
graded complement.  We also fix from now on a parabolic $\gp$ and write $\gq$ for 
its natural complement. 

Next we set $F=\sum _i\, f_i$, the sum of all generators with $cdegree =1$.
 We also note that $ad\,E$ and $ad\, F$ are invertible on $\gu$. 
\medskip
\noindent
\Theorem {{\bf (3.7.2).}\quad Let $g\in Q$ and assume $ e^{tE}ge^{-tE}\in
U_-$ for all $t$ in some neighborhood of $t=0$.  Then
$g\in\Gamma _-$.
}\medskip
\noindent
\Proof. We will use for our algebras and groups the coordinates as in \S 14.3 of [10]. We would like to remark also that the proof of semisimplicity 
of $E$ in [10, Prop.14.2] carries over to the extension of $ad E$ to the loop
algebra $g\l(n)$. The nice thing about that particular choice of coordinates is
that the canonical generators $e_i,f_i,h_i$ become homogeneous functions of $\lambda$ of degree 1, in particular $E=\lambda \circE $.  Consequently, expanding in terms of powers of $\lambda$ we can
write $g=a_{0}+a_{-1}/\lambda
+a_{-2}/{\lambda ^2}\dots$.  The hypothesis implies that $(ad(E))^m g$ is analytic in $1/{\lambda}$
for all $m\ge 0$. In particular, we get $ad(E)a_0=0, (ad(E))^2 a_{-1}=0,\dots 
(ad(E))^{(k+1)} a_{-k}=0, \dots$. We claim that 
this implies that $a_{-k}\in Ker\, ad(E)$.  For $k\ge 2$ we write $ad(E)^{(k+1)}a_{-k}=0$ as 
$ad(E)(ad(E)^{k}a_{-k})=0$.  Thus $ad(E)^{k}a_{-k}\in Im\, ad(E) \cap
Ker\,ad(E)=\{0\}$ hence by the remark above $(ad(E))^{k}a_{-k}=0$.  Iterating this procedure we
obtain that $ad(E)a_{-k}=0$ for any positive integer $k$.    \blsq
\medskip
Next we prove similar to the last theorem:
\medskip
\noindent
\Theorem {{\bf (3.7.3).}\quad Let $g\in P$ and assume $e^{tF}\
ge^{-tF}\in P$ for all $t$ in some neighborhood of $t=0$.
Then $ g\in \Gamma \cap P$.
}\medskip
\noindent
\Proof .  We use the same coordinates as in the previous theorem.  In particular, $F=\lambda ^{-1}\circF$, where $\circF$ is $\lambda$ independent.  Furthermore, there exists a positive integer $k$ such that for every $g \in P, \, \lambda ^k g$ is analytic in $\lambda$.  Thus the hypothesis of the present theorem
can be restated as $\lambda ^k (e^{tF}\
ge^{-tF})=e^{tF}\
(\lambda ^kg)e^{-tF}$ is analytic in $\lambda$.  The rest of the argument is the same as in the previous theorem with $\lambda$ replacing $1/\lambda$ and 
$F$ replacing $E$.  \blsq
\medskip
{\bf\S 4.\quad  More Banach Lie groups and subgroups}
\medskip

In this Section  we want to introduce the ``ingredients'' which we will
use later for
factorization.

\noindent
\noindent
{\bf 4.1 }\quad Let $G$ be a Banach loop group with Lie algebra $\gg $.  In
particular, $G$ and
$\gg $ consist of functions defined on the unit circle $ S^1$ with values
in 
$\C$$ ^{ n \times n }$ for
some $n$.

For $ R>0$ we set
$$
S^R = \{ z \in {\Bbb C} : |z| = R\} \leqno(4.1.1)
$$
and define
$$
G^R =\{ g^R:S^R\to\circG , \,
\lambda \mapsto g^R(R\lambda) \in G\quad \hbox{for}\,\lambda \in S^1 \}
\leqno(4.1.2)
$$
Similarly we define $ G^r$, $\gg ^R $ and $\gg ^r$.  We will always assume $ 0
< r \le 1 \le R < \infty $. It is easy to see that $G^R \mapsto G, \, g^R\mapsto g(\lambda)=g^R(R\lambda),\, \lambda \in S^1$, is an isomorphism of groups.  
This way $G^R$ inherits naturally a Banach Lie group structure from $G$. The 
need for using two distinct circles is demonstrated clearly in Proposition 5.2
and Theorem 5.3.1 in [15].  

Let $\gp$ be a standard parabolic subalgebra of $\gg $ and
$\gq$ its natural
complement.  By $ P$ and $Q$ we denote the corresponding connected Banach
Lie subgroups.  Via
(4.1.2) we thus obtain Banach Lie groups $P^R, P^r, Q^R$ and $Q^r$ as well.

For this paper the following Banach Lie groups are particularly important:
$$
\HH = G^R \times G^r, \leqno(4.1.3)
$$
$$ \HH_- = Q^R \times P^r, \leqno(4.1.4)$$
$$ 
\HH_+ =\left\{ \eqalign{(g_1, g_2)\in \HH :g_1\in G^R\hbox {and}\, g_2\in
G^r\,\hbox{have the same} \cr
\hbox{holomorphic extension 
in the annulus}\quad r < \vert z\vert < R}\right\}.\leqno(4.1.5)
$$
In particular, if $ r = R$, then $\HH_+$ is just the diagonal
in $ \HH = G\times G$.  The
corresponding Lie algebras will be denoted by $ \bf {h} , \bf {h}_-$ and $ \bf{h}_+$
respectively. On Fig.1 below we indicate the regions of the complex plane $z$
where the respective groups live.
\vfill
\eject

\input epsf
\centerline{\epsfxsize=8cm \epsfbox{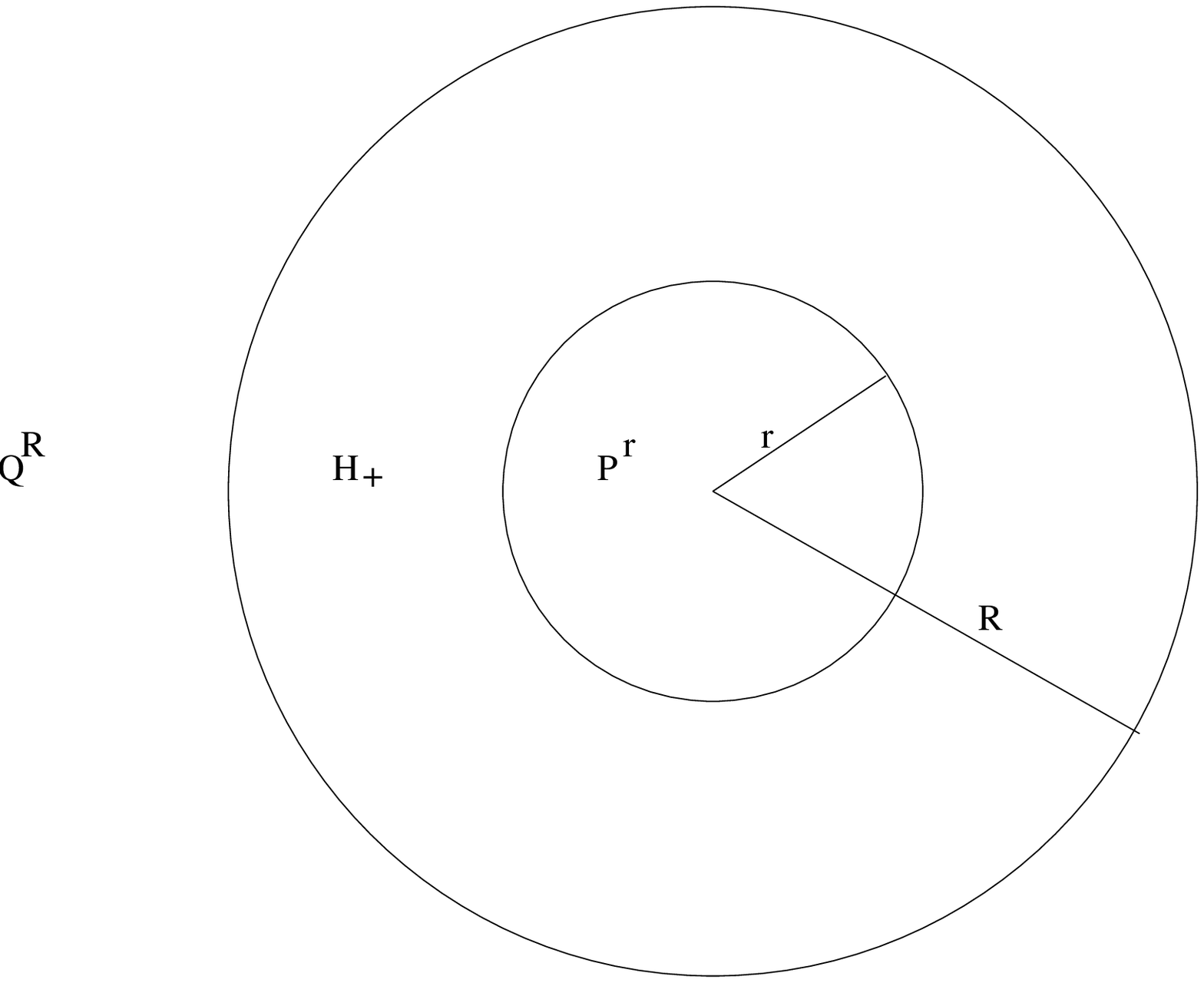}}
\centerline{Fig.1}
\medskip
\noindent
{\bf 4.2 }\quad In what follows we will frequently use the facts listed  below.
\medskip

\noindent
\Theorem {{\bf (4.2.1).}\quad $ \HH_+$ and $ \HH_-$ are connected Banach Lie subgroups of $\HH$.  Moreover, $\HH_-$ and $\HH_+$ have the following properties:
\item{(a)} $ \HH_- \HH_+$ is open and dense in $ \HH$.
\item{(b)} $ \HH_- \HH_+$ is analytically diffeomorphic with $
\HH_-\times \HH_+$.
\item {(c)}$ \HH _+$ and $\HH _-$ are closed in $\HH$. }\medskip
\noindent
\Proof .\quad The first part of the theorem and Item (c)  follow from Corollary (3.7.4) and the lemma below.  
\medskip
To prove (a) and (b) we observe that $ \bf {h} _- + \bf {h} _+ = \bf {h} $ and 
$ \bf {h}_-
\cap \bf{h}_+ = \{ 0\}$
holds [15].  From this it follows that the map
$$
\cases{ \bf {h}_- +\bf{h} _+ &$\to \HH$\cr
(h_-,h_+) &$\mapsto \exp h_- \exp h_+$
}$$
is a local diffeomorphism at 0; therefore $ \HH_- \HH_+$ is open in $\HH$
and, locally at the
identity, analytically diffeomorphic with $ \HH_- \times \HH_+$.

For general $ h_- h_+\in \HH_- \HH_+$, consider a neighborhood $ U$
of $ h_- h_+$ such
that $ h^{-1}_- Uh^{-1}_+$ can be mapped diffeomorphically into $ \HH_-\times
\HH_+$.  From this, (b) follows,

To see that $ \HH_- \HH_+$ is dense in $\HH$, we pick an arbitrary $ g
= (g^R,g^r)\in \HH$.
Since $Q P$ is open and dense in $G$, in every neighborhood of $g$ we
can find a $\tilde g$
such that
$$
(\tilde g^r)^{-1} \in Q^r P^r\ .
$$
Hence we may assume that $ g^r\in P^r Q^r$.  Write $g^r = p^r
q^r$.  Then
$$
g = \big(g^R(\mu ),\, g^r(\lambda )\big)  = \big( g^R(\mu ) q^r(\mu
)^{-1},\, p^r(\lambda )\big)
\big(q^r(\mu ),\, q^r(\lambda )\big)\ .
$$
If necessary, replace $g^R$ by $\hat g^R$ arbitrarily close to $g^R$ such that
$$
\hat g^R (\mu )q^r(\mu )^{-1} = \hat q^R (\mu )\hat p^R (\mu )\in P^RP^R\ .
$$
Then we obtain
$$
\eqalign{
&\big(\hat g^R(\mu ) q^r(\mu )^{-1},\, p^r (\lambda )\big) = \big(\hat
q^R (\mu ) \hat p^R
(\mu ),\, p^r(\lambda )\big)\cr
&= \big(\hat q^R (\mu ),\, p^r (\lambda ) \hat p^R (\lambda
)^{-1}\big) \big(\hat p^R (\mu
),\, \hat p^r(\lambda )\big)\ ,
}$$
where we used that $P^R$ naturally embeds in $P^r$.  
Thus 
$ \big(\hat g^R (\mu ),\, g^r (\lambda )\big)\in \HH_- \HH_+$
and the proof is complete.    \blsq
\medskip
\noindent
\Lemma {{\bf (4.2.2).}\quad $\HH_-\cap \HH_+ = \{ I\}$.
}\bigskip
\noindent \Proof .\quad 
This is a coordinate dependent proof.  We use the coordinates introduced earlier following \S 14.3 of [10].  Let $ (g_1, g_2) \in \HH_-\cap \HH_+$.  Since $ g_1 \in
Q^R$, the Laurent
expansion of $ g_1(\lambda ) $ about $\lambda = 0$ contains only
nonpositive exponents.
Therefore, there exists some $h=\sum h_n z^n$ such that $\sum h_nR^n \lambda ^n=g_1 \in Q$ and $\sum h_n r^n \lambda ^n =g_2 \in P$.  Since $h_n=0$ for $n \ge 0$, $\sum h_n z^n $ converges for $|z| \ge r$.  Moreover, $\sum h_n R^n \lambda ^n \in P\cap Q=\{I\}$. 
\blsq  
\medskip
\noindent
\Remark.\quad The above  section generalizes the setup in [15] to general loop
groups as well as more general
subgroups.

\noindent
{\bf\S 5.  Factorization}
\medskip

\noindent
{\bf 5.1 }\quad In this section we recall some facts about the cyclic
element of $\GG $, all of
which can be found in [10].  By the \underbar{cyclic element} we mean
$E:=\sum e_i \in \GG$, the
sum of all canonical generators of canonical degree 1.

Its centralizer, $\gs := Cent_{\GG }E = Ker\, ad\, E = \{F\in\GG:[E,F] =
0\}$, is an abelian
subalgebra of $\GG $, which is graded with respect to the canonical grading:
$$
\gs = \bigoplus\limits_{k\in \ints } \gs _k\ .\leqno{( 5.1.1)}
$$
The integers $k$, for which $ \gs _k\not= \{ 0\}$, are called the
\underbar{exponents} of $\GG $.  If
$k$ is an exponent of $\GG $, then $ \dim _{\ssubC } \gs _k$ is called the
\underbar {multiplicity}
of $k$.

It turns out that the multiplicity is always one, with the exception of $
D^{(1)}_{2m}, m\ge 2$
[10, Ch.14].  Nevertheless the exponent 1 has always multiplicity one and the
space $ \gs _1$ is spanned
by $E$.  (Remark 14.2 and table $E_0$ in [10]).

An important feature of $E$ is that $ ad\, E: \GG\to\GG $ is bijective on
its image, or
equivalently,
$$
\GG = Im\, ad\, E\oplus Ker\, ad\, E\ .  \leqno{(5.1.2)}
$$

An easy proof of this can be deduced by extending the argument given by Kac
in [10,
Prop. 14.2] for the case of $\GG $ simple and finite dimensional.
Finally, we
remark that $ Im\, ad\, E$ is graded with respect to the canonical grading.
\bigskip

\noindent
{\bf 5.2 }\quad For every integer $k$ we choose a nonzero element $ E_k\in
\gs _k$, if possible,
and let $ E_k = 0$ otherwise.  We will always assume $ E_1 = E$.  Then
$$
\{\cdots , E_{-2}, E_{-1}, E_0, E_1, E_2,\cdots\}   \leqno{(5.2.1)}
$$
spans $\gs $ in case $\gg $ is not of 
type $ D^{(1)}_{\ell},  \, \ell \ge 4,\, \ell
$ even.  In that case we choose two linearly independent elements $ E_{\ell
-1}, E_{\ell -1} ^{\pri }$.
Note that $0$ is never an exponent; thus $ E_0 = 0$.  Nevertheless we include it
for the sake of
simplicity of notation.

We set
$$
\gt : = (\cdots , t_{-2}, t_{-1}, t_0, t_1,t_2, \cdots )  \leqno {(5.2.2)}
$$
for $ t_k \in\reals $ and $ k \in \ints $,
$$
{\Bbb R ^{\Bbb Z}} 
= \{ \gt: \sum_{k\in \ints } t_k E_k \in \GG \}.\leqno
{(5.2.3)}
$$
As above, we will double the coefficient $ t_s$ if $ \GG $ is of type $
D^{(1)}_{\ell }$ and the
exponent $s$ is of multiplicity two.

Define $E^R_k\in \GG ^R $ as usual via
$$
E^R_k := E_k({\mu})\quad\hbox{for}\quad \mu \in
S^R ,\leqno{(5.2.4)}
$$
Similarly we define\quad $ E^r_k \in \GG ^r$.

For this paper the following action of $ \reals ^{\ints }$ on $ \HH$ will be
of particular importance.
$$
\left\{ \eqalign{
{\reals ^{\ints}} \times \HH &\to \HH\cr
\big(\gt , (h_1, h_2)\big) &\mapsto \gt  (h_1,h_2), \cr} \right. \leqno(5.2.5)
$$
$$
\gt  (h_1, h_2) = \left(\exp\left( \sum_{k\in\ints } t_k
E^R_k\right) h_1,\ \exp\left(
\sum_{k\in\ints} t_k E^r_k\right) h_2\right)\ .  \leqno {(5.2.6)}
$$
Note that $ E^R_k$ and $h_1$ depend on $ \mu \in S^R$, whereas $ E^r_k$ and
$ h_r$ depend on $
\lambda \in S^r$.
\bigskip

\noindent
{\bf 5.3 }\quad We recall from Theorem(4.2.1) that $ \HH_-  \HH_+$ is open
and dense in $\HH$.
Therefore, if $ (h_1,h_2)\in \HH_- \HH_+$, then also $ \gt  (h_1,
h_2)\in \HH_- \HH_+$ for
all $ \gt$ in an open neighborhood of $\gt = \underline {0}$. The proof of the following Proposition is in immediate adaptation of that in [15, Proposition 3.6]. 
\medskip
\noindent 
\Proposition {{\bf (5.3.1)} If $ (h_1,h_2)\not\in \HH_- \HH_+$, then $ \gt 
(h_1,h_2)\in \HH_- \HH_+$
if $\gt $ is contained in some open subset of $ \{\gt;\, t_1\not= 0\}$.}
\medskip
\noindent
So from now on we assume that $ (h_1, h_2)\in \HH_- \HH_+$.
Keeping this in mind we consider the Riemann-Hilbert splitting
$$
\gt (h_1,h_2) = \big(
g^-(\gt, \mu )^{-1},\ g^+(\gt,\lambda )^{-1}\big)\big( b_R(\gt
,\mu),\, b_r (\gt,\lambda
)\big),  \leqno (5.3.1)
$$
such that the first factor lies in $ \HH_-$, the second one in $ \HH_+$.  Note
that $ b_R (\gt, \mu )$
and $ b_r(\gt, \lambda )$ are analytic in $\gt $, since $ \gt
(h_1,h_2)$ is analytic and the
map $\HH_- \HH_+\to \HH_-\times \HH_+$ is an analytic diffeomorphism.

Let $ \partial _j$ denote ${\partial\over \partial tj}$ for any $ j\in
\ints $.

Differentiating (5.3.1) and using that $ \gs$ is abelian, we get
$$
(\partial j\, b_R) b^{-1}_R = \Omega ^R_j\ ,\quad \hbox{where}  \leqno
{(5.3.2)}
$$
$$
\Omega ^R_j = (\partial _jg^-) (g^-)^{-1} + g^- E^R_j
(g^-)^{-1}\quad\hbox{and}\leqno
(5.3.3)
$$
$$
(\partial_j b_r) b^{-1}_r = \Omega ^r_j\ ,\quad\hbox{where}  \leqno(5.3.4)
$$
$$
\Omega ^r_j = (\partial_j g^+) (g^+)^{-1} + g^+  E^r_j
(g^+)^{-1}\ .\leqno(5.3.5)
$$

We call $ \Omega _j$ the $\underline{j^{\rm th}\ {\rm potential}}$ and call
it \underbar{positive}
(or \underbar{negative}) iff $ j > 0$ (or $ j< 0$).  The equality of the
mixed derivatives of $ b_R$,
i.e. $\partial_{ij} b_R = \partial _{ji} b_R$, yields the equation
$$
\partial _i\Omega ^R_j - \partial _j\Omega ^R_i = [\Omega ^R_i,\Omega
^R_j],  \leqno{(5.3.6)}
$$
$$
\partial _i\Omega ^r_j - \partial _j\Omega ^r_i = [\Omega ^r_i,\Omega
^r_j].  \leqno{(5.3.7)}
$$
The conditions (5.3.6) and (5.3.7) are usually called
\underbar{Zero-Curvature Conditions (ZCC)}.
\medskip
Recall from Proposition (2.2.2) that every $\gp$ is defined by choosing a subset $X$ of simple roots which in turn defines a subset of the positive root system 
called $\tilde{\triangle }_+$.  Let us define $s$ to be the maximal height 
of the  roots in $\tilde{\triangle }_+$. Since $\tilde{\triangle }_+$ is 
finite, s is a finite, positive integer.  We set $s=0$ if $X$ is empty.  With this notation we can describe   
another important feature of the $ \Omega ^*_j ,\ * = R$ or $r$:

\noindent
\Proposition {{\bf(5.3.2).}
\item {(a)}  If $j> 0$, then there exists a
Laurent polynomial $\Omega
_j$ in $ \lambda $ such that $ \Omega _j\vert S^r = \Omega ^r_j$ and $
\Omega _j\vert S^R = \Omega
^R_j $.   Moreover, $ \Omega _j\in \gp$ and $\Omega _j$ contains only
components of the $ p$-degree between $0$ and $j$, and the canonical degree between $-s$ and $j$.
\item {(b)} If $ j< 0$, then there exists a polynomial $\Omega _j$ in
$\lambda ^{-1}$ such that $
\Omega _j\vert S^r = \Omega ^r_j$ and $ \Omega _j\vert S^R = \Omega ^R_j$.
Moreover, $ \Omega _j \in
\gq$ and $\Omega _j$ contains only components of the $p$-degree between $j$ and
$-1$, and the canonical
degree between $j -s$ and
$-1$.
}\medskip

\noindent
\Proof . (a)  From (5.3.5) we see that $ \Omega ^r_j$ is in $ \gp ^r$ for
$r > 0$.  From (5.3.4) we
know that $ \Omega ^r_j$ can be extended to the region between $ S^r$ and $
S^R$.  Now (5.3.3) and
(5.3.2) show that only finitely many $p$-degrees (resp. canonical degrees)
can occur in $ \Omega
^r_j$.  The rest is a matter of comparing $\Omega ^r_j$ with 
$\Omega ^R_j$.  For example, in order to count the canonical degrees we count
the powers of $\lambda$ appearing simultaneously in $\Omega ^r_j$ and
$\Omega ^R_j$.  For example, for $j>0$, $\Omega ^R_j$ has degrees $\le j$ whereas $\Omega ^r_j$ has degrees $\ge -s$ on account of the first term in (5.3.5).  All other cases are dealt with analogously.  \blsq
\vfill
\eject
\noindent
\medskip
{\bf\S 6. Systems of Partial Differential Equations obtained from Factorization}
\medskip

\noindent
{\bf 6.1 }\quad In this section we derive systems of PDE's from the
\underbar{zero-curvature} \underbar{conditions} (5.3.6) and (5.3.7).  We
treat here the case $ j > 0$ for
which we show that all potentials $ \Omega _j, j> 0$, can be expressed in
terms of a certain number
of functions parametrizing $ \Omega _1$.
We give an example of the generalized Drinfeld-Sokolov system
generalizing the results of
Drinfeld and Sokolov [3] for maximal parabolic subalgebras.  We would like to 
add that the possibility of this extension is already mentioned in [3, p.2014] where it is referred to as the ``partially modified'' generalized KdV equations.

In this chapter we will always assume  $ (h_1, h_2)\in \HH _- \HH _+$ in the splitting equation 
(5.3.1).  In view of
Proposition (5.3.1) this is a very mild restriction.

The assumption above implies that there exists some $(c_R,c_r)\in \HH _+$ such that
$$
h_1 = h^- c_R\quad\hbox{for some}\quad h^-\in Q^R  \leqno (6.1.1)
$$
$$
h_2 = h^+ c_r\quad\hbox{for some}\quad h^+\in P^r\ .  \leqno (6.1.2)
$$

Then (5.3.1) shows
$$ g^- (\un0 ,\mu ) = h^- (\mu )\quad\hbox{and}\quad g^+(\un0 ,\lambda
) = h^+ (\lambda )\
.  \leqno(6.1.3)
$$
Therefore, for sufficiently small $ \gt $, there exist
$$
q (\gt ,\mu )\in \gq^R\quad\hbox{and}\quad p (\gt ,\lambda )\in \gp^r
$$
such that
$$
g^-(\gt, \mu ) = h^-(\mu)\exp \big(q(\gt,\mu )\big)  \ ,
\leqno (6.1.4)
$$
$$
g^+(\gt, \lambda ) = h^+(\lambda)\exp \big(p(\gt,\lambda )\big) \ .\leqno (6.1.5)
$$

Using this, the potentials defined in (5.3.3) and (5.3.5) can be expressed as
$$
\Omega ^R_j =Ad(h^-)( (\partial _j e^q) e^{-q}+ e^q E^R_j
e^{-q})\ ,
\leqno(6.1.6)
$$
and 
$$
\Omega ^r_j = Ad(h^+)((\partial _j e^p) e^{-p} + e^pE^r_j  e^{-p}),
\leqno(6.1.7)
$$
\medskip

\noindent
{\bf 6.2 }\quad As outlined above, we are trying to express all potentials
$ \Omega ^R_j$ resp. $
\Omega ^r_j$, $j>0$ in terms of a set of basic, ``independent'', functions.  To
achieve this, we will make
use of the following.
\medskip

\noindent
\Lemma {{\bf (6.2.1).}\quad  Let $h(\gt )\in \GG $.  Then
$$
(\partial _j e^h) e^{-h} = \sum_{n\ge 1} {1\over n!}(ad\, h)^{n-1}
\partial_jh\ . \leqno (6.2.1)
$$
In other words, $ (\partial _j e^h) e^{-h} = \psi (ad\, h)\partial _j\,
h$, where $ \psi :\C\to
\C $ is the entire function defined by
$$
\psi(z) = \cases{{e^z - 1\over z} ,&\ \ $z\not= 0$\cr
\ \ \ 1 ,&\ \ $z = 0$\ .
}$$
}\medskip

\noindent
\Remark .\quad For matrix groups, (6.2.1) can be proved by a
straightforward induction.  For a
general proof, see [25, chapter II, Theorem 1.7] or 
[23, Ch.3, \S 6, Proposition
12].  Note that for any Lie group, $ (\partial _j e^h) e^{-h}$ is
contained in its Lie algebra.
In particular we have
$$
(\partial _j e^q) e^{-q}\in\gq\quad\hbox{and}\quad (\partial _j e^p)
e^{-p}\in \gp\ .
\leqno(6.2.2) $$
\medskip

\noindent
{\bf 6.3 }\quad In this section we will show how every positive potential $
\Omega _j$ can be
expressed as a differential polynomial in the components of $ \Omega _1$.
To accomplish this, we
need the following:
\medskip
\noindent
\Lemma {{\bf (6.3.1).} For every sufficiently
small $ q\in \gq $
there exist uniquely determined
$$
q_I\in (Im\, ad\, E)_-=(Im\, ad\,E)\cap \gg _- 
\quad\hbox{and}\quad q_K\in (Ker\, ad\, E)_-=(Ker\, ad\, E)\cap \gg _-
$$
such that
$$
e^q = e^{q_I} e^{q_K}  \leqno(6.3.1)
$$
and $q_I+q_K \in \gq$.
}\medskip

\noindent
\Proof .\quad Recall (5.1.2):\ \ $\GG = Im\, ad\, E\oplus Ker\, ad\, E$.  Since
 $Im\, ad\, E, $ and $Ker\, ad\, E$ are canonically graded, we have\hfil\break
$ \GG _- = (Im\, ad\, E)_-\oplus (Ker\, ad\, E)_-$.  On the other hand $\gq \subset \gg_-$, therefore the map
$$
\cases {(Im\, ad\, E)_- + (Ker\, ad\, E)_- &$\rightarrow U_-$\cr
(q_I, q_K) &$\mapsto e^{q_I} e^{q_K}$
}\leqno(6.3.2)
$$
is a local diffeomorphism.  The last statement of the theorem
is obtained by considering the curve $e^{tq}$ and using the Baker-Hausdorff
formula for small $t$.  \blsq
\medskip

We are now ready to prove
\medskip

\noindent
\Theorem {{\bf (6.3.2).}\quad For $ j > 0$ the potential $ \Omega _j$ is a universal $
\partial _1$-differential
polynomial in $ \Omega _1$ with rational coefficients.
}\medskip

\noindent
\Remarks .\quad By ``$y$ is a $\partial _k$-differential polynomial in $x$''
we shall mean that each
component of $y$ is a polynomial in the components of $x$ and its
derivatives with respect to $x$.

\noindent The word ``universal polynomial'' means that the coefficients of the polynomial are independent of $\Omega_1$. 

\medskip

\noindent
\Proof . (a)  We choose any $h^- \in Q$ such that 
$g^- (\gt)=h^- \exp(q)$
and \hfill\break \hbox{$\exp(q)=\exp(q_I)\exp(q_K)$}.  
We then substitute $g^- (\gt)=h^- \exp(q)$ 
into (6.1.4) and set $ j = 1$ in (6.1.6) 
to get
$$
\Omega ^R_1 = Ad(h^-)\big\{(\partial _1 e^{q_I}) e^{-q_I} +
e^{q_I}(\partial _1 e^{q_K})
e^{-q_K} e^{-q_I} 
+ e^{q_I} E^R e^{-q_I}\big\}. \leqno (6.3.3)
$$
Since $Ker\, ad\, E^R$ is abelian, formula (6.2.1) shows
$$
(\partial _1 e^{q_K}) e^{-q_K} = \partial _1 q_K\ .  \leqno (6.3.4)
$$
Using this and (6.1.7) we get
$$
h^-.\Omega ^R_1 = (\partial _1 e^{q_I}) e^{-q_I} + e^{q_I}
(\partial _1 q_K + E^R) e^{-q_I}
\leqno (6.3.5)
$$
where $h^-.\Omega ^R_1 = Ad^{-1}(h^-)\Omega ^R_1$, and where we have 
used that $q_K$ and $E^R$ commute.
\medskip

\noindent (b)  We will show that $ q_I$ and $ \partial _1q_K$ are determined by
$h^-.\Omega _1$.  To see
this, we rewrite (6.3.5) as 
$$
(\partial _1 e^{-q_I}) e^{q_I} + e^{-q_I} h^-.\Omega ^R_1 e^{q_I} =
\partial _1 q_K + E^R\ . \leqno
(6.3.6)
$$
In the next step we  compare terms of the same canonical degree.  
For $ q_I, q_K\in\GG _- $ we may write:
$$
q_I = q_{I,-1} + q_{I,-2} + \ldots , \leqno (6.3.7)
$$
$$
q_K = q_{K,-1} + q_{K,-2} + \ldots . \leqno (6.3.8)
$$
Similarly, by Proposition (5.3.2), we have $$
h^-.\Omega ^R_1 = (h^-.\Omega ^R_1)_1 + (h^-.\Omega ^R_1)_0 +
(h^-.\Omega ^R_1)_{-1} + \ldots
\leqno (6.3.9)
$$
We will prove by induction on the canonical degree $m$ that $ q_I$ and $\partial _1q_K$ are determined by $ h^-.\Omega
^R_1$.  For $ m = 1$ we obtain from (6.3.6) that
$(h^-.\Omega ^R_1)_1= E^R$
holds.  For $ m = 0$ we get
$$
[-q_{I,-1}, E^R] + (h^-.\Omega ^R_1)_0 = 0
$$
Since $ (h^-.\Omega ^R_1)_1= E^R$ and $ad\, E^R$ is bijective on its
image, $ q_{I,-1}$ is
uniquely determined by $ h^-.\Omega ^R_1$.  To
illustrate the procedure
we consider the case $ m = -1$ separately.  Here we have
$$
\eqalign{
&-\partial _1 q_{I,-1} + [-q_{I,-2},E^R] + {1\over
2}\big[ q_{I,-1},[q_{I,-1},
E^R]\big]  \cr
&-[q_{I,-1},(h^-.\Omega ^R_1)_0] + (h^-.\Omega ^R_1)_{-1} = \partial _1
q_{K,-1} \ .
}$$
The sum
$$
[q_{I,-2}, E^R] + \partial _1 q_{K,-1}
$$
is therefore some $ \partial _1$-differential polynomial in $(h^-.\Omega ^R_1)_0$ and $(h^-.\Omega ^R_1)_{-1}$. 
By projecting on $ Im\, ad\, E^R$ along $ Ker\, ad\, E^R$ and vice
versa, we conclude that each term in the sum is a $\partial
_1$-differential polynomial in $(h^-.\Omega ^R_1)_0$ and $(h^-.\Omega ^R_1)_{-1}$.  Thus $ \partial _1
q_{K,-1}$ is a $\partial _1$-differential polynomial in $(h^-.\Omega ^R_1)_0$ and $(h^-.\Omega ^R_1)_{-1}$.
Since $ ad\, E^R$ is bijective when restricted to its image, $ q_{I,-2}$ is
a $ \partial _1$-differential polynomial in $(h^-.\Omega ^R_1)_0$ and $(h^-.\Omega ^R_1)_{-1}$.

By the same token, for $ m \le -1$, we obtain from (6.3.6) that
$$
[q_{I,-i-1},E^R] + \partial _1 q_{K,-i}
$$
is some $ \partial _1$-differential polynomial in $ (h^-.\Omega ^R_1)_0,
(h^-.\Omega ^R_1)_{-1}, \cdots (h^-.\Omega ^R_1)_{-i}$, and by
repeating the argument
given for $ m = -1$ we see that
$$
q_{I,-i-1}\quad\hbox{and}\quad \partial _1 q_{K,-i}\quad\hbox{are}
$$
$\partial _1$-differential polynomials in $ (h^-.\Omega ^R_1)_0,
(h^-.\Omega ^R_1)_{-1}, \cdots (h^-.\Omega ^R_1)_{-i}$. From the definition of $h^-.\Omega _1^R$ we 
immediately see that we can separate off the explicit dependence on
$h^-$.  As a result of this operation 
$$
q_{I,-i-1}\quad\hbox{and}\quad \partial _1 q_{K,-i}\quad\hbox{become}
$$
$\partial _1$-differential polynomials in $ (\Omega ^R_1)_0,
(\Omega ^R_1)_{-1}, \cdots (\Omega ^R_1)_{-i} \quad \hbox{and polynomials 
in the entries of $h$}$. Thus we can write $q_{I,-i}(h^-,\Omega ^R_1)$  etc.

\noindent (c) After these preparations we are able to prove the theorem.  
From Proposition 
(5.3.2) we know that for $ j> 0$, $ \Omega _j$ has only components of
nonnegative $p$-degree.  Thus 
$$
\Omega ^R_j = \big( (h^-) e^{q_I(h^-,\Omega ^R_1)} E^R_j e^{-q_I(h^-, \Omega ^R_1)} 
(h^-)^{-1}\big)^{(+)}\ ,  \leqno
(6.3.10)
$$
since $ q_K$ and $ E^R_j$ commute.  Note that $ \Omega^R_j$ is uniquely
determined by $ q_I$, which in
turn is uniquely determined by $ \Omega ^R_1$.  To see that   
$ \Omega ^R_j$
depends polynomially on $\Omega ^R_1
$, we use the fact that $\gg^{1}\oplus\gg^{0}$ is a finite dimensional vector space
and therefore  any $x \in \gg_-$, such that $cdeg(x)$ is sufficiently 
negative, must be 
in
$\gq$.  Now, let us apply this remark to (6.3.10).
We obtain that 
$$
\leqalignno{\Omega ^R_j &= \big( (h^-) \big(\sum_{i=0}^{N}ad^i(q_I(h^-,\Omega ^R_1))( E^R_j)  
\big)(h^-)^{-1}\big)^{(+)}\cr &= 
\big( (h^-)\big( \sum_{i=0}^{N}ad^i(\sum_{l=1}^{N}q_I(h^-,\Omega ^R_1)_{-l})( E^R_j) \big) 
(h^-)^{-1}\big)^{(+)}, &(6.3.11)} 
$$  
for sufficiently large $N$.  Thus $\Omega ^R_j$ depends on finitely many 
$q_I(h^-,\Omega ^R_1) _{-i}$, all of which are differential polynomials in $\Omega ^R_1$ and polynomials in the entries of $h^-$.  
However, by repeating the whole argument  for $h^-$ in 
a neighborhood  of the identity we conclude that $q_I(\Omega ^R_1), \partial _1 q_{K}(\Omega ^R_1)$, in other words, both are constants as functions 
of $h^-$. Since the dependence of $q_{I,-i}$ on $h^-$ is polynomial, thus analytic, we obtain that $q_{I, -i}$ is a differential polynomial
in $\Omega ^R_1$, with no explicit dependence on $h^-$.  
Moreover, we used only linear operations (projections, commutators) to determine $q_I$, consequently, the coefficients in  $q_I(\Omega ^R_1)_{-i}$
are all rational numbers independent of $\Omega ^R_1$.  This completes the proof.
\blsq
\medskip
Below we will not use the superscripts $R$ or $r$ to distinguish
the two circles appearing in our discussion.  The context will clearly tell the reader if
one needs to make a distinction. In particular we will not attach a superscript to $E$.  
\medskip 
\noindent
\Corollary {{\bf (6.3.3).}\quad The ZCC
$$
\partial _j\Omega _i - \partial _i\Omega _j = [\Omega _j,\Omega _i ],\quad
i,j > 0\ ,\leqno (6.3.12)
$$
is a system of equations for the scalar components of $ \Omega _1$.  In
particular, for $i = 1$ we
obtain a system of evolution equations for the scalar components of $
\Omega _1$.
}\medskip

\noindent
We will discuss (6.3.12) for $ i = 1$ in more detail later.  There, it will
be important to know a
priori how many equations we expect to obtain.  To this end we consider
$$
\unU = [\gq^{(-1)},E^{(1)}],\leqno (6.3.13)
$$
where the superscript denotes as usual the $p$-degree.  We note that
$$
(h^- E(h^-)^{-1})^{(1)} = E^{(1)}
$$
since $ h^-\in Q$.  Thus, by (6.1.6), $\Omega _1 -E \in \unU$ 
so essentially $ \unU$
is the space where
$ \Omega ^R_1$ naturally lives.  
\medskip
\noindent
\Proposition {{\bf (6.3.4).}\quad Let $q$ be any smooth function of $\gt$ with values in
$\gq$ and set $\Omega _1 =Ad(h^-)(
(\partial _1 e^q) e^{-q} + e^qE e^{-q})$ and $ \Omega _j = (Ad(h^-e^q)
E _j)^{(+)}
,\ j > 0$.  Then
$$
\partial _j\Omega _1 - \partial _1\Omega _j - [\Omega _j, \Omega _1]\in \unU\ .
$$
}\medskip

\noindent
\Proof .\quad We know that $ \Omega _1 -E$ and all its partial derivatives are
contained in $ \unU$.  Next we
note
$$
\eqalign{
\partial _1 (Ad(h^-e^q)
E _j) &= Ad(h^-)([(\partial _1 e^q) e^{-q},\,
Ad(e^q)E _j ])\cr
&= [\Omega _1,\ Ad(h^-e^q)E _j]\ .}
$$
Hence
$$
\eqalign{
\partial _j\Omega _1 - \partial _1\Omega _j - [\Omega _j, \Omega _1] &=
\partial _j\Omega _1 -
[\Omega _1, Ad(h^-e^q) E _j]^{(+)} + [\Omega _1, \Omega _j]\cr
&= \partial _j \Omega _1 - \big[\Omega _1, \Omega _j + (Ad(h^-e^q)E _j
)^{(-)}\big]^{(+)} +
[\Omega _1, \Omega _j]  \cr
&= \partial _j \Omega _1 - \big[\Omega _1, (Ad(h^-e^q)E _j)^{(-)}\big]^{(+)} \cr
&= \partial _j\Omega _1 - \big[ E^{(1)},\ (Ad(h^-e^q)E _j)^{(-1)}\big] \in \unU, 
} 
$$
where we used that $\Omega _1 ^{(1)} =E^{(1)}$.  \blsq
\medskip

\noindent
{\bf 6.4 }\quad In the next section we intend to show that the entries 
of $ \Omega _1$,
relative to a certain basis, are all $ \partial _1$-polynomials in $ \dim\GG
_0$ ``basic
functions.''  The present section collects all necessary algebraic facts 
needed in the proof of the forthcoming Theorem (6.5.1).   The main 
idea of this theorem is to exploit (6.3.3) by projecting both sides of that 
equation onto a subspace of $\unU$ given by the kernel of 
certain operator ($ Ker (B)$ 
in the notation appearing in the proof of Theorem (6.5.1)).  The dual of 
this subspace is $\unW\slash \widetilde \unW$ in the notation of this section.  
It is exactly the relation between $\unU,\unW \hbox { and }  \widetilde \unW$ 
that we study in this section.  

The discussion in this section deals exclusively with $\gg^{fin}$ .  Therefore we omit the superscript ``fin'' in this section.  
First, we introduce the following notation.  
Let $(.|.)$ denote 
the canonical nondegenerate symmetric
bilinear form
on $ \GG $ as in [10, Theorem 2.2].  We furthermore define
$$
\gb = \GG ^{(0)} \cap (\GG_- \oplus \GG_0) = \gb _0 \oplus \gb _{-1}
\oplus\ldots \oplus \gb _{-s}\
,\leqno (6.4.1)
$$
where $s$ is the maximal integer such that $\gb _{-s} \not= \{ 0\}$.
$$
\gb ^* = \GG ^{(0)} \cap (\GG _0\oplus\GG_+) = \gb _0 \oplus \gb _1\oplus
\ldots \oplus \gb _s\
,\leqno (6.4.2)
$$
$$
\gU _{-k} = [\GG ^{(-1)}_{-k-1},E^{(1)}]\ ,\ \ 0\le k \le s\leqno (6.4.3)
$$
$$
\gU = \gU _0 \oplus \gU _{-1} \oplus \ldots \oplus \gU _{-s}\ ,  \leqno (6.4.4)
$$
$$
\gW _k = \big\{A_k \in \GG _k :\big( [\gq_{-k-1}, E]\vert A_k\big) =
0\big\}\ ,  \leqno (6.4.5)
$$
$$
\gW = \gW _0 \oplus \gW _1\oplus \ldots  \oplus \gW _s\ ,\leqno (6.4.6)
$$
$$
\widetilde \gW _k = \big\{ A_k\in \gW _k : (A_k\vert \gU_{-k}) = 0\big\}\
,\leqno (6.4.7)
$$
$$
\widetilde \gW = \widetilde \gW _0\oplus \ldots \oplus \widetilde \gW _s\ .
\leqno (6.4.8)
$$
We also define
$$
\gr _{-k} :=(Ker\, ad\, E)\cap \gq_{-k}\quad\hbox{for}\quad k = 1,\ldots , s,  
\leqno (6.4.9a)
$$
$$
\gr _{k} :=(Ker\, ad\, E)\cap \gq^*_{-k}\quad\hbox{for}\quad k = 1\ldots 
, s,  \leqno (6.4.9b)
$$
where $ \gq ^*_{-k}$ is the natural dual of $ \gq_{-k}$ with respect to
$(.|.)$.  Note that $ \dim \gr _{-k} = \dim \gr _k $ for
$ k = 1,\ldots , s$.

We collect some information about the dimensions of the vector spaces $ \gU
, \gW $ and $\widetilde
\gW$.  We will use the following notation.  For $\alpha \in \triangle,\, 
\alpha =\sum k_i \alpha \buildrel \vee \over \alpha _i$ we set $ht(\alpha)=
\sum k_i$.  More generally, for a fixed $X \subset \Pi$ we set $pht(\buildrel
\vee \over \alpha _i)=0$ if $i\in \Pi$, $1$ otherwise.  
\medskip

\noindent
\Theorem {{\bf (6.4.1).}\quad
$$
\gW _k = \big\{ A_k \in \GG _k : [E,A_k]\in \gb _{k+1}\big\}\ \ \hbox{for}\ \
k = 0,\ldots , s\
.\leqno {\rm (a)}
$$
$$
\dim \gW = \dim \gb - \dim \GG _0 + \sum ^s_{k=1} \dim \gr_{-k}\leqno{\rm (b)}
$$
$$
\dim \gU + \dim\widetilde \gW = \dim \gb + \sum ^s_{k=1} \dim \gr _k .\leqno
{\rm (c)}
$$
}\medskip

\noindent
\Proof . \quad (a)  We know that 
$$
\eqalign{
\gW _k &= \big\{ A_k\in \GG _k : \big( \gq _{-k-1}\ \big\vert\ [E,
A_k]\big) = 0\big\}\cr
&= \big\{ A_k\in \GG _k : [E, A_k]\in \GG ^{(0)} \cap \GG_{k+1} =
\gb_{k+1}\big\}\ .
}$$
The latter equality holds because, for $ 1\le k\le s,\ (\gq _{- k})^{\bot}
\cap \gg _k =\gg ^{(0)}\cap
\gg_k$.  Indeed, if $x_{\alpha} \in \gq _{- k})^{\bot}\cap \gg_k$ then the height of $\alpha$ , $ht(\alpha)$, equals $k$. 
On the other hand the corresponding $pht(\alpha)$, when only roots from $X$ are counted, satisfies $pht(\alpha) \ge 0$.  We want to show that $pht(\alpha)=0$.
Assume therefore that $pht(\alpha)>0$. Then $pht(-\alpha) <0$ and $x_{-\alpha}
\in \gq _{-k}$.  Since $(x_{-\alpha},x_{\alpha})\ne 0$ we get a contradiction
as $x_{\alpha} \in (\gq _{- k})^{\bot}$.  This proves that $pht(\alpha)=0$.  
For $ k = s + 1$, on the other hand, $  \gq_{ -(s +
1)}^{\bot}\cap \gg_{s+1} = 0\equiv
\gb_{s+1}$.
\medskip

(b)  Since $(.|.)$ is nondegenerate on $ \GG_{-k} \times \GG _k$, we get from
(6.4.5) the
relation 
$$
\dim \unW _k = \dim \GG _k - \dim [\gq_{-k-1}, E]\ .
$$

Furthermore 
$$
\dim \unW _k = \dim\GG _k - \dim \gq _{-k-1} + \dim \gr_{-k-1}\quad\hbox
{for}\ \ k = 0, 1,\ldots,
s-1\ .\leqno (6.4.10)
$$
For $ \unW _s$ we note that $ \gq_{-s-1} = \GG_{-s-1}$, thus:
$$
\leqalignno{
\unW _s &= \big\{ A_s\in \GG _s :\big( [\GG _{-s-1}, E]\ \vert\ A_s\big) =
0\big\} &(6.4.11)\cr
&= (Ker\, ad\, E)_s\ .
}$$

From (6.4.10) and (6.4.11) , we get
$$
\eqalign{
\dim \unW &= \dim \GG _0 + \sum ^{s-1}_{k=1} (\dim\GG _{-k} - \dim
\gq_{-k}) - \dim \gq_{-s} \cr
&+ \dim(Ker\, ad\, E)_s + \sum ^s_{k=1} \dim \gr _{-k} \ .}
$$
Since $\gg_{-k}=\gq _{-k}\oplus\gb_{-k}$, $\dim \gb _{-k}=\dim\GG _{-k} - \dim
\gq_{-k}$, $k=1\ldots s$. Therefore
$$
\dim \unW=\dim \GG _0 + \left(\sum ^s_{k=1} \dim \gb _{-k}\right) - \big( \dim \GG
_{-s} - \dim (Ker\,
ad\, E) _s\big) + \sum ^s_{k=1} \dim \gr_{-k}.
$$
Since $\dim \gg_0 =\dim \gb_0$, we obtain that
$$
\dim \unW = \dim \gb - \big( \dim \GG _s - \dim (Ker\, ad\, E)_s\big) + \sum
^s_{k=1} \dim \gr _{-k}.  \leqno (6.4.12)
$$
By virtue of [10, Proposition 14.3a],
$$
\dim \GG _s - \dim (Ker\, ad\, E)_s = \dim\GG _0\ ,  \leqno (6.4.13)
$$
thus
$$
\dim \unW = \dim \gb - \dim \GG _0 + \sum ^s_{k=1} \dim \gr_{-k}\ .  \leqno
(6.4.14)
$$
(c)  Let $ \unV _k = (\unU _{-k})^{\bot}\cap \gb_k$.  Then the space $ \unV
_k\subset \gg _k^{(0)}$ satisfies for $k=0,1 \ldots , s$
$$
\leqalignno{
&\unV _k \subseteq \gb _k\ ,  &(6.4.15)\cr
&\dim \unU _{-k} + \dim \unV _k = \dim \gb _k\quad\hbox{and}\cr
&(\unU _{-k}\ \vert\ \unV _k) = 0, \ .
}$$
We claim that $ \widetilde\unW _k = \unV _k\oplus \gr _k$ holds.

To verify $ \unV _k\subset \widetilde\unW _k$, we decompose $ E = E^{(0)} +
E^{(1)}$ and derive
from the definitions of $ \unV _k$ and $\unU _k$ immediately that 
$$
[\unV _k, E^{(1)}] = 0  \leqno (6.4.16)
$$
Moreover, since $ E^{(0)}\in \GG
^{(0)}_1$, we have
$$
[\unV _k, E^{(0)} ]\subset \gb_{k+1}\ .  \leqno (6.4.17)
$$
This implies
$$
[\unV _{k}, E] \subset \gb _{k+1}\ .  \leqno (6.4.18)
$$
From (a) we now obtain $ \unV _k\subset \unW _k$.  This together with the
definition of $ \unV _k$ and (6.4.8) 
shows
$$
\unV _k \subset \widetilde\unW _k\ .  \leqno (6.4.19)
$$
Since $ \gr_k\, \bot\, \gp$ we obtain that $\gr_k \subseteq \widetilde \unW _k$.  Since $
\unV _k \subset \GG ^{(0)}$ and $
\gr _k \in \GG ^{(+)}$, we have $\unV _k +\gr_k=\unV _k \oplus \gr_k$.

Now let $ A_k\in \widetilde\unW _k$.  Note that
$$
\gb _k \oplus (\gq_{-k})^* = \GG _k\ .  \leqno (6.4.20)
$$
Let $ B_k\in \gb _k,\ C_k\in (\gq_{-k})^*\subseteq \GG^{(+)}$ such that $
A_k = B_k + C_k$.  Then
$$
(A_k\, \vert\, \unU_{-k}) = (B_k\,\vert\, \unU_{-k}) + (C_k\,\vert\, \unU
_{-k} )\ .\leqno (6.4.21)
$$
Here the second term vanishes, since $ C_k\in \GG^{(+)}$ and $ \unU_{-k}\in
\GG^{(0)}$.  Therefore,
also $ B_k\in\unV _k$,  whence $ B_k\in\widetilde W_k$ by (6.4.19).  This
in turn means that
$C_k\in\widetilde \unW _k$ by (6.4.8), and in particular $ C_k\in \unW _k$.  Hence, by
virtue of (a),
$$
[C_k, E]\in \gb_{k+1}\ .  \leqno (6.4.22)
$$
Since $ C_k \in \GG^{(1)}$ and $ E = E^{(0)} + E^{(1)}$,
$$
[C_k, E]\in  \gb _{k+1}\cap \GG^{(+)} = \{ 0\}\ .\leqno (6.4.23)
$$
Thus
$$
C_k\in (Ker\, ad\, E)\cap (\gq _{-k})^*\ , \leqno (6.4.24)
$$
whence $ \widetilde \unW _k\subseteq \unV _k + \gr _k$.

This proves that $ \widetilde\unW _k = \unV _k\oplus \gr _k$.

Now we have
$$
\dim \widetilde\unW _k = \dim\unV _k + \dim \gr _k   \leqno(6.4.25)
$$
Hence (6.4.15) implies
$$
\leqalignno{
\dim \unU_{-k} + \dim \widetilde\unW _k &= \dim\unU_{-k} + \dim\unV _k +
\dim \gr _k &(6.4.26)\cr
&=\dim \gb _k + \dim \gr _k
}$$
Summation over $ k = 0,\ldots, s$ completes the proof.\blsq
\medskip
\noindent
{\bf 6.5. }\quad In this section we will roughly show that among the scalar
components of $ \Omega _1$, a
subset of cardinality rank $ \GG = \dim \GG _0$ can be chosen, such that
the remaining functions
are $ \partial _1$-differential polynomials in those.

More precisely, we show
\medskip

\noindent
\Theorem {{\bf (6.5.1).}\quad  There exists a basis $c_1,\ldots, c_m$ of $
\unU$, such that in the
expansion
$$
\Omega _1 = E^{(1)} + \sum ^m_{k=1} u_k c_k
$$
the coefficient functions $ u_{l+1},\ldots, u_m$ are $\partial
_1$-differential polynomials in
$u_1,\ldots , u_{\ell}$, where $ \ell = $ rank $\GG = \dim\GG_0$.
}\medskip

\noindent
\Proof .\quad From Proposition (5.3.2) we know that
$$
\Omega  _1 = \Omega ^{(0)}_1 + \Omega ^{(1)}_1 \ , \leqno (6.5.1)
$$
where the superscript denotes as usual the $p$-grading.  From (6.1.6) 
we obtain
$$
\Omega ^{(1)}_1 = E_1^{(1)}  \leqno (6.5.2)
$$
and we write $ \Omega _1 = E + U_0 + \ldots + U_{-s}$, where $ U_{-k}$ is
homogeneous of canonical
degree $ -k$.  Next we write (6.3.3) in the form
$$
\Omega _1 = Ad(h^-) \big((\partial _1 e^{q_I}) e^{-q_I} + e^{q_I}(\partial _1
q_K + E )e^{-q_I}\big).
\leqno (6.5.3)
$$

Decomposition into components of canonical degree $-j$ yields:
$$
U_0 = [q_{I,-1}, E] +R_0(h^-), \leqno (6.5.4a)
$$
where $R_0(h^-)=[h^-_{-1},E]$ for some $h^-_{-1} \in \gg$,
$$
\leqalignno{
\quad U_{-j} = [{q_{I,-(j+1)},E}] + &\partial _1 q_{K,-j}& \cr + &R_j\big(
q_{I,-1},\ldots, q_{I,-j}, q_{K,-1},\ldots, q_{K,-(j - 1)}, h^-\big), \, \quad 1\le j\le s, &(6.5.4b) \cr}$$
where $R_j$ is a differential polynomial in its arguments.  From Theorem
A.1.2, it follows that all
dependence on $ q_K$ in (6.5.4b) drops out, thus (6.5.4b) simplifies to
$$
U_{-j} = \big[ q_{I, -(j+1)},E\big] + R_j(q_{I,-1},\ldots, q_{I,-j},h^-)\
.\leqno (6.5.4c)
$$
We can, furthermore, simplify (6.5.4c) by eliminating $ q_{I,-1},\ldots
q_{I,-j}$.  Indeed, using
(6.5.4a) and solving (6.5.4c) for $ q_{I,-k}, 0\le k\le j$, we can
write (6.5.4c) as:
$$
U_{-j} = \big[q_{I,-(j+1)},E\big] + \tilde R_j (U_0,\ldots, U_{-j+1}, h^-),\quad 1\le
j\le s\ .\leqno (6.5.5)
$$
Note, however, that in the neighborhood of the identity, $\tilde R_j $ is 
independent of $h^-$.  Since $\tilde R_j $ is analytic in $h^-$ we get that
$\tilde R_j=\tilde R_j(U_0,\ldots, U_{-j+1})$.  
Relation (6.5.5)  cut out a subset $\SS $ of $ \unU\otimes \RR$, where $\RR $
is the germ of 
holomorphic maps  at $x = 0$.  We
proceed now to describe $\SS$.  We apply $(\cdot |A_j)$ to (6.5.5), where $ A_j\in \unW
_j\slash \widetilde
\unW _j$.

By (6.4.6), (6.4.8) and (6.5.4a) we get 
$$
(U_{-j}\vert A_j) = \big(\tilde R_{-j} (U_0,\ldots, U_{-j+1})\vert A_j\big)
,\quad 0\le j\le s\ .
\leqno (6.5.6)
$$
Thus we get $ \dim \unW _j - \dim \widetilde\unW _j$ relations for the
components of $ (\Omega
_1)_{-j} = U_{-j}$.  Let us now choose a basis $ b_1, \ldots, b_m$ of $
\unU$, which is consistent
with the canonical grading, i.e.
$$
\unU _0 = \sum_{0 < j\le l_0} u_jb_j,\, \unU_{-1} = \sum_{l _0 <j\le
l_1} u_j b_j, \ldots, \,
U_{-s} = \sum_{l_{s-1}< j\le l_s = m} u_j b_j
$$
and the $u's$ are in $\RR$.  Then the system (6.5.6) takes the form
$$
\pmatrix{ B_0 &0 & & & &\cr
 0 &B_1 & & & &0  \cr
& &\ddots & & &\cr
&&&\ddots &&\cr
 0 & & & & &B_s}\quad\pmatrix{ &\hat u_0 &\cr
                               &\hat u_1 &\cr
                               &\vdots &\cr
                               &&\cr
                               &\hat u_s&\cr} = \pmatrix{ &\hat v_0 &\cr
                                                          &\vdots &\cr
                                                           & &\cr
                                                           &\vdots &\cr
                                                           &\hat v_s &\cr}\leqno
(6.5.7)
$$
where $B_j$ is a $ (\dim \unW _j - \dim \widetilde\unW _j)\times \dim
\unU_{-j}$ matrix with entries
in $ \C $, and
$$
\hat u_j = \pmatrix{ &u_{l_{j-1+1}} &\cr
                     &\vdots\qquad\ \  &\cr
                     &u_{l_j}\qquad\ \  &
}.
$$
Note that by (6.5.4) $\hat v_0=0$.  Moreover, in $\hat v_j, j>0$,
only terms which are differential polynomials in $\hat u_{0},\ldots,\hat u_{l _{j-1}}$
can occur.  By construction rank $ B_j = \dim \unW _j - \dim \widetilde\unW _j\equiv
d_j$.  Thus, in order to
solve (6.5.7) we first solve $ B_0\hat u_0 = 0$ whose solution $ \hat
u^0_0$ is an arbitrary
element in $ Ker\, B_0$.
Clearly,
$$
B_1 \hat u_1 = v_1 (\hat u_0)
$$
and
$$
\hat u_1 = \hat u^0_1 + B^{-1}_1 v_1 (\hat u^0_0),\ \hat u^0_1\in Ker B_1\ .
$$
An easy proof by induction shows that
$$
\hat u_j = \hat u^0_j + B^{-1}_j v_j(\hat u^0_0 ,\hat u^0_1,\ \hat
u^0_{j-1}),\ \hat u^0_j\in Ker\,
B_j\ .  \leqno (6.5.8)
$$
Let us denote by $B$ the linear map $ \unU \to \unU$, whose matrix of
coefficients is given by $
B_0, B_1, \ldots, B_s$ as in (6.5.7).

Then we consider the map $\psi : Ker B\otimes \RR \to \unU\otimes \RR$,
$$
\hat u = \sum ^s_{i=0} \hat u^0_i\ \ {\buildrel\psi\over\longmapsto }\ \
\sum ^s_{i=0} \hat u_j
\leqno (6.5.9) $$
where $ \hat u_j$ is given by (6.5.8).

Note that $ \dim Ker\, B = \dim \unU - (\dim \unW - \dim\widetilde\unW ) =
\dim \GG _0$, the latter
following from Theorem (6.4.1).

The claim of the theorem is proven by choosing a basis
$$
\{ c_1,\ldots , c_{\ell }\}\ \ \hbox{of}\ \ Ker\, B\ \ \hbox{and}\ \ \{
c_{l+1},\ldots , c_m\}\ \
\hbox {of}\ \ \unU /Ker\, B\ .
$$\blsq
\medskip
\noindent
{\bf 6.6 }\quad  In this section we illustrate how Theorem (6.5.1) works 
on three examples.  First we analyze the well known case of the KdV equation 
or rather the potential KdV equation. 
The appropriate Kac-Moody algebra is $\gg =A^{(1)}_1$.  
We choose the Chevalley generators to be: 
$$
\eqalign{
e_0 &= \lambda  E_{21} ,\ e_1 = E_{12},\cr
f_0 &= \lambda ^{-1} E_{12} ,\ f_1 = E_{21}\cr
h_i &:= [e_i, f_i]\quad\hbox{for}\quad i = 0, 1.} 
$$
The parabolic algebra $\gp$ is generated by $\{ e_0,e_1,f_1\}$.  
Our goal is to determine the form of $\Omega _1$.  
First, we find that 
$$
\gq ^{(-1)}=\left \{\pmatrix{{\alpha \over \lambda} &{\beta\over \lambda}\cr
                  {\gamma  \over \lambda}&{\delta \over \lambda}}
: \alpha, \beta, \gamma, \delta \in \Bbb C \right \}
 $$
and 
$$
E^{(1)}=\pmatrix{0&0 \cr
                \lambda&0}.
$$
By (6.3.13) we get 
$$
\frak U=\left \{\pmatrix{\alpha&0 \cr
                \beta&-\alpha}: \alpha, \beta \in \Bbb C\right\}
$$
Consequently,
$$
\frak U_{0}=\left \{\pmatrix{\alpha&0 \cr
                0&-\alpha}:\alpha \in \Bbb C\right \}
$$
and 
$$
\frak U_{-1}=\left \{\pmatrix{0&0 \cr
                \beta&0} :\beta \in \Bbb C\right \}.
$$

Thus $\Omega _1$ is of the form:
$$
\Omega _1=\pmatrix{\alpha&1\cr
                   \lambda +\beta&-\alpha}.
$$
Since the rank of $\gg$ is $1$, we expect to have only one 
function parametrizing $\Omega _1$.  In order to see how Theorem (6.5.1) works 
in this case we essentially go through the main steps of the proof of 
that theorem. First we have to setup equations (6.5.4a) and (6.5.4b).  
To this end we observe that 
$$
U_{0}=\pmatrix{\alpha&0\cr
               0&-\alpha}
$$ and
$$
U_{-1}=\pmatrix{0&0\cr
               \beta &0}.
$$ 
We are allowed to set $h^-=I$ in computations.  Thus, instead of (6.5.3), we 
use 
$$
\Omega _1 = (\partial _1 e^{q_I}) e^{-q_I} + e^{q_I}(\partial _1
q_K + E^R)e^{-q_I}
$$
where, as in Lemma (6.3.1), $q_I+q_K \in \gq$.  We get 
$$
E+U_0+U_{-1}=\partial _1( q_{I,-1}+q_K)+E+[q_{I,-1},E]+[q_{I,-2},E]
+{1 \over 2}[q_{I,-1},[q_{I,-1},E]]
$$
from which we get
$$
\eqalign{U_0&=[q_{I,-1},E], \cr
         U_{-1}&=\partial _1( q_{I,-1}+q_K)+[q_{I,-2},E]
+{1 \over 2}[q_{I,-1},[q_{I,-1},E].}
$$
Now we solve the first equation for $q_{I,-1}$.  The result is:
$$
q_{I,-1}={U_0\over 2} E^{-1}.
$$
To get $q_{K,-1}$ we use $q_I+q_K \in \gq$ which implies:
$$
q_{K,-1}={[U_0]_{11}\over 2}E^{-1}, \quad [U_0]_{11}=\alpha.
$$
This brings us to (6.5.5), where all terms except for the very 
first one, which depends on $q_{I,-2}$, depend on $U_j, j=0,1$.  
Now we would like to recover (6.5.6).  
To this end we apply $(\cdot,A_1), A_1\in \frak W _1 /\widetilde \frak W_1$ 
to both sides of the last equation above.  The symmetric, invariant, 
bilinear form 
$(\cdot,\cdot)$  is defined as follows: $(x\lambda ^n,y \lambda ^m)=
\delta_{m,-n}\,trace (xy),\, x,y \in \gg$.  One directly checks that 
$\frak W _1 /\widetilde \frak W_1 ={\Bbb C} E$ so that the result of applying 
$(\cdot,E)$ to the equation for $U_{-1}$ is:
$$
\beta=\partial _1 \alpha -\alpha ^2,
$$ 
and the final result for $\Omega _1$ is:
$$
\Omega _1 =E+ \pmatrix {\alpha & 0 \cr
                     \partial _1 \alpha -\alpha ^2 & -\alpha }.  
$$
This result should be contrasted with the situation appearing in 
other approach to a generalization of the Drinfeld-Sokolov systems 
[20] in which $U=\Omega _1 -E$ is in a fixed 
complementary subspace to $[E^{(0)},\gg ^{(0)}\cap \gg _-]$ in 
$\gb $.  This is not the case with us. 
To finish this example we note that 
the resulting hierarchy of equations on $\alpha$ is 
the potential KdV hierarchy. 

So far we have looked at an example of a $2 \times 2$ system.  Below we 
present  two examples of systems of PDEs which can be obtained
by using the results of Theorems (6.3.2) and (6.5.1), but which are related to 
$3 \times 3$ systems.  The resulting
equations are in some sense generalizations of the Boussinesq system
studied usually in connection with the spectral problem [26] for the
operator:
$$
L=D^3 +pD+q, \quad \hbox{where}\quad D={d\over dx}.
$$ 
Then the first Boussinesq flow is given by the following equations:
$$
p_t=2q_x-p_{xx} \leqno (6.6.1a)
$$ 
and 
$$
q_t=q_{xx}-{2\over3}p_{xxx}-{1\over 3}(p^2)_{xx},\leqno (6.6.1b)
$$
or, after one differentiation  of the former, 
$$
p_{tt}={1\over 3}p_{xxxx}-{2\over3}(p^2)_{xx}.\leqno (6.6.2)
$$
The latter equation is often called the Boussinesq equation.  

In our set-up, we first choose a Kac-Moody algebra and one of its parabolic 
subalgebras. The appropriate choice for the Boussinesq equation is that of $\gg=A_2^{(1)}$, the 
Kac-Moody algebra associated to $s\ell (3,{\C})$. A set of Chevalley generators is chosen as:
$$ 
\eqalign{
e_0 &= \lambda  E_{31} ,\ e_1 = E_{12},\ e_2 = E_{23},\cr
f_0 &= \lambda ^{-1} E_{13} ,\ f_1 = E_{21},\ f_2 = E_{32},\cr
h_i &:= [e_i, f_i]\quad\hbox{for}\quad i = 0, 1, 2 .
}$$
where we denote by $E_{ij}$ the $3\times
3$-matrix with 1 in the $(i,j)$-position and $0$'s elsewhere.
In the first example the parabolic subalgebra $\gp$ is generated by $\GG
_0\oplus\GG _+, f_1$.
As the first step we find a general form of $\Omega _1$.  To this end we 
use (6.3.13), where 
$$
E=\pmatrix{0 &1&0\cr
           0 &0&1\cr
           \lambda & 0 &0}, \quad E^{(1)}=\pmatrix{0 &0&0\cr
           0 &0&1\cr
           \lambda & 0 &0}
$$
and
$$
q^{(-1)}=\pmatrix{0 & 0 & {\alpha_1\over \lambda} \cr
                  0  & 0 & {\alpha _2\over \lambda}\cr
                  \alpha _3 & \alpha _4 & 0 }
$$
respectively.
Thus (6.3.13) implies that $\Omega _1 $ is of the form:
$$
\Omega _1 =\pmatrix{u& 1& 0 \cr
                    \alpha & -v&1 \cr
                   \lambda & 0 & v-u}
$$
where $u$, $v$, $\alpha$ are parameters.  Moreover, since the rank of $\circgg$ is 2, we will use $u$ and $v$ to parametrize $\Omega _1$ and consequently all other (at least all positive) potentials.
Thus in the first place we have to see how $\alpha$ is expressed in terms of 
$u$ and $v$.  We observe that $\alpha E_{21}$ carries the canonical degree 
-1 so we have to know $q_{I,-1}$ and $q_{I,-2}$.  We proceed by implementing the proof of Theorem (6.5.1), in particular, (6.5.4a) and (6.5.4b). In the subsequent computations we take $h=I$. The results are:
$$
q_{I,-1}=\pmatrix{0 & 0& {1 \over 3}{2u-v \over \lambda} \cr 
               -{1 \over 3}v-{1 \over 3}u& 0 & \cr
                0 & -{1 \over 3}v+{2\over 3}u& 0}
$$
and 
$$
q_{I,-2}=\pmatrix{0 & {1 \over 6}{-2u_x+2v_x+u^2-v^2 \over \lambda}& 0\cr 
               0& 0 & {1 \over 6}{2v_x+4u_x+4uv -3u^2-2v^2 \over \lambda}\cr
               -{1 \over 3}u_x-{2 \over 3}v_x -{2 \over 3}uv+{1 \over 3}u^2 +{1 \over 2}v^2 & 0 & 0}.
$$
For $q_K$ part we get:
$$
q_{K,-1}=\pmatrix{0 & 0& {1 \over 3}{u_x+v_x \over \lambda} \cr 
      {1 \over 3}(u_x+v_x)& 0         & 0 & \cr
                0 &{1 \over 3}(u_x+v_x) & 0}.
$$
Now, it is easy to compute $\Omega _1$ or rather $\alpha$ in terms of $u$ and $v$. The final result is:
$$
\Omega _1=\pmatrix{u & 1 &0 \cr
                   \alpha & -v &1 \cr
                   \lambda & 0 & v-u } \quad \hbox {where}\quad  \alpha = u_x+v_x +uv -u^2 -v^2.\leqno (6.6.3)
$$

In a similar manner we can compute potentials corresponding to the higher flows.  For example the second flow, i.e. $j=2$, gives rise to 
$$
\Omega _2=\pmatrix{v_x+uv-v^2&u-v& 1 \cr
                   \alpha &u_x+uv-u^2&-u \cr
                   \lambda v& \lambda&-u_x -v_x -2uv +u^2+v^2}\leqno (6.6.4) 
$$
where
$$
\alpha = uu_x-vv_x +{1\over 3}v^3-{1\over 3}u^3 -{1\over 3}u_{xx}
+{1\over 3}v_{xx}.
$$ 
The ZCC yields:
$$
u_t={2\over 3}v^3-{2\over 3}u^3+2u^2v-2uv^2 +2uv_x+{2\over 3}v_{xx}+{1 \over 3}u_{xx}-2vv_x, \leqno (6.6.5a)
$$
$$
v_t={2\over 3}v^3-{2\over 3}u^3+2u^2v-2uv^2 -2vu_x-{2\over 3}u_{xx}-{1 \over 3}v_{xx}+2uu_x.\leqno (6.6.5b)
$$
These equations can be thought of as a version of 6.6.1a and 6.6.1b.  This assertion is 
further supported by the fact that after one differentiation with respect to 
$t$ we get two copies of the potential Boussinesq equation 6.6.2, 
which is to say that $u$ and $v$ satisfy:
$$
z_{tt}=-{1 \over 3} z_{xxxx}+4z_xz_{xx}. \leqno (6.6.6)
$$
For the sake of comparison we would like to quote the formulas for a maximal
parabolic.  In that case the only generator which has a nonzero p-degree is $e_0$.  The remainder of computations is almost identical, except for the choice of the parametrization of $\Omega _1$.  It is not sufficient to choose 
functions from
the diagonal, as there is only one function there while we need two functions.
We make the following choice of functions following [28]:
$$
u=(\Omega _1) _{11}, \quad v=(\Omega _1) _{32} -(\Omega _1)_{21}.\leqno (6.6.7)
$$
  Thus we arrive at the formulas:
$$
\Omega _1 =\pmatrix {u & 1 & 0 \cr
 -{1 \over 2}u^2 +{1 \over 2}u_x -{1 \over 2}v & 0 & 1 \cr
uv -v_x +\lambda & -{1 \over 2}u^2 +{1 \over 2}u_x +{1 \over 2}v & -u }, 
$$
$$
\Omega _2 =\pmatrix {{1 \over 2}u^2 -{1 \over 2}u_x -{1 \over 2}v & u &1 \cr
\alpha & -u^2 +u_x & -u\cr
\beta & \gamma &{1 \over 2}u^2 -{1 \over 2}u_x +{1 \over 2}v },
$$ 
where 
$$
\alpha = {3 \over 2}u u_x +{1 \over 2}uv -{1 \over 2}u^3 -{1 \over 2}u_{xx}
-{1 \over 2}v_x +\lambda ,
$$
$$
\beta = -{1 \over 4} v^2 -{3 \over 2}u^2 u_x +{1 \over 4}u^4 -{1 \over 3}
u_{xxx} + {5 \over 4}(u_x)^2 +uu_{xx},
$$
and 
$$
\gamma=-{3 \over 2}u u_x +{1 \over 2}uv +{1 \over 2}u^3 +{1 \over 2}u_{xx}
-{1 \over 2}v_x +\lambda .
$$
As the result of the ZCC we obtain:
$$
u_t=-v_x, \leqno (6.6.8a)
$$
$$
v_t={1 \over 3}v_{xxx}-2(u_x)^2.\leqno (6.6.8b)
$$
These results should be compared with (6.6.1a), (6.6.1b) and their counterparts (6.6.5a) and (6.6.5b). After one more differentiation with respect to $t$ we obtain that $u$ satisfies the 
potential Boussinesq equation 6.6.6.  On the other hand $v$ satisfies:
$$
v_{tt}={1 \over 3}v_{xxxx} +4u_xv_{xx}. \leqno (6.6.9)
$$

\medskip
\noindent
{\bf\S 7.  Negative Potentials}
\medskip

In this section we will investigate the negative potentials, $\Omega _j, \,
j<0$.
First, we outline the main
results.  We prove that every potential (both positive and negative ones)
can be expressed in terms
of the component of $p$-degree zero of $ g^+(\gt ,\lambda ); $ cf. Section
5.3.  The Zero-Curvature
Condition (ZCC) gives systems of PDEs that include the so called two-dimensional
Toda lattice (e.g. the
Sinh-Gordon equation) in the case where $\gp $ is the minimal standard
parabolic subalgebra, i.e.
the standard Borel subalgebra.

Our second result is that if $ \gp$ is not minimal, and some additional
condition are satisfied (see Proposition (7.3.2)) then $ \Omega _1$ uniquely 
determines every potential $
\Omega _j,\ j\in \ints$. 
\medskip

\noindent
{\bf 7.1 }\quad Let $ a \in G$ be the component of $p$-degree $0$ of $g^+$, i.e.
$$
g^+ = a \tilde g^+\ ,\leqno (7.1.1)
$$
where $ \tilde g^+ \in Q ^+_X $ (see Proposition 3.3.1).
Applying this to (5.3.5), we get
$$
\Omega ^r_j = (\partial _j a) a^{-1} + a\big((\partial _j \tilde g^+)
 (\tilde g^+)^{-1}
+ \tilde g^+ E_j^r (\tilde g^+)^{-1}\big) a^{-1}\ .  \leqno (7.1.2)
$$
First we look at $ j = 1$;  (5.3.3) reveals:
$$
\Omega ^R_1 = (E^R_1)^{(1)} +\ \ \hbox{terms of nonpositive}\
p\hbox{-degree}\ ,  \leqno (7.1.3)
$$
whereas from (5.3.5) we derive
$$
\Omega ^r_1 = (\partial_1 a) a^{-1} + a (E^r)^{(0)} a^{-1} + \ \
\hbox{terms of positive }
p\hbox{-degree}\ ,  \leqno (7.1.4)
$$
thus by Proposition (5.3.1)
$$
\Omega _1 = E^{(1)} _1 + (\partial _1 a) a^{-1} + a E^{(0)}_1
a^{-1}\ .\leqno (7.1.5)
$$
Similarly one gets
$$
\Omega _{-1} = E^{(0)}_{-1} + a E^{(-1)}_{-1}  a^{-1}\ .  \leqno
(7.1.6)
$$
From Theorem (6.3.2) we know that $\Omega _1$ determines every positive
potential; we see therefore that $a$ determines every positive potential.
Moreover, since $det(a)=1$, both $\Omega _1 $ and $ \Omega _{-1}$ are $\partial _1 $- differential polynomials in the entries of $a$.  
To see whether $\Omega _{-1}$ determines all negative potentials, we proceed as
follows.  Since $a\in
G^{(0)}$, $a$ is a Laurent polynomial in $\lambda \in S^r$.  Therefore it
can be holomorphically
extended to $\mu\in S^R$.  We use this to ``swap'' the original splitting
(5.3.1).  Let
$$
\tilde g^- = a^{-1} g^-\ ,  \leqno (7.1.7)
$$
where $a$ is the same as in (7.1.1).  Then we get
$$
\leqalignno{
\gt (h_1,h_2) &= \big( (g^-)^{-1}, (g^+)^{-1}\big)  (b_R, b_r)\cr
&= \big( (\tilde g^-)^{-1} a^{-1}, (\tilde g^+)^{-1} a^{-1}\big)
 (b_R, b_r)\cr
&= \big( (\tilde g^-)^{-1}, (\tilde g^+)^{-1} \big)  (a^{-1}
b_R,\, a^{-1} b_r)
}$$
The second factor is still in $ \HH _+$, whereas the first one now is in
$$
\tilde \HH _- := P^R_{opp} \times Q^r_{opp}\ ,  \leqno (7.1.9)
$$
meaning the ``opposite'' standard parabolic subgroup and its complementary
subgroup.  In terms of
Lie algebras this means:
$$
\gp _{opp} = \sum _{k\le 0} \GG ^{(k)},\leqno (7.1.10)
$$
$$
\gq _{opp} = \sum _{k> 0} \GG ^{(k)}.\leqno (7.1.11)
$$
Roughly speaking, everything we did for positive potentials remains true if
replace ``(non)
negative'' by ``(non) positive'' and vice versa.  In particular, we see 
that the negative potentials $ \tilde\Omega _{-k}$
(corresponding to the
splitting into $ \HH _+$ and $\tilde \HH _-$) are $ \partial _{-1}$-differential polynomials in $
\tilde\Omega _{-1}$.
From (5.3.5) we know how to express them:
$$
\leqalignno{
\tilde\Omega ^r_j &= (\partial _j\tilde g^+) (\tilde g^+)^{-1} +\tilde g^+
E^r_j (\tilde g ^+)^{-1}
 & (7.1.12)\cr
&= (\partial _j (a^{-1}))a + a^{-1} \big((\partial _jg^+) (g^+)^{-1} + g^+
E^r_j(g^+)^{-1}\big) a\ ,
}$$
or
$$
\Omega _j = (\partial _j a) a^{-1} + a\tilde\Omega _j\, a^{-1}\ .
\leqno (7.1.13)
$$
Now we have the following chain:
$$
a\to \Omega _{-1} \to \tilde\Omega _{-1}\to \tilde\Omega _{-k} \to \Omega _{-k},\quad k
\ge 1\ ,
$$
where ``$\to$'' stands for ``determines''.

We summarize the results:
\medskip

\noindent
\Theorem {{\bf (7.1.1).}\quad Every positive potential $\Omega _k$ is a $
\partial _1 $ differential polynomial in $a$.  Every negative potential $\Omega
_{-k}$ is a $\partial _{-1}-\partial _{-k}-$ differential polynomial in a.
More precisely,  $\Omega _{-k}$ is gauge equivalent to $\tilde\Omega _{-k}$
as in $(7.1.13)$, where $\tilde\Omega _{-k}$  is a 
$\partial _{-1}- $ differential polynomial in $a$.
}\medskip

\Example \quad In the case where $ \gp$ is the minimal parabolic subalgebra, there is a
natural choice of
functions parametrizing $a$.  In this case, $ G^{(0)}$ is an abelian subgroup
and there is a $ w\in
\GG _0$ such that $ a = e^w$.  Then (7.1.5) and (7.1.6) yield:
$$
\Omega _1 = E + \partial _1 w  \leqno (7.1.14)
$$
$$
\Omega _{-1} = e^w E _{-1} e^{-w}\ , \leqno (7.1.15)
$$
and the ZCC is equivalent to
$$
\partial_{1,-1} w = [e^w E_{-1} e^{-w}, E]\ ,  \leqno (7.1.16)
$$
the ``\underbar{two-dimensional Toda lattice}.''

\noindent
\big(for $ \GG = A^{(1)}_1$, this gives the \underbar{Sinh-Gordon equation}
$\partial _{1,-1} w = 2
sinh (2w)$.\big)
\medskip
\noindent
{\bf 7.2 }\quad Before we attack the main question as to which extent $ \Omega _1$
determines $ \Omega _{-1}$ in
general, we prove the following partial result:
\medskip

\noindent
\Lemma {{\bf (7.2.1.)}\quad $ \partial _1\Omega _{-j}$ is a universal $\partial _1
-\partial_{-j}-$differential
polynomial in $ \Omega _1$, for $ j\ge 1$.}
\medskip

\noindent
\Proof .\quad Along the lines of the proof of Theorem 6.3.2., we obtain
$$
h^-.\Omega _{-j} = (\partial_{-j} e^{q_I}) e^{-q_I} + e^{q_I}((\partial
_{-j} q_K) + E_{-j})
e^{-q_I}\ , \leqno (7.2.1.) 
$$
where
$$
h^-.\Omega _{-j} = Ad\, (h^-)^{-1}\Omega _{-j} \hbox { and }h^-\in Q.
$$
We single out the part of the above expression which lies in $Ker\,ad E$, namely,
$$
(\partial _{-j} e^{-q_I}) e^{q_I} + e^{-q_I}h^-.\Omega _{-j}e^{q_I} =
\partial _{-j}q_K +
E_{-j}\ .  \leqno (7.2.2)
$$
Then we decompose (7.2.2) with respect to the canonical grading, using
$$
h^-.\Omega _{-j} = (h^-.\Omega _{-j})_{-1} +( h^-.\Omega _{-j})_{-2} +
\cdots .  \leqno (7.2.3)
$$
and $ (h^-.\Omega _{-j})_{-1} = \Omega _{-j,-1}$,
to obtain for cdeg =  -1 :
$$
 -\partial_{-j} q_{I,-1} + \Omega _{-j,-1} = \partial _{-j}
q_{K,-1} +(E_{-j})_{-1}\ .
$$
From the proof of Theorem (6.3.2) we know, that $q_I$ and $ \partial _1 q_K$
are $\partial
_1 -$ differential polynomials in $h ^-.\Omega _1$, therefore $\partial
_1(h^-.\Omega _{-j,-1})$ is a
$\partial _1 - \partial_{-j}- $ differential polynomial in $h^-.\Omega _1$.

Now we inspect the component of (7.1.2) of degree $-i$:
$$
(h^-.\Omega _{-j})_{-i} = \partial _{-j} q_{K,-i} +(F_{-j})\left(q_I, (h^-.\Omega _{-j})_{-1},\cdots (h^-.\Omega _{-j})_{-i+1}\right ) _{-i}
$$
where $(F_{-j})_{-i}$ is a polynomial in its arguments. Inductively, we conclude, that $ \partial _{1}((h^-.\Omega _{-j})_{-i})$ is a
differential polynomial in
$ h^-.\Omega _1$ of the desired form.  In fact, $F_{-j}$ is also a polynomial
in $\Omega _1$ with coefficients depending analytically on $h^-$.  However, 
in the neighborhood of the identity, there is no explicit dependence on $h^-$ and thus $F_{-i}$ is $h^-$ independent.      \blsq
\medskip

\noindent
{\bf 7.3 }\quad We want to investigate more closely to which extent $
\Omega _1$ determines the
quantities involved in the splitting procedure.

To this end, let  $h$ and $ k \in \HH$ be such that the associated potentials
$ \Omega ^h_1$ and $
\Omega ^k_1$ are equal.  Denote by $g^{\pm}_h$ and  $ g^{\pm}_k$ the
splitting components
associated with $ h$ and $k$ respectively.  With this notation we can show
\medskip
\noindent
\Proposition  {{\bf (7.3.1).}
\medskip
\noindent
(a)
$$
\Omega ^h_1 = \Omega ^k_1\ \ iff
$$
$$
g^-_h = g^-_k  \gamma ^- ,\ \gamma ^-\in \Gamma ^R \cap Q ^R  \leqno
(7.3.1)
$$

and
$$
g^+_h = g^+_k e^{t_1E^r} m^+ e^{-t_1E^r},\ m^+\in P^r\quad and\quad
\partial _1 \gamma ^- =
\partial _1 m^+ = 0\ .  \leqno (7.3.2)
$$
\medskip

\noindent
(b)  If $ \Omega ^h_1 = \Omega ^k_1$\ \ then\ \ $ \Omega ^h_j = \Omega
^k_j$,\ \ for all\ \ $ j> 0$.
\medskip

\noindent
(c)\quad $\Omega ^h_{-1} = \Omega ^k_{-1}\ \ iff$
$$
g^-_h = g^-_k e^{-t_{-1}E^R_{-1}}m^- e^{t_{-1} E^R_{-1}} \quad , m^-\in Q ^R
\leqno (7.3.3)
$$

and
$$
g^+_h = g^+_k \gamma ^+ ,\ \gamma ^+\in \Gamma ^r\cap
P^r\quad\hbox{and}\quad \partial
_{-1}\gamma ^+ = \partial _{-1} m^- = 0\ .  \leqno (7.3.4)
$$
\medskip

\noindent
(d)  If $ \Omega ^h_{-1} = \Omega ^k_{-1} $ then $ \Omega ^h_j = \Omega
^k_j$ for all $ j < 0$.
\medskip

\noindent
(e)  If $ \Omega ^h_1 = \Omega ^k_1$ and $ \Omega ^h_{-1} = \Omega ^k_{-1}$
then $ \Omega ^h_j =
\Omega ^k_j$ for all $ j\in \ints $.
\medskip

\noindent
(f) If $ \Omega ^h_{-1} = \Omega ^k_{-1}$ then $\Omega ^h_1 = \Omega ^k_1$
\medskip

\noindent
(g) If $ \Omega ^h_{-1} = \Omega ^k_{-1}$ then $\Omega ^h_j = \Omega ^k_j$
for all $ j \in \ints$.
\medskip

\noindent
(h) If $ \Omega ^h_{-1} = \Omega ^k_{-1}$ and $\Omega ^h_1 = \Omega ^k_1$
then $g_h^-=g_k^- \gamma ^-$ and $g_h^+=g_k^+ \gamma ^+$ with 
$\gamma ^+$
and $\gamma ^-$ independent of $t_1$ and $t_{-1}$.}
\medskip
\noindent
\Proof .

\noindent
(a)  From \S {5.3} we know
$$
\leqalignno{
(\Omega ^h_1)^R &= (\partial _1 g^-_k)  (g^{-}_h)^{-1} + g^-_h  E^R
(g^-_h)^{-1} &(7.3.5)
\cr
&= \partial _1 (g^-_h  e^{t_1 E^R}) (g^-_h  e^{t_1 E^R})^{-1}, \cr
(\Omega ^h_1)^r &= \partial _1 (g^+_h e^{t_1 E^r}) (g^{+}_h  e^{t_1
E^r})^{-1}.  & (7.3.6)
}$$

It is straightforward to show that
$$
\leqalignno{
(\partial _j g) g^{-1} &= (\partial _j \tilde g)\tilde
g^{-1}\quad\hbox{if and only if}
&(7.3.7)\cr
g &= \tilde g  \hat g\quad\hbox{and}\quad \partial _j\hat g = 0\ .
}$$

Thus:
$$
g^-_h = g^-_k  e^{t_1 E^R} \gamma ^- e^{-t_1E^R}  \leqno (7.3.8)
$$
$$
g^+_h = g^+_k  e^{t_1 E^r} m^+  e^{-t_1 E^r}  \leqno (7.3.9)
$$
for some $ \gamma ^- \in G^R,\ m^+\in G^r$ independent of $t_1$.  Setting $
t_1 = 0$ shows $ \gamma
^-\in Q^R$ and $ m^+\in P^r$.
Moreover,
$$
(g^-_k)^{-1} g^-_h = e^{t_1 E^R} \gamma ^- e^{-t_1 E^R}\in Q^R\
,\leqno (7.3.10)
$$
thus $ \gamma ^-\in \Gamma ^r$ by Theorem (3.7.2).
This settles the ``only if'' part.

The ``if'' part is straightforward.
$$
\leqalignno{
(\Omega ^h_1)^R &= \partial _1 (g^-_k \gamma ^- e^{t_1 E^R})
(g^-_k \gamma ^- e^{t_1
E^R})^{-1}  &(7.3.11)\cr
&= \partial _1 g^-_k (g^-_k)^{-1} + g^-_k (\partial _1\gamma ^-)
(\gamma ^-)^{-1}
(g^-_k)^{-1}  \cr
&+g^-_k \gamma ^- E^R (\gamma ^-)^{-1} (g^-_k)^{-1}\cr
&= (\Omega ^k_1)^R,\ \ \hbox{since}\ \ \partial _1\gamma = 0\ \ \hbox{and}\
\ [\gamma ^-, E^R] = 0\
.}$$
$$
\leqalignno{
(\Omega ^h_1)^r &= \partial _1 (g^r_k e^{t_1 E^r} m^+) (g^+_k e^{t_1
E^r} m^+)^{-1}  &(7.3.12)\cr
&= (\partial _1 g^+_k)  (g^+_k)^{-1} + g^+_k E^r(g^+_k)^{-1} =
(\Omega ^k_1)^r\ ,
}$$
since $ \partial _1 m^+ = 0$.
\medskip

\noindent
(b)  Again, from 5.3 and (a) we know
$$
\leqalignno{
(\Omega ^h_j)^R &= \partial _j (g^-_h e^{t_j E^R_j}) (g^-_h  e^{t_j
E^R_j})^{-1}
&(7.3.13)\cr
&= \partial _j (g^-_k  \gamma ^- e^{t_j E^R_j}) (g^-_k \gamma
^-  e^{t_j E^R_j})^{-1}
, \cr
&= (\Omega ^k_j)^R + g^-_k  (\partial _j \gamma ^-) (\gamma
^-)^{-1} (g^-_k)^{-1}
}$$
This shows
$$
(\Omega ^h_j)^R - (\Omega ^k_j)^R \in \gq ^R  \leqno (7.3.14)
$$

On the other hand, we know from Proposition (5.3.1) that $ (\Omega ^*_j)^R \in \gp ^R $, for $j>0$.  Whence 
$$
(\Omega ^h_j)^R - (\Omega ^k_j)^R \in \gp^R\ .
$$
This shows $(\Omega ^h_j)^R = (\Omega ^k_j)^R$.  
Note that the latter condition defines $ \Omega _j$ uniquely, since two
Laurent polynomials that
coincide on the circle are equal.  
Thus:\quad $ \Omega ^h_j = \Omega ^k_j$.
\medskip
\noindent
(c)  As in (a) one concludes
$$
g^-_h = g^-_k e^{t_{-1} E^R_{-1}}  m^- e^{-t_{-1}
E^R_{-1}}  \leqno (7.3.15)
$$
$$
g^+_h = g^+_k e^{t_{-1} E^r_{-1}}  \gamma ^+ e^{-t_{-1}
E^r_{-1}}  \leqno
(7.3.16)
$$
for some $ m^-
\in Q^R,\ \gamma ^+\in P^r$, such that $\partial _{-1}\gamma ^+=\partial _{-1}
m^-=0$.

Moreover,
$$
(g^+_k)^{-1} g^+_h =e^{t_{-1}E^r_{-1}}  \gamma ^+ 
e^{-t_{-1}E^r_{-1} } \in P^r\ ,
\leqno (7.3.17)
$$
thus $ \gamma ^+\in \Gamma ^r \cap P^r$ by Theorem (3.7.3).
This settles the ``only if'' part.  The ``if'' part is straightforward.
\noindent
(d)  As in (b) one computes
$$
\leqalignno{
(\Omega ^h_j)^r &= \partial _j(g^+_h e^{t_j E^r_j})(g^+_h e^{t_j
E^r_j})^{-1} &(7.3.18)\cr
&= \partial _j(g^+_k \gamma ^+  e^{t_j E^r_j})(g^+_h\gamma
^+ e^{t_j E^r_j})
^{-1}\cr
&= (\Omega ^k_j)^r + g^+_k (\partial _j\gamma ^+) (\gamma
^+)^{-1} (g^+_k)^{-1}.
}$$

This shows
$$
(\Omega ^h_j)^r - (\Omega ^k_j)^r \in \gp ^r\ .  \leqno (7.3.19)
$$
On the other hand we see from (5.3.3) that
$$
\Omega ^R_j - (E^R_j)^{(0)} \in \gq ^R\ ,  \leqno (7.3.20)
$$

thus
$$
(\Omega ^h_j)^R - (\Omega ^k_j)^R \in \gq  ^R\ .  \leqno (7.3.21)
$$
This shows that $ \Omega ^h_j = \Omega ^k_j$.

\noindent
(e)  follows from (b) and (d)
\medskip

\noindent
(f)  Using (c) we derive
$$
\leqalignno{
(\Omega ^h_1)^r &= \partial _1 (g^+_h e^{t_1 E^r})(g^+_h e^{t_1 E^r})^{-1}
&(7.3.22)  \cr
&= \partial _1 (g^+_k \gamma ^+  e^{t_1 E^r})(g^+_k \gamma
^+  e^{t_1
E^r})^{-1}\cr
&= (\Omega ^k_1)^r + g^+_k  (\partial _1\gamma ^+ ) (\gamma
^+)^{-1} (g^+_k)^{-1}.
}$$
We need to show that the second term vanishes.  From (5.3.3) and 
(7.3.13) we know that
$$
\Omega ^R_1 - E^R \in (\GG^{(0)}_-\oplus \GG _0)^R  \leqno (7.3.23)
$$
thus
$$
(\Omega ^h_1)^r - (\Omega ^k_1)^r = Ad(g_k^+)((\partial _1\gamma ^+)(\gamma ^+)^{-1})\in (\GG^{(0)}_- + \GG _0
)^r\ .  \leqno (7.3.24)
$$
Now $ (\partial _1\gamma ^+) (\gamma ^+)^{-1} \in (Ker\, ad\, E \cap \gp)^r$.

At this point we apply Theorem (B.1.2) from the Appendix B stating that 
$Ker\, ad\, E
\cap\GG^{(0)} = 0$.  Thus  $min \, pdeg(Ad(g_k^+)((\partial _1\gamma ^+)(\gamma ^+)^{-1})\ge 1$ which implies the claim.
\medskip

\noindent
(g)  obvious from (b), (d), (f),  

\noindent
(h) follows from (a) and (c). \blsq
\medskip

\noindent
\Remark .\quad The results of Theorems (7.1.1) and (7.3.1) give the 
following
picture (where
``$\longrightarrow $'' means ``determines uniquely''):
$$
\matrix{
      &         &\Omega _1    &\longrightarrow & &\Omega _j (j>0) \cr
      &\nearrow &              &               & &          \cr
a     &         & \uparrow     &               & &          \cr
      &\searrow &              &               & &            \cr
      &         &\Omega _{-1} &\longrightarrow & &\Omega _j (j< 0)
}$$
\medskip

\noindent
\Proposition {{\bf 7.3.2}.\quad  If $ Ker\, ad\, E\cap \unGG^{(-1)} = \{
0\}$, then $ \Omega _1^h=\Omega _1^k$ implies 
$ \Omega _{-1}^h=\Omega _{-1}^k$.
}
\medskip

\noindent
\Proof .\quad Assume $ \Omega ^h_1 = \Omega
^k_1$.  Then by Lemma (7.2.1) we
have $\partial _1 \Omega ^h_{-1} = \partial _1\Omega ^k_{-1}$, hence
$$
\Omega ^h_{-1} = \Omega ^k_{-1} + C_{-1}\ ,  \leqno (7.3.8)
$$
where $ C_{-1}$ depends on $ t_{-1}$ only.  (We ignore variables different
from $ t_1$ and $ t_{-1}$
here.)

By Proposition (7.3.1 a)
$$
g^-_h = g^-_k  \gamma^-(t_{-1})\quad\hbox{and}\quad \gamma^- (t_{-1})\in \Gamma
_-\cap Q\ .  \leqno
(7.3.9)
$$
This implies (setting $ g^-_h (0,0) = (h^-)^{-1} , \, g^-_k (0,0) 
= (k^-)^{-1}_-$)
$$
\gamma ^- (0) = k^- (h^-)^{-1} \in \Gamma _-\ .  \leqno (7.3.10)
$$
Next we evaluate the $(-1)$ - potentials, using the formulas 
of \S {5.3}:
$$
\eqalign{
(\Omega ^h_{-1})^R &= (\partial _{-1} g^-_k)  (g^-_k)^{-1} + g^-_k 
(\partial _{-1} \gamma^-)
(\gamma^-)^{-1} (g^-_k)^{-1} + g^-_k E^R_{-1} (g^-_k)^{-1}\cr
&=(\Omega ^k_{-1})^R + g^-_k (\partial _{-1}\gamma^-) (\gamma^-)^{-1} (g^-_k)^{-1}
}$$
From (7.3.8) we conclude
$$
(\partial _{-1} \gamma^-) (\gamma^-)^{-1} = (g^-_k)^{-1}  C_{-1}
g^-_k\ .  \leqno (7.3.11)
$$
From (7.1.6) we see that
$$
C_{-1} \in \gq ^{(-1)}\ .  \leqno (7.3.12)
$$
Upon decomposing $C_{-1}$ with respect to the canonical grading we get
$$
C_{-1} = c_{-1} +\cdots + c_{-m},\quad  \hbox {for some finite $m$.}  \leqno (7.3.13)
$$
Let
$$
(g^-_k)^{-1} = k^- \exp (r_{-1} + r_{-2} + \cdots),  \leqno (7.3.14)
$$
where $k^-\in Q$ is $t$ and $x$ independent.  Thus (7.3.11) turns into
$$
(\partial _{-1} \gamma^-)  (\gamma^-)^{-1} = k^- \Big(\exp \big( ad (r_{-1}
+ r_{-2} +s )\big)
C_{-1}\Big) (k^-)^{-1}.  \leqno (7.3.15)
$$
Let
$$
(\partial _{-1} \gamma ^-) (\gamma ^-)^{-1} = s_{-1} + s_{-2} + \cdots \ .  \leqno
(7.3.16)
$$
Now compare in (7.3.15) and (7.3.16) components of the same
 canonical degree.  Note that $ k^- \in Q$).  First, we consider
$$
\eqalign{
cdeg=-1 : &\hbox{ we obtain }s_{-1}. \hbox{ Now (7.3.12) and (7.3.9) 
imply } \cr 
&s_{-1} = c_{-1} \in Ker\, ad\, E\cap \unGG ^{(-1)} = \{ 0\},
\hbox{thus}\quad s_{-1} = c_{-1} = 0. }
$$
Next, we consider  
$$
\eqalign{
cdeg= -2 : &\hbox{ Since  } s_{-1}=0 \hbox{ and } c_{-1}=0,\cr 
&\hbox{ we 
obtain } s_{-2}=c_{-2}.\hbox{  As above we conclude } s_{-2}=c_{-2}=0.
 }
$$
The remainder of the proof goes by induction on the canonical degree,
thus yielding  $C=0$.   \blsq
\medskip

To see that the condition of the previous Proposition is nontrivial we
briefly discuss the
following two examples:

\noindent
{\bf Example 1:}\quad In the case of minimal parabolics the $p$-grading
coincides with the
canonical grading.  Thus the condition $ Ker\, ad\, E \cap \gq^{-1} = \{
0\}$ does not hold and $
\Omega _{-1} $ is not uniquely determined by $ \Omega _1$.  To see that one
only needs to examine
(7.1.14) and (7.1.15).  However, one has a natural parametrization of $ \Omega
_1$ in terms of $ w\in
\gg _0$, namely $ \Omega _1 = E + \partial _1 w$, which gives $ \Omega
_{-1} = e^w E_{-1} e^{-w}
$.  It remains an interesting open question whether one can find a natural
parametrization in terms
of elements of $ \gg ^{0}$ in each case $ Ker\, ad\, E\cap \gq^{(-1)} \not= \{ 0\}$.

\noindent
{\bf Example 2:}\quad Let $\GG ^{fin}$ be of type $ A^{(1)}_3$, and $\GG =
s\ell _4 (A_w)$ where $
A_w$ is the Wiener Algebra w.r.t. some weight $ w$ (cf. 1.1).  Denote by $
E_{ij}$ the $4\times
4$-matrix with 1 in the $(i,j)$-position and $0$'s elsewhere.  A set of
Chevalley generators is
given by
$$
\eqalign{
e_0 &= \lambda  E_{41} ,\ e_1 = E_{12},\ e_2 = E_{23},\ e_3 = E_{34} ,\cr
f_0 &= \lambda ^{-1} E_{14} ,\ f_1 = E_{21},\ f_2 = E_{32},\ f_3 =
E_{43} ,\cr
h_i &:= [e_i, f_i]\quad\hbox{for}\quad i = 0, 1, 2, 3.
}$$
Let $\gp $ be the standard parabolic subalgebra generated by $\GG
_0\oplus\GG _+, f_1$ and $ f_3$,
i.e.
$$
\gp = \left\{\pmatrix{ \sum\limits _{n\ge 0} \lambda ^n A_n
&\sum\limits_{n\ge 0} \lambda ^n  B_n
\cr
\sum\limits _{n\ge 0} \lambda ^n C_n &\sum\limits _{n\ge 0} \lambda ^n  D_n
}: A_n,D_n \in sl_2({\C}),\quad
 B _n , C_n \in gl_2 (\C )\right \} .
$$
It is easy to see that
$$
\gq^{(-1)} = \left\{\pmatrix{ 0 &\lambda ^{-1} A\cr
B &0} : A, B \in gl_2 (\C )\right\}\ .
$$
Thus, $ \gq ^{(-1)} \cap Ker\, ad\, E \not= \{ 0\} $, since
$$
E_{-2} = E^{-2} = \pmatrix{0 &\lambda ^{-1} I_2\cr
I_2 &0} \in \gq ^{(-1)} \cap Ker\, ad\, E\ ,
$$
where $I$ denotes the $2\times 2$ identity matrix.

For \lq\lq nontwisted\rq\rq $\GG $, i.e. $\GG $ of type $ X^{(1)}_{\ell}$ we are
able to determine all
the cases where the condition of Proposition 7.3.2. is violated, i.e. where
$$
Ker\, ad\, E \cap \unGG ^{(-1)} \not= \{ 0\}\ .
$$
The result is as follows [27].
\medskip
\noindent
\Theorem {{\bf (7.3.3).}\quad Let $\GG$ be of type $ X^{(1)}_{\ell}$, and $h$ the
Coxeter number of
$\circgg $, where $ \circgg $ is finite dimensional and simple of type $
X_{\ell }$.  Let $k$ be a
positive exponent of $ \GG $.

The following are equivalent:

\item {(a)} $k$ divides $h$

\item {(b)} There is a parabolic subalgebra $\gp $ of $\GG $ such that
$$
(Ker\, ad\, E)_{-k} \cap \GG^{(-1)} \not= \{ 0\}\ .
$$}
\medskip
\Remark \quad The relations between $h,k$ and $\gp$ for which $(Ker
\, ad\, E)_{-k}\cap \gg ^{-1} \ne 0$ have been investigated in detail in
[27].  The following table summarizes the main result of that paper.
\medskip
\input table.mac
\midinsert{\centerline{%
          \table{|l|c|c|c|} \hline
$X_{\ell }$ & $h$ & exponents $k$ & $k\vert h$? \\
\hline\hline
$_{\ell}$ & $\ell + 1$ & $1, 2, \cdots , \ell $  & depends on $\ell $ \\
\hline
$B_{\ell}$ & $2\ell $ & $1,3,\cdots, 2\ell - 1$ & depends on $\ell $ \\
\hline
$C_{\ell}$ & $2\ell $ & $1,3,\cdots 2\ell - 1 $ & depends on $\ell $  \\
\hline
$D_{\ell}$ & $2\ell - 2$ & $1, 3, \cdots , 2\ell - 3, \ell - 1$  & depends
on $\ell $  \\
\hline
$E_6$ & 12 & $1,4,5,7,8,11$ & $k = 4$  \\
\hline
$E_7$ & 18 & $ 1,5,7,9,11,13,17$  & $k = 9$  \\
\hline
$E_8$ & 30 & $1,7,11,13,17,19,23,29$  & none  \\
\hline
$F_4$ & 12 & $ 1, 5, 7, 11$  & none \\
\hline
$G_2$ & 6  & $1, 5$  & none  \\
\hline
\endtable
}}\endinsert
\medskip
This shows in particular that the situation is fairly easy for the exceptional Lie algebras.  We will not pursue this any further in this paper.  
\vfil\eject

\medskip

\noindent
{\bf 7.4 }\quad We present below an example of a ``negative potential'' and the associated differential equation.
We use the following notation: the derivatives with respect to $x$ and $t$ are denoted either by $\partial _{x,t} $ or by using $x$ or $t$ as a subscript. 

We will follow the procedure described in the last two chapters.  The most 
important point we would like to emphasize is that by results of 
Section 6 it suffices to work in the neighborhood of the identity of the group $G$.
This allows one to carry out computations on a Lie algebraic level.
In this example, we want to closely investigate to which extent $
\Omega _1$ determines the
quantities involved in the splitting procedure. 
We consider the case of $A_1^{(1)}$.  Let $\gg ^{fin}=s\ell _2(A_w)$ where 
$A_w$ is the Wiener Algebra w.r.t. some weight $ w$ (cf. 1.1).  Denote by $
E_{ij}$ the $2\times
2$-matrix with 1 in the $(i,j)$-position and $0$'s elsewhere.  A set of
Chevalley generators is
given by
$$
\eqalign{
e_0 &= \lambda  E_{21} ,\ e_1 = E_{12},\cr
f_0 &= \lambda ^{-1} E_{12} ,\ f_1 = E_{21},\cr
\alpha_i^{\vee} &:= [e_i, f_i]\quad\hbox{for}\quad i = 0,1.
}$$
Let $\gp $ be the standard parabolic subalgebra generated by $\GG
_0\oplus\GG _+, f_1$,
i.e.
$$
\gp = \left\{\pmatrix{ \sum\limits _{n\ge 0} \lambda ^n a_n
&\sum\limits_{n\ge 0} \lambda ^n  b_n
\cr
\sum\limits _{n\ge 0} \lambda ^n c_n &- \sum\limits _{n\ge 0} \lambda ^n a_n
}: a_n,b_n ,c_n\in {\C}\right \} .
$$
The p-grading in this case agrees with the grading according 
to the powers of $\lambda$, i.e. p-deg$(E_{ij}\lambda ^k)=k$.  
To further simplify the computation we observe that by Theorem (6.3.2) we can use 
the formula 
$$
\leqalignno{
\Omega ^R_{j} &= (\partial_{j} e^{q_I}) e^{-q_I} + e^{q_I}(\partial
_{j} q_K + E^R_{j})
e^{-q_I}\ . &(7.4.1)}
$$
In fact we can ignore the size of the contour and set $R=1$.
Our goal is to compute $\Omega _{-1}$.  To this end we will have to 
find the relevant contributions from both $q_{I}$ and $q_K$.  We first decompose these two with respect to the canonical grading.  This for the case at hand
amounts to setting c-deg$(E_{ij}\lambda ^k)=j-i+2k$. This makes $E$ homogeneous
of degree 1, consequently $E_{j}$ has degree $j$.   We write 
$$
 q_I = q_{I,-1} + q_{I,-2} + \ldots,   
$$
similarly
$$
q_K = q_{K,-1} + q_{K,-2} + \ldots. 
$$
By (7.1.6) the essential part of $\Omega _{-1}$ lives in $\gg ^{(-1)}$.  In the present case this gives 
$$
\gg^{(-1)}=\lambda ^{-1}\pmatrix{a & b \cr c &-a},\quad a,b,c \in {\C}.\leqno(7.4.2)
$$
We observe that the minimal canonical degree appearing in  $\Omega _{-1}$ is $-3$.
We will therefore use the following variables $\{q_{I,-i},q_{K,-i},i=1..3\}$.
In fact $q_{K,-2}=0$ as $E^{-2} \not \in \gg$.  Furthermore, it is useful to 
observe that $v \in Im\,ad(E)\cap \gg _i$ if and only if
there exists a diagonal matrix $a=diag(\alpha,-\alpha),\alpha \in {\C}$,  
such that
$v=aE^{i}$. A similar result for $ker \,ad(E)$ states that $a$ has to be 
a multiple of the identity.  Thus we can parametrize $\{q_{I,-1},q_{I,-2}q_{I,-3},q_{K,-1},q_{K,-3}\}$ in terms of 5 scalars $\{h_1,h_2,h_3,w_1,w_3\}$ by forming the appropriate diagonal matrices.
Using (7.1.6) we can now express
$\Omega _{-1}$ in terms of these variables.
This is a purely Lie algebraic
computation and the result is:

$$
\Omega _{-1}=\pmatrix{ {2 h_1+\partial _{t} h_2 +2 \partial _{t} w_1 h_1 \over
 \lambda } &
{1+\partial _{t} h_1 +\partial _{t} w_1 \over \lambda }\cr
(1-\partial _{t} h_1 +\partial _{t} w_1 )+a &-{2 h_1+\partial _{t} h_2 +2 \partial _{t} w_1 h_1 \over
 \lambda }} \leqno (7.4.3)
$$
where
$$
a=-\lambda ^{-1}(2h_1^2+h_1\partial _{t}h_2-h_2\partial _{t}h_1+2\partial _{t} w_1 h_1^2
+2\partial _{t}w_1 h_2 +2h_2-\partial _{t}w_3+\partial _{t}h_3).
$$
We parametrize the potential $ \Omega _1$ as follows 
$$
\Omega _1 =\pmatrix{u & 1 \cr
 \lambda + u_x -u^2 & -u }. \leqno (7.4.4)
$$
Now we use (7.4.1) to express $\{h_1,h_2,h_3 \}$ in terms of $u$ and its $x$
derivatives.  The result is :
$$
h_1={1 \over 2}u, \, h_2=-{1 \over 2}u_x +{1\over  4}u^2, \, 
h_3={1 \over 4}u _{xx} -{1 \over 2}uu_x +{1 \over 12}u^3. \leqno (7.4.5)
$$
From the proof of Theorem (6.3.2) we know that $\partial _x q_K $ is a differential polynomial in $\Omega _1$, in the case of the present example we 
obtain:
$$
w_{1,x}={1\over 2}u_x, \quad 
w_{3,x}=-{1\over 4}(u_x)^2 -{1\over 8}u^2 u_x+
{1 \over 4} u u_{xx} . \leqno (7.4.6)
$$
We would like to point out that in fact $w _1$ is uniquely determined.  This is
clear from the fact that if $\exp(q_I) \exp(q_K)$ and $\exp(q_I) \exp(q_K)\exp(c E_{-1})$ satisfy  (7.4.1) and both are in $Q$ then this implies 
$\exp(c E_{-1}) \in Q$ for some $c\in {\C}$. Note, however, that in the present example 
$\gg ^{(-1)} \cap ker\, ad(E)=0$, in particular $E _{-1} \not \in \gq$. Moreover 
the group $Q={\overline Q^{fin}}$ has the property that $Q-I$ consists of functions analytic
around $\infty$ in $\lambda$ and being zero there.  This implies $w_1={1 \over2}u$. Substituting (7.4.5) and $w_1$ back into $\Omega _{-1}$ yields:
$$
\Omega _{-1}=\pmatrix{-{u_{x,t} -2u
u_t -2u\over 2\lambda } & {1+u_t\over \lambda} \cr
1+b & {u_{x,t} -2u
u_t -2u\over 2\lambda }}
$$
where 
$$
b=-\lambda ^{-1}{1 \over 8}(-8u_x +8u^2 +7u^2u_t +2u_{xxt}
-6u_x u_t -6uu _{xt}-8w_{3,t}).
$$
In the next step we use the ZCC to express $\partial _t w_3$ in terms of
$\partial _x -\partial _t$ derivatives of $u$.
The ZCC reads:
$$
\partial _t \Omega _1 -\partial _x \Omega _{-1} +[\Omega _1,\Omega _{-1}]=0.
\leqno (7.4.7)
$$
The final formula for $\Omega _{-1}$ takes now a simple form:
\vfill
\eject
$$
\Omega _{-1}=\pmatrix{-{u_{xt} -2uu_t -2u\over 2\lambda } & {1+u_t\over 
\lambda }\cr
1-{-4u_x +2u^2 +2u^2u_t +u _{xxt}
-4u_x  u_t -2uu _{xt} \over 2\lambda } & { u_{xt} -2u
u_t -2u\over 2\lambda } }. \leqno (7.4.8)
$$
What now remains to do is to use the latter form of  $\Omega _{-1}$ to compute
the ZCC. A straightforward computation leads to a simple 
ZCC:
$$
{1\over \lambda }\pmatrix{0 & 0\cr
u _{xxxt} -4u _{xx} -8u_{xt}u_t -
4u_t u _{xx} &0}=0.
$$
Summarizing, the $\Omega _{-1}$ and $\Omega_1$- generated flows give rise to 
the following partial differential equation:
$$
u_{xxxt}=4u_{xx}+8u_{xt}u_t+4u_tu_{xx}.  \leqno (7.4.9)
$$
With the help of the simple substitution $z(x,t)=u(4x,t)+t$ 
we can reduce this 
equation to 
$$
z_{xxxt}=z_{xx}z_t +2z_x z_{xt}.  \leqno (7.4.10)
$$
We would like to add that from the point of
view of the KdV theory, to which one passes by setting $u=z_x$, equation 
(7.4.10) is an integro-differential equation.  Thus, if one thinks of it as 
a member of the KdV hierarchy, then one is confronted with the question of 
its utility for the KdV theory.  This point remains to be clarified.  
Another remark is in order here.  Equation (7.4.10) is a special case of the Bogoyavlenskii equation [1], namely,
$$
z_{xv}=z_{xxxt}-z_{xx}z_t -2z_x z_{xt},  \leqno (7.4.11)
$$
where $v$ is the third independent variable.The factorization problem for this equation was formulated in [21].

Equation (7.4.10) is one of many reductions of (7.4.11), namely the one corresponding to the symmetry generator $\partial _v$ ($z$ is $v$ independent).  Another reduction which has been studied is the one corresponding to the symmetry generator $\partial _t -\partial_ v$.  
The resulting equation was studied by Hirota and Satsuma 
[2] as a 
model for shallow water waves.  In particular the aforementioned authors 
found its soliton solutions.

We would like to conclude the discussion of this section with the remark that, since the time the first draft of this paper was written, a theory of `negative` potentials has been further advanced by G. Haak [29].

\medskip
\vfill
\eject
\noindent
{\bf Appendix A:  Proofs of Propositions (3.3.1) and (3.5.2)}
\medskip
\noindent
\Proposition {{\bf (3.3.1)}.\quad Let $ P = P_X$ be a parabolic subgroup of $G$ and
$w\in W$.  Then
\item{(a)}  $P = P_X = \circG _X\, Q^+_X = Q^+_X \circG _X\cong \circG _X\times Q^+_X. $
\item{(b)}  The stabilizer $ U^w_-$ in $U_- $ of $wP\in G/P$ is $ U^w_- =
U_-\cap wPw^{-1}$.
Moreover, $ \dim U^w_- < \infty $.
\item{(c)} There exists a closed subgroup $ V^w_-$ of $U_-$ such
that group
multiplication induces a diffeomorphism $ U_-\cong U^w_- \times V^w_-$.}
\medskip

\noindent
\Proof .\quad (a)   First we note that $\gg ^{(0)}$ and $\gg ^{(+)}$ are 
closed complementary subalgebras of $\gp$.  Let $G^{(0)}$ and $G^{(+)}$ denote
the corresponding integral subgroups of $P$.  Arguing with the gradings as before, we see $Ad\,(G^{(0)} \cap G^{(+)})=\{I\}$.  But this implies $G^{(0)} \cap G^{(+)}=\{I\}$ as in the proof of Theorem 2.4.1.  Therefore  $P = P_X=
G^{(0)} \,G^{(+)}\cong G^{(0)} \times G^{(+)} $ by Corollary 3.1.4.

To prove $P = P_X=\circG _X\, Q^+_X=Q^+_X \circG _X\cong \circG _X\times Q^+_X $ it now suffices to show $G^{(0)}\cong \circG _X \times A_Q$, where $A_Q=H\cap
Q^+_X$.  We want to apply Corollary (3.1.4).  To this end we note that 
$\circgg$ and $\frak a  _Q$ are closed complementary subalgebras of $\gg ^{(0)}$.  Therefore, it suffices to prove $\circG _X \cap  A_Q=\{I\}$.  In view of the Iwasawa
decomposition [11, \S 5] of $G\cong K\times A\times U_+$ and $\circG _X
\cong K_X \times \circA \times (\circG _X)_+$ we see that it suffices to show
$A_Q \cap \circA =\{I\}$.  We know that $[\circgg _X,\frak a _Q]=0$, so
$Ad\, A_Q\vert \circG _X =I$.  Therefore for $g \in A _Q \cap \circA$ we have
$Ad\, g \vert \circG _X =I$.  Since $g\in \circA \in \circG _X$, $g=I$ follows.

\item {(b) }  The first statement is easy to verify.  To see that $\dim
U^w_- <\infty $ it suffices
to note $ l(w) = \#\{\alpha\in\triangle _+ ;\ w(\alpha ) < 0\}$.

\item {(c) }  The following proof is an adaptation of [12;8.6].  
Set $ \gu ^w_{\, -} = \gu _{\, -} \cap w\gp  w^{-1}$.  In
the overlaying Kac-Moody
algebra we see that $ \gu ^w_-$ is invariant under the action of the Cartan
algebra.  Therefore $
\gu ^w_{\, -}$ is a direct sum of root spaces.  Moreover, $\gv ^w_{\, -} =
\gu_{\, -} \cap w \gp
^{opp} w^{-1}$, $ \gp ^{opp}$ being the opposite 
parabolic, is
a closed subalgebra of
$ \gu_{\, -}$ such that $ \gu_{\, -} = \gu ^w_{\, -}\oplus \gv^w_{\, -}$.
We note that $\gp ^{opp}$ can be obtained by applying the Chevalley involution
$\omega$ of $\gg$ to $\gp$ [10;\S 1.3].  Therefore, by [23,
Ch.3, \S 6, Theorem 2], there exists a connected subgroup $ V^w_{-}$
of $U_-$ with Lie
algebra $ \gv ^w_{\, -}$.  More precisely, we have the following closed,
connected subgroups
$$
U^w_- = U_- \cap w\, Pw^{-1}\quad\hbox{and}\quad V^w_- = U_-\cap w\,
P^{opp} w^{-1}
$$
with Lie algebra $ \gu ^w_{\, -}$ and $ \gv ^w_{\, -}$ respectively.  Let $
u^{(k )}_-$ denote all
elements in $ \gu _{\, -}$ of sufficient high degree $\ge k$.  Then $
\gu^{(k )}_{\, -}\subset
\gv ^w_{\, -}$ and $ \gu ^w_{\, -}$ is contained in the natural complement
of $ \gu ^{(k)}_{\,
-}$ in $ \gu_{\, -}$. Moreover, $\gu _-^k$ is an ideal in $\gu_-$, whence 
$\hat \gu_-=\gu_-$ $mod\gu_-^{(k)}$ is a finite dimensional, nilpotent,
Lie algebra, with complementary subalgebras $\hat \gu_-^w$ and $\hat \gv_-^w$.
  It is not difficult to see that for the corresponding groups $\hat U_-^w\,
,\hat U_+ ^w$ and $\hat V_-^w$ we have $\hat U_- =\hat U_-^w\hat V_-^w$.  Hence $ U_-= U_-^w\,V_-^w$ and the multiplication map $m:\quad 
U_-^w\times V_-^w\rightarrow U_-$ is surjective.  It is well known that $m$ is everywhere regular.  Therefore it suffices now to show that $m$ is injective.  This is equivalent with $U_-^w\cap V_-^w=\{I\}$.  
Let $g\in \, U_-^w\cap V_-^w$.  
Then we map $g$ to $\hat g\in \hat U_-^w \cap \hat V_-^w$.  Since 
$\hat U_-$ is finite dimensional we know $\hat g=\exp(\hat u)=\exp(\hat v),\,
u\in \gu _-^w\, ,\,v\in\gv_-^w$.  Moreover, $\hat u_-$ has a faithful representation consisting of nilpotent matrices.  Hence $\hat u =\hat v$; i.e. 
$u=v+q$, $q\in\gu _-^{(k)}$.  But this implies $u=0$, whence $g\in U_-^{(k)}$.  Since we can choose $k$ arbitrary large, $g=I$ follows.
\medskip
\noindent
\Proposition {{\bf (3.5.2).}\quad For every $w\in W/W_X$ we have
$$
\overline {C_X (w)} = \bigcup _{M} C_X (w^{\prime}) 
$$
where $ \cal M = \{ w^{\prime}\in W/W_X,\ w^{\prime} \succeq w\}$.
}\medskip

\noindent
\Proof .\quad At first we follow  [24].
 Let $w^{\prime}$ be $X$-reduced and $w^{\prime} \succeq w$.  Then for some
$i$
$$
 w^{\prime} = w_1 r_iw_2,\ \ \hbox{where}\ \ w_1 (\alpha _i
) > 0\ \ \hbox{and}\ \
w^{-1}_2 (\alpha _i) > 0\, \hbox{ and } w=w_1 w_2.
$$
$$
\hbox{Then}\quad B_- w_1w_2 P= B_- (w_1 U_{-\alpha _i} w^{-1}_1) w_1w_2
(w^{-1}_2 U_{\alpha _i}
w_2)P= B_-w_1 U_{-\alpha _i} H_i U_{\alpha _i}w_2 P\ .
$$
We know that the closure of $ U_{-\alpha _i} H_i U_{\alpha _i}$, where $H_i=
\exp\,{\C}h_i$,  is a
subgroup $G_i$ of $G$ which is
isomorphic with $ SL(2,\C )$.  In particular, $ r_i\in G_i$, whence
$$
\eqalign{
B_- w^{\prime}P&= B_- w_1 r_i w_2 P\subset B_-w_1 \overline{U_{-\alpha _i} H_i
U_{\alpha }} w_2 P\cr
&\subset\overline{B_- w_1 U_{-\alpha _i} H_i U_{\alpha } w_2 P} \cr
&= \overline{B_- w_1 w_2 P}\ .
}$$
This proves $C_X(w^{\prime})\subset \overline {C_X(w)}$.  Conversely,
$\overline{C_X(w)} =\bigcup _{w^{\prime} \in J}C_X(w^{\prime})$, where $J$ is 
some 
subset of the set $W^{\prime}$ of $X$-reduced elements.  Assume first
that $X=\emptyset$, i.e. $B=P_X$.  Then we consider the representation of 
$\pi$ of $\gg$ in $gl_{res}$ described in [12;Ch.6].  Our assumptions on the weights 
$w$ imply that $\pi $ is continuous.  In particular, 
$$
\pi(\overline{C_{\emptyset}})\subset \overline{\pi(C_{\emptyset})}.
$$
From [12,Proposition (7.3.3)] we can now conclude (see also [12,
 chap.8,
in particular Theorem (8.7.2)]) that $w^{\prime} \succeq w$ for all $w ^{\prime}
\in J$.  Consider now the general case.  We note that
$$
C_X (w) = \bigcup _{{\tilde w}\in W_X} 
C_{\emptyset} (w {\tilde w}). 
$$
It is important to note that here $W_X$ is finite.  Therefore,  
$\overline{C_X(w)}=\bigcup _{{\tilde w}\in W_X}
\overline{C_{\emptyset} (w {\tilde w})}$.  This implies 
$$
\overline{C_X(w)}=
\bigcup _{{\tilde w}\in W_X}\, \bigcup _{w^{\prime}\succeq w \tilde {w}}
C_{\emptyset} (w^{\prime})=\bigcup _{w^{\prime} \succeq w \atop 
  X-reduced}
C_X(w^{\prime}). 
$$ \blsq
\vfill
\eject
\noindent
{\bf Appendix B:  Injectivity Problems}
\medskip

In the following we prove two theorems concerning the
map $ad\ E$ and its
restriction to $\frak g ^{(0)}$.  
\medskip

\noindent
\Lemma {{\bf B.1.1.}\quad $Ker\, ad\, E = Ker\, ad\, E_{-1}$\ \ for any\ \
$ E_{-1}\in (Ker\, ad\,
E)_{-1}\backslash \{ 0\}$.
}\medskip

\noindent
\Proof .  Using the Chevalley involution $\omega$, we get that 
$\omega (Ker\,ad\, E) \subset Ker \, ad\, E _{-1}$.  Moreover, by Proposition
14.3 a in [10] we get that $\omega (Ker\,ad\, E)\quad =Ker\,ad\, E $. So 
$Ker\,ad\, E \subset Ker \, ad\, E _{-1}$.  By the same argument we get $Ker\,
ad\, E_{-1} \subset Ker\, ad\, E$.  \blsq
\medskip

\noindent
\Theorem {{\bf B.1.2.}\quad Let $\frak g $ be an affine Kac-Moody algebra
and $ \frak p \not= \frak g$ a
standard parabolic subalgebra of $\frak g$.  Then:  $Ker\, ad\, E\cap \frak
g^{(0)} = \{ 0\}$.
}\medskip

\noindent
\Proof .\quad The proof is broken down into nine subclaims:
\medskip
\noindent
$Claim\, 1:Ker\, ad\, E\cap \frak g _0 =\{ 0\}\ . $ 
\medskip

\noindent
Indeed the integer $0$ is never an exponent [6; \S 14].
\medskip
\noindent
$Claim\, 2:
Ker\, ad\, E \cap \frak g ^{(0)}\subseteq Ker\, ad\, E^{(0)} \cap \frak g
^{(0)}\ . 
$
\medskip

\noindent
Let $ S\in \frak g^{(0)}$ such that $ [S,E] = 0$.  Since $E =
E^{(0)} + E^{(1)},\ [S,
E^{(0)}] + [S, E^{(1)}] = 0$.  The first term is in $\frak g^{(0)}$,
whereas the second in $\frak g
^{(0)} $, thus $[S,E^{(0)}] = 0$.
\medskip
\noindent
$Claim\, 3: \hbox{The map}\quad ad\, E^{(0)} : \frak g^{(0)} \to \frak
g^{(0)}\quad\hbox{is nilpotent.}$
\medskip

\noindent
Clear, since $ cdeg\, E^{(0)}=1$ and $\frak g^{(0)}$ is 
finite dimensional.
\medskip

\noindent
$Claim\, 4:\hbox{There are elements}\quad F^{(0)},\ H^{(0)}\in \frak g
^{(0)}\quad\hbox{such that}\quad \frak Q
:\, = {\Bbb C}F^{(0)}{\Bbb C}\oplus  H^{(0)} \oplus {\Bbb C} 
E^{(0)}
\hbox{is isomorphic to}\,sl(2,{\Bbb C})\hbox{, i.e.}\,
H^{(0)} = [E^{(0)},\ F^{(0)}],\ [H^{(0)}, E^{(0)}] = 2E^{(0)},\
[H^{(0)}, F^{(0)}] = -2F^{(0)}\ .$
\medskip

\noindent
Since $ ad E^{(0)}$ is nilpotent, this follows from the Theorem
of Jacobson-Morozov.
Here, however, we can define the elements $ F^{(0)}$ and $ H^{(0)}$ 
explicitly.
Let 
$$
\eqalign{
F^{(0)} :\, = \sum c_i f_i,\ \ \hbox{then}\ \ H^{(0)} &=
\sum c_i [e_if_i] = \sum
c_i h_i\ ,\cr
[H^{(0)}, E^{(0)}] = \sum c_i [h_i,e_j] &= \sum c_ia_{ij} e_j\cr
\hbox{and}\ \ [H^{(0)},\ F^{(0)}] = \sum c_i c_j [h_i,f_j] &= -\sum _{j}
c_j \left(\sum _i c_i
a_{ij}\right) f_j,}
$$
where $(a_{ij}) = :A $ is the Cartan matrix of ,
$\tilde{\frak g}^{(0)} :=
[\frak g ^{(0)},\ \frak g ^{(0)}]\ ,$
the semisimple part of $\frak g^{(0)}$.  Since $ \tilde{\frak g}^{(0)}$
is finite dimensional,
$A$ is an invertible matrix, whence the system of equations
$$
\sum _j \left(\sum_i (c_ia_{ij})\right) e_j = \sum _j 2e_j
$$
has a unique solution.  It is easy to verify that this gives $ F^{(0)}$ and $
H^{(0)}$ as required.
\medskip
\noindent
$Claim \, 5: \alpha (H^{(0)}) = 2(ht\,\alpha )\quad\hbox{for every root }
\,
\alpha \, \hbox{of}\,\frak g ^{(0)}. 
$
\medskip

\noindent
This statement is a direct consequence of $[H^{(0)}, e_i] = 2e_i,\ [H^{(0)},
f_i] = -2fi$.
\medskip
\noindent
$Claim \, 6: 
Ker\, ad\, E^{(0)} \cap \frak g^{(0)}_- = \{ 0\}\ . $
\medskip

\noindent
Since $\frak g^{(0)}$ is an $\frak a  $-module, by Weyl's Theorem
it is the direct
sum of a finite number of irreducible $\frak a $-modules:
$$
\frak g^{(0)} = V_1 \oplus \cdots \oplus V_r\ .
$$
Consider one of these irreducible $\frak a  $-module $ V=V_s$.  
Since $\frak a \ \ \cong
s\ell (2, \Bbb C)$, there is a basis of $V$ consisting of eigenvectors of $
ad\, H^{(0)}: v_{-m},
v_{-m+2}, \cdots , v_{m-2}, v_m$ such that $ [H^{(0)} , v_j] = j\cdot v_j$.
From (5) we see that $m$ is always an even number, thus $ \dim V$ is odd.
Since also $ [E^{(0)}, v_j] = v_{j+2}$, the kernel of $ad\, E^{(0)}$ in $V$
is spanned by the
highest-weight vector $v_m$.  Since, by (5), this vector has a nonnegative canonical
degree we
conclude that $Ker\, ad\, E^{(0)} \cap \frak g^{(0)}_{-} = \{ 0\}$\ .
\medskip
\noindent
$Claim \, 7: Ker\, ad\, E\cap \frak g^{(0)}_- = \{ 0\}\ .   
$
\medskip

\noindent
This is a direct consequence of (2) and (6).
\medskip

\noindent
$Claim \, 8 : Ker\, ad\, E_{-1} \cap \frak g^{(0)}_+ = \{ 0\}\ .  $
\medskip

\noindent
Copy the proof of (7), interchanging $ e_i$ and $ f_i$.
\medskip
\noindent
$ Claim \, 9: Ker\, ad\, E\cap \frak g ^{(0)}_+ = \{ 0\}\ .  $
\medskip

\noindent
This claim follows from (8) and Lemma 6.4.1.

Altogether, (1), (7) and (9) yield  the claim.\blsq

\vfill
\eject

\centerline{\bf References}
\smallskip

\item{1. } Bogoyavlenskii, O.I., Breaking solitons in 2+1 - dimensional
integrable equations, {\it Russian Math. Surveys} {\bf 45}, (1990), no.4, 1-86.
\medskip
\item{2. } Hirota, R., Satsuma, J., N-soliton solutions of Model Equations for Shallow Water Waves, {\it J.Phys.Soc.Japan} {\bf 40}, (1976), 611
\medskip
\item{3. }  Drinfel'd, V.G.,and Sokolov,V.V.,
Lie algebras and equations of Korteweg-deVries type,
{\it Journal of Soviet Mathematics}, {\bf 30}, (1985), 1975-2036.  
\medskip 
\item{4. } Wilson, G., The modified Lax and two-dimensional Toda lattice equations associated with simple Lie algebras, {\it Ergodic Theory and Dynamical Systems}, {\bf 1}, (1981), 361-380.
\medskip
\item{5. } Sato M., Soliton equations as dynamical systems on infinite dimensional Grassmann manifolds, {\it RIMS Kokyuroku} {\bf 439}, (1981), 30-46
\medskip
\item{6. } Segal, G., Wilson, G., Loop groups and equations of KdV type, {\it Publ.Math. I.H.E.S.} {\bf 63}, (1984), 1-95
\medskip
\item{7. } Garland, H., The arithmetic theory of loop groups, {\it Publ.Math. I.H.E.S.} {\bf 52}, (1980), 5-136.
\medskip
\item{8. } Peterson, D.H., Kac, V.G., Infinite flag varieties and conjugacy theorems, {\it Proc. Natl. Acad. Sci. USA} {\bf 80}, (1983), 1778-1782.
\medskip
\item{9. } Tits, J., Resum\'{e} de cours, Coll\`{e}ge de France, Paris, (1981)
\medskip
\item{10. } Kac, V., {\it Infinite dimensional Lie algebras}, Cambridge University
Press, (1985). 
\medskip
\item{11. } Goodman, R., Wallach, N. R., Structure and 
unitary cocycle representations of loop groups and the group of diffeomorphisms of the circle, {\it J.Reine Angew. Math.}, {\bf 347}, (1984), 69-133.
\medskip 
\item{12. } Pressley, A.,Segal, G.,{\it Loop groups and their representations},
Oxford University Press, (1986).
\medskip  
\item{13. } Dorfmeister, J., Szmigielski, J., Riemann-Hilbert Factorizations and Inverse Scattering for the AKNS-Equations with $L ^1$-Potentials I, 
{\it Publ. RIMS, Kyoto University} {\bf 29}, (1993), 911-958.  
\medskip 
\item{14. } Szmigielski, J., Infinite dimensional manifolds with translational
symmetry and nonlinear partial differential equations, Ph.D Dissertation, University
of Georgia-Athens, GA, (1987).
\medskip
\item{15. } Dorfmeister, J., Wu, H., Constant mean curvature surfaces and loop
groups, {\it J.reine angew. Math.} {\bf 440}, (1993), 43-76. 
\medskip
\item{16 } Dorfmeister, J., Haak, G., 
On symmetries of constant mean curvature surfaces, Part II: Symmetries in 
a Weierstrass type representation, to appear
\medskip 
\item{17} Dorfmeister, J., Haak, G., On constant mean curvature surfaces with
periodic metric, to appear in Pacific J.Math.
\medskip 
\item{18. } Szmigielski, J., On the soliton content of the self dual Yang-Mills
equations, {\it Phys.Lett.A} {\bf 183}, (1993), 293-300.
\medskip
\item{19. } Dorfmeister, J., Banach Manifolds of Solutions to Nonlinear Partial Differential Equations, and Relations with Finite-Dimensional Manifolds,
{\it Proceedings of Symposia in Pure Mathematics} {\bf 54}, (1993), 121-139.
\medskip
\item{20} de Groot, M.F., 
Hollowood, T.J., and Miramontes, J.L., Generalized Drinfeld-Sokolov hierarchies
{\it Comm. Math. Phys.} {\bf 145}, (1992), no.1, 57-84.
\medskip
\item{21} Burroughs, N.J., de Groot, M.F., 
Hollowood, T.J., and Miramontes, J.L., Generalized Drinfeld-Sokolov hierarchies. II. The Hamiltonian structures, 
{\it Comm. Math. Phys.} {\bf 153}, (1993), no.1, 187-215. 
\medskip
\item{22} Feh\`{e}r, L., Harnad, J., and Marshall, I., Generalized Drinfeld-Sokolov reductions and KdV type hierarchies,
{\it Comm. Math. Phys.} {\bf 154}, (1993), no.1, 181-214. 
\medskip  
\item{23. } Bourbaki, B., {\it Groupes et alg\`{e}bres de Lie}. Ch. 1, Hermann, Paris, 1964; Ch. 2, et 3, Hermann, Paris, 1972; Ch.4,5 and 6, Hermann, Paris, 1968;
Ch. 7 et 8, Hermann, Paris, 1975; Ch. 9, Masson, Paris, (1982).
\medskip
\item{24.} Kac, V.G., Peterson, D. K., Regular functions on certain infinite-dimensional groups, in {\it Arithmetic and Geometry}, 
Progress in Math. {\bf 36}, 141-166, Birkh\"{a}user, Boston, (1983).
\medskip
\item{25. } Helgason, S., {\it Differential geometry, Lie groups and symmetric spaces.} Academic Press, New York, (1978).
\medskip
\item{26. } Beals, R., Deift, P., Tomei, C., {\it Direct and Inverse Scattering
 on the Line.} Amer.Math. Soc.., Providence, RI, (1992).
\medskip
\item{27. } Gradl, H., A result on exponents of finite-dimensional simple
Lie algebras and its application to Kac-Moody algebras, {\it Linear Algebra 
Appl} {\bf 233}, (1996), 189-206. 
\medskip
\item{28. } Sattinger, D.H., Szmigielski, J., Factorization and the dressing method for the Gelfand-Dikii hierarchy, {\it Physica D} {\bf 64}, (1993), 1-34.
\medskip
\item{29. } Haak, G., Negative flows of the potential KP-hierarchy, 
{\it Trans. Amer. Math. Soc.} {\bf 348}, (1996), no.1, 375-390.
\medskip

\vfill
\end